\documentclass[aps,prb,twocolumn,superscriptaddress]{revtex4-1}
\usepackage{graphicx}
\usepackage{bm}
\usepackage{amsmath,amssymb,amsfonts,bm}
\usepackage{epsfig}
\usepackage{epstopdf}
\usepackage{dcolumn}
\usepackage{grffile}
\usepackage{verbatim}
\usepackage{mathrsfs}
\usepackage{appendix}
\usepackage{xcolor}
\usepackage{txfonts}
\usepackage[normalem]{ulem}
\usepackage[colorlinks=true,linkcolor=blue,citecolor=blue, urlcolor=blue]{hyperref}

\definecolor{alizarin}{rgb}{0.82, 0.1, 0.26}

\begin{document}

\title{
Collective effects in an incompressible electronic liquid
}
\date{\today}

\author{Jian-Jian Miao}
\affiliation {Department of Physics, The Chinese University of Hong Kong, Shatin, New Territories, Hong Kong, China}

\author{Hui-Ke Jin}
\affiliation {Department of Physics TQM, Technische Universit\"{a}t M\"{u}nchen, James-Franck-Straße 1, D-85748 Garching, Germany}

\author{Yi Zhou}
\email{yizhou@iphy.ac.cn}
\affiliation {Institute of Physics, Chinese Academy of Sciences, Beijing 100190, China}
\affiliation{Songshan Lake Materials Laboratory, Dongguan, Guangdong 523808, China}
\affiliation{Kavli Institute for Theoretical Sciences $\&$ CAS Center for Excellence in Topological Quantum Computation, University of Chinese Academy of Sciences, Beijing 100190, China}

\begin{abstract}
Starting from the Landau's kinetic equation, we show that an electronic liquid in $d=2,3$ dimensions depicted by a Landau type effective theory will become incompressible on condition that the Landau parameters satisfy either (i) $1+F_{1}^{s}/d=0$ or (ii) $F_{0}^{s}\to{}+\infty$. The condition (i) is the Pomeranchuk instability in the current channel and suggests a quantum spin liquid (QSL) state with a spinon Fermi surface; while the condition (ii) means that the strong repulsion in the charge channel and leads to a conventional charge and thermal insulator.
In the collisionless regime ($\omega\tau\gg{}1$) and the hydrodynamic regime ($\omega\tau\ll{}1$), the zero and first sound modes have been studied and classified by symmetries, including the longitudinal and transverse modes in $d=2,3$ and the higher angular momentum modes in $d=3$. The sufficient (and/or necessary) conditions of these collective modes have been revealed. It has been demonstrated that some of these collective modes will behave in quite different manners under two incompressibility conditions (i) or (ii).
Possible nematic QSL states and a hierarchy structure for gapless QSL states have been proposed in $d=3$.
\end{abstract}

\maketitle

\tableofcontents

\section{Introduction}
The past few decades have witnessed a proliferation of research interests in quantum spin liquids (QSLs). As a novel phase of matter, QSL was proposed as a pristine Mott insulator which carries an odd number of electrons per unit
cell and hosts paramagnetic ground states~\cite{Anderson73,Anderson87,Lee08,Balents10,QSLRMP,Savary2016,Knolle2019,Broholm2020}. Such a ``featureless" Mott insulator is characterized by fractional spin excitations and long-ranged quantum entanglement, which are beyond Landau's symmetry-breaking paradigm and naturally associated with the celebrated idea of resonating valence
bond (RVB)~\cite{Anderson73}. Moreover, high-temperature superconductivity has been considered to arise from doping such an RVB state~\cite{Anderson87,rmp06}.

While a decisive experiment for identifying fractional spin excitations is still missing, this exotic phase of matter has been well established theoretically via exactly solvable models together with other analytical and numerical analyses~\cite{QSLRMP,Savary2016,Knolle2019,Broholm2020}.
Mathematically, the Lieb-Schultz-Mattis-Oshikawa-Hastings theorem~\cite{LSM,LSM2,Oshikawa2000,Hastings2004} states that when the lattice translational symmetry and the $O(2)$ [or $U(1)$] spin rotation conservation are preserved, the ground state of a quantum spin system is either gapless or gapped with topological degeneracy, if the $U(1)$ charge (i.e., the spin quantum) per unit cell is a half-odd integer. This puts a strong constraint on QSL ground states and low energy excitations on top of them: a QSL state is either gapped with topological order, or gapless. In contrast to gapped QSLs that have been extensively studied by topological orders and various numerical tools, systematic studies on gapless QSLs are relatively rare. 

On the experiment side, a variety of QSL candidates has been found in realistic materials, and many, if not most of them exhibit gapless features~\cite{QSLRMP,Broholm2020}. Indeed, gapless QSLs are expected to display many-body electronic excitations inside the Mott charge gap and manifest themselves in various experiments that include but are not limited to thermodynamic and magnetic measurements, thermal transports, neutron scattering, nuclear magnetic resonance (NMR), muon spin resonance ($\mu$SR), optical conductivity, Raman scattering, thermal Hall effect, sound attenuation, and angle-resolved photo-emission spectroscopy (ARPES)~\cite{QSLRMP}.

Most theoretical studies on QSLs and/or relevant metal-to-Mott-insulator transition start with a Mott insulator, while the dual approaches from the metallic side are relatively less known~\cite{Mross11,Zhou13}. 
On the other hand, metallic states have been well understood by using the classic Landau's Fermi liquid theory~\cite{LLBook,FLBook1,FLBook2}. Collective modes in a Fermi liquid represent a kind of elementary excitations, which involves a coherent motion of the system as an entirety. In this paper, we shall study electronic collective modes in a gapless QSL state with a spinon Fermi surface that are naturally conveyed from quasiparticle Fermi surface in an incompressible (i.e., insulating) limit.
In particular, we shall focus on density fluctuations of chargeful quasiparticles on the Fermi surface, and study these collective modes in the gapless QSL state in three dimensions (3D) and two dimensions (2D).

The rest of this paper is organized as follows: In Section~\ref{sec:revisit}, we revisit a Landau type effective theory for gapless QSLs. In Section~\ref{sec:LKE}, we introduce Landau's kinetic equation and formulate collective modes for a spherical or circular Fermi surface. In Section~\ref{sec:ZS3D} and Section~\ref{sec:FS3D}, we study zero sound modes and first sound mode in 3D respectively. In Section~\ref{sec:2D}, collective modes in 2D have been studied in parallel with those in 3D. Section~\ref{sec:sum} is devoted to summary and discussions.

\section{An effective theory for QSL: revisit}\label{sec:revisit}

Landau's Fermi liquid theory describes a system of interacting fermions, which evolves from a non-interacting Fermi gas in a continuous way, namely, there is no phase transition as long as fermion-fermion interactions are turned on adiabatically. Thus, the low energy excited states of interacting fermions remain one-to-one correspondence to those of a non-interacting Fermi gas, which are called ``quasiparticles'' and labelled by fermion occupation number. Landau's Fermi liquid theory is built with these quasiparticles and interactions between them~\cite{LLBook,FLBook1,FLBook2}.

An effective theory was proposed for gapless $U(1)$ QSLs with a spinon Fermi surface in Ref.~[\onlinecite{Zhou13}], which takes almost the same form as Landau's Fermi liquid theory. The building blocks in the effective QSL theory are assumed to be ``chargeful'' quasiparticles, even though these quasiparticles are \emph{not} protected by adiabaticity.
The quasiparticles in these QSLs are labeled by the same occupation numbers as free fermions. Consequently, by defining $\delta n_{\mathrm{p\sigma }}=n_{\mathrm{p\sigma }}-n_{\mathrm{p\sigma }}^{0}$ the departure of the fermion distribution function from the ground state distribution $n_{\mathrm{p}}^{0}=\theta (-\xi _{\mathrm{p}})$,
the excitation energy $\Delta E=E-E_{G}$ for these states are also given by a Landau-type expression as follows,
\begin{equation}
\Delta E=\sum_{\mathrm{p\sigma }}\xi _{\mathrm{p}}\delta n_{\mathrm{p\sigma }}+\frac{1}{2}\sum_{\mathrm{pp}^{\prime }\sigma \sigma ^{\prime }}f_{\mathrm{pp}^{\prime }}^{\sigma \sigma ^{\prime }}\delta n_{\mathrm{p}\sigma }\delta{}n_{\mathrm{p}^{\prime }\sigma ^{\prime }}+O(\delta n^{3}),  \label{eq:F}
\end{equation}
where $\xi_{\mathrm{p}}=\frac{\mathrm{p}^{2}}{2m^{\ast }}-\mu$ is the single-particle energy measured from the chemical potential $\mu$, $m^{\ast}$ is the effective mass, $\sigma$ ($\sigma^{\prime}$) denotes spin index, and $f_{\mathrm{pp}^{\prime}}^{\sigma\sigma^{\prime}}$ is the interaction energy between excited quasiparticles.
Then the single quasiparticle excitation energy reads
\begin{equation}
\varepsilon_{\mathrm{p\sigma}}\equiv\frac{\delta{}\Delta{}E}{\delta{}n_{\mathrm{p}\sigma}}=\frac{\mathrm{p}^{2}}{2m^{\ast }} -\mu + \sum_{\mathrm{p}^{\prime } \sigma ^{\prime }}f_{\mathrm{pp}^{\prime }}^{\sigma \sigma ^{\prime }}\delta n_{\mathrm{p}^{\prime }\sigma ^{\prime }}. \label{eq:ep}
\end{equation}
Here a spherical Fermi surface is assumed for simplicity and $f_{\mathrm{pp}^{\prime }}^{\sigma \sigma ^{\prime }}$ can be written in terms of spin symmetric and spin anti-symmetric components $f_{\mathrm{pp}^{\prime
}}^{\sigma \sigma ^{\prime }}=f_{\mathrm{pp}^{\prime }}^{s}\delta _{\sigma\sigma ^{\prime }}+f_{\mathrm{pp}^{\prime }}^{a}\sigma \sigma ^{\prime }$, where $f_{\mathrm{pp}^{\prime }}^{s(a)}$ depends only on the angle $\theta $ between $\mathrm{p}$ and $\mathrm{p}^{\prime }$. In 3D, it can be expanded as
\begin{subequations}\label{eq:flsa}
\begin{equation}
f_{\mathrm{pp}^{\prime }}^{s(a)}=\sum_{l=0}^{\infty}f_{l}^{s(a)}P_{l}(\cos \theta),
\end{equation} with $P_{l}$'s the Legendre polynomials and in 2D, it reads
\begin{equation}
 f_{\mathrm{pp}^{\prime}}^{s(a)}=\sum_{l=0}^{\infty}f_{l}^{s(a)}\cos (l\theta).
\end{equation}
\end{subequations}
The Landau parameters, defined by $F_{l}^{s(a)}=N(0)f_{l}^{s(a)}$, provide a dimensionless measures of the strengths of the interactions between quasiparticles, where $N(0)$ is the Fermi surface density of states. The low temperature properties of the QSLs are completely determined by the effective mass $m^{\ast}$ and the interaction $f_{\mathrm{pp}^{\prime}}^{\sigma\sigma^{\prime}}$ (or $F_{l}^{s(a)}$) as those of the Fermi liquid theory.

Extra conditions have to be imposed to ensure that the excitations in the effective QSL theory carry zero charge and finite entropy, whereas quasiparticles are chargeful.
It turns out that an electrical insulating but simultaneously thermal conducting state, i.e., a QSL with a spinon Fermi surface, can be achieved by putting a constraint on the Landau parameter $F_1^s$. This can be seen as follows. First, the charge current $\mathbf{J}$ carried by quasiparticles is given by 
\begin{subequations}
\begin{equation}
\mathbf{J}=\frac{m}{m^{\ast }}\left(1+\frac{F_{1}^{s}}{d}\right)\mathbf{J}^{(0)},
\label{Je}
\end{equation}
where $\mathbf{J}^{(0)}$ is the charge current carried by the corresponding non-interacting fermions and $d$ is the dimensionality.
Meanwhile, the thermal current $\mathbf{J}_{Q}$ is only renormalized by the effective mass and reads 
\begin{equation}
\mathbf{J}_{Q}=\frac{m}{m^{\ast }}\mathbf{J}_{Q}^{(0)},  \label{JQ}
\end{equation}
where $\mathbf{J}_{Q}^{(0)}$ is the corresponding thermal current carried by non-interacting electrons. Therefore, when
\begin{equation}
1+\frac{F_{1}^{s}}{d}\rightarrow 0~\text{while}~\frac{m^{\ast }}{m}\neq{}0, \label{eq:F1s}
\end{equation}
\end{subequations}
we have $\mathbf{J}\rightarrow 0$ and $\mathbf{J}_{Q}\neq 0$, suggesting that the electronic system is in a special state where spin-1/2 quasiparticles do not carry charge due to interaction but they still carry entropy.

It is crucial that $F_{1}^{s}$ is independent of $\frac{m^{\ast}}{m}$ for this mechanism to work. Indeed, it cannot happen in translational invariant systems, where the charge current carried by quasiparticles is not renormalized since $\frac{m^{\ast}}{m}=1+\frac{F_{1}^{s}}{d}$.
However, there does exist a way out for electrons in crystals, where the Galilean invariance is lost. So that $\frac{m^{\ast }}{m}\neq 1+\frac{F_{1}^{s}}{d}$ and $\mathbf{J}$ is renormalized by interactions. 

Note that the quasiparticles should be distinguished from elementary excitations in the effective theory. The quasiparticles are chargeful and described by $\delta n_{p\sigma }$, whereas elementary excitations are eigenstates of Landau's kinetic equation with charges renormalized by $1+F_{1}^{s}/d$ and become zero in the limit $\mathrm{q}=0$ and $\omega =0$.

\section{Landau's kinetic equation and collective modes}\label{sec:LKE}

Landau's kinetic equation, which is more than a Boltzmann equation, can be utilized to study both equilibrium and non-equilibrium properties of Fermi liquids. It describes the change of quasiparticle distribution function $n_{\mathrm{p\sigma }}(\mathrm{r},t)$ in the phase space, 
\begin{equation}
\frac{\partial n_{\mathrm{p}}(\mathrm{r},t)}{\partial t}+\nabla _{\mathrm{p}}\varepsilon _{\mathrm{p}}(\mathrm{r},t)\cdot \nabla _{\mathrm{r}}n_{\mathrm{p}}(\mathrm{r},t)-\nabla _{\mathrm{r}}\varepsilon _{\mathrm{p}}(\mathrm{r},t)\cdot \nabla _{\mathrm{p}}n_{\mathrm{p}}(\mathrm{r},t)=I[n_{\mathrm{p}^{\prime }}],  \label{LTE}
\end{equation}
where $I[n_{\mathrm{p}^{\prime }}]$ is the collision integral. This equation is considerably richer than the usual Boltzmann equation used to describe weakly interacting gas, since ``quasiparticles" are not bare particles. Instead, they are combined with characteristics from the background, i.e., other particles and/or excitations in a quantum fluid.
In particular, the difference between bare particles and quasiparticles can be viewed by adding a force term to $\nabla _{\mathrm{r}}\varepsilon _{\mathrm{p}}(\mathrm{r},t)$. To be specific, let us assume that an external scalar potential $U(\mathrm{r},t)$ is applied to the system as
\begin{equation}
\nabla _{\mathrm{r}}\varepsilon _{\mathrm{p}\sigma}(\mathrm{r},t) = \nabla _{\mathrm{r}}U(\mathrm{r},t)+\sum_{\mathrm{p}^{\prime }\mathrm{\sigma }^{\prime }}f_{\mathrm{pp}^{\prime }}^{\sigma \sigma ^{\prime }}\nabla_{\mathrm{r}} \delta{}n_{\mathrm{p}^{\prime }\sigma^{\prime }}(\mathrm{r},t).
\end{equation}
Thus, the Landau's kinetic equation allows us to study collective effects in such a quantum fluid.

The collective modes are coherent motions of the quasiparticles on the Fermi surface. For the charge sector, the collective modes are given by the density fluctuation $\delta n_{\mathrm{p}}$, which can be written as 
\begin{equation}
\delta n_{\mathrm{p}}=-\frac{\partial n_{\mathrm{p}}^{0}}{\partial \varepsilon _{\mathrm{p}}}\nu _{\mathrm{p}}. \label{eq:density1}
\end{equation}
Here $\nu _{\mathrm{p}}$ is the energy by which the density wave shifts the quasiparticle distribution in the direction of $\mathrm{p}$, and can be expanded in terms of spherical harmonics in 3D,
\begin{equation}
\nu _{\mathrm{p}}=\sum_{l}\sum_{m=-l}^{l}Y_{l}^{m}(\theta _{\mathrm{p}},\phi_{\mathrm{p}})\nu _{l}^{m}. \label{eq:nu}
\end{equation}

In this paper, we are particular interested in two types of collective modes: zero sound modes in the collisionless regime with $\omega\tau \gg 1$ and first sound modes in the opposite regime with $\omega\tau \ll 1$.

\section{Zero sound in the collisionless regime: $\omega\tau\gg{}1$}\label{sec:ZS3D}

Zero sound is density fluctuation in the collisionless regime, $\omega\tau \gg 1$. In this limit, the quasiparticles no longer have time to relax to equilibrium in one period of the sound. Thus, the liquid no longer remain in local thermodynamic equilibrium, the character of the sound propagation begins to change.

In the collisionless regime, we can safely neglect the collision integral $I[n_{\mathrm{p}^{\prime }}]$ in the kinetic equation,
\begin{equation}
\frac{\partial \delta n_{\mathrm{p}}}{\partial t}+\vec{v}_{\mathrm{p}}\cdot\nabla _{\mathrm{r}}\left( \delta n_{\mathrm{p}}-\frac{\partial n_{\mathrm{p}}^{0}}{\partial \varepsilon _{\mathrm{p}}}\delta \varepsilon _{\mathrm{p}
}\right) =0.  \label{LTEZS}
\end{equation}
The variation of local energy comes from applied scalar field $U$ as well as the quasiparticle interaction,
\begin{equation*}
\delta \varepsilon _{\mathrm{p}}(\mathrm{r},t)=U(\mathrm{r},t)+\sum_{\mathrm{p}^{\prime }}f_{\mathrm{pp}^{\prime }}^{s}\delta n_{\mathrm{p}^{\prime }}.
\end{equation*}
Assuming $U(\mathrm{r},t)=Ue^{i(\mathrm{q}\cdot \mathrm{r}-\omega t)}$ and putting the above into the kinetic equation [Eq.~\eqref{LTEZS}] leads to
\begin{equation*}
(\omega -\mathrm{q}\cdot \vec{v}_{\mathrm{p}})\delta n_{\mathrm{p}}+\frac{\partial n_{\mathrm{p}}^{0}}{\partial \varepsilon _{\mathrm{p}}}(\mathrm{q}\cdot \vec{v}_{\mathrm{p}})\left( U+\sum_{\mathrm{p}^{\prime }}f_{\mathrm{pp}
^{\prime }}^{s}\delta n_{\mathrm{p}^{\prime }}\right) =0.
\end{equation*}
Then the fluctuating modes $\nu _{\mathrm{p}}$ defined in Eq.~\eqref{eq:nu} can be solved from the following equation,
\begin{equation}
\nu _{\mathrm{p}}+\frac{\mathrm{q}\cdot \vec{v}_{\mathrm{p}}}{\omega - \mathrm{q}\cdot \vec{v}_{\mathrm{p}}}\sum_{\mathrm{p}^{\prime }}f_{\mathrm{pp}^{\prime }}^{s}\frac{\partial n_{\mathrm{p}^{\prime }}^{0}}{\partial
\varepsilon _{\mathrm{p}^{\prime }}}\nu _{\mathrm{p}^{\prime }}=\frac{\mathrm{q}\cdot \vec{v}_{\mathrm{p}}}{\omega -\mathrm{q}\cdot \vec{v}_{\mathrm{p}}}U.  \label{LTEZS1}
\end{equation}
Note that Eq.~\eqref{LTEZS1} is invariant under the rotation along the axis of $\mathrm{q}$, and this $O(2)$ rotational symmetry allows us to decouple Eq.~\eqref{LTEZS1} in accordance with the quantum number $m$ given in Eq.~\eqref{eq:nu}.

\subsection{Longitudinal ($m=0$) mode}

For simplicity, we begin with $m=0$ modes, i.e., longitudinal sound modes in such a quantum liquid, and denote the angle between $\mathrm{p}$ and $\mathrm{q}$ as $\theta _{\mathrm{p}}$. So that the expansion in Eq.~\eqref{eq:nu} can be simplified as,
\begin{equation}
\nu _{\mathrm{p}}=\sum_{l}P_{l}(\cos \theta _{\mathrm{p}})\nu _{l}, \label{eq:nu0}
\end{equation}
where $\nu_{l}\equiv\nu_{l}^{m=0}$. With the help of the  addition theorem for spherical harmonics, orthogonal relations, and Bonnet's recursion formula for Legendre polynomials, we find two sets of algebraic equations as follows,
\begin{subequations}
\begin{equation}
\frac{\nu _{l}}{2l+1}+\sum_{l^{\prime }}F_{l^{\prime }}^{s}\Omega_{ll^{\prime }}(s)\frac{\nu _{l^{\prime }}}{2l^{\prime }+1}=-\Omega_{l0}(s)U,  \label{LTEZS3}
\end{equation}
which is referred to as ``type I" hereafter, and
\begin{equation}
\nu _{l}s-\left( 1+\frac{F_{l-1}^{s}}{2l-1}\right) \frac{l}{2l-1}\nu_{l-1}-\left( 1+\frac{F_{l+1}^{s}}{2l+3}\right) \frac{l+1}{2l+3}\nu_{l+1}=\delta _{l1}U.  \label{LTEZS4}
\end{equation}
\end{subequations}
which is referred to as ``type II". Here,
\begin{equation}
s\equiv \frac{\omega }{qv_{F}}~~\text{and}~~\mu =\cos \theta _{\mathrm{p}} \label{eq:s}
\end{equation}
are two dimensionless parameters, $q=|\mathrm{q}|$ and $v_F$ is the Fermi velocity, and $\Omega _{ll^{\prime }}(s)=\Omega _{l^{\prime }l}(s)$ is defined as
\begin{equation}
\Omega _{ll^{\prime }}(s)=\frac{1}{2}\int_{-1}^{1}d\mu P_{l}(\mu )\frac{\mu}{\mu -s}P_{l^{\prime }}(\mu ). \label{eq:omega}
\end{equation}
The first few $\Omega_{ll^{\prime }}(s)$ can be found in Table \ref{tab:omega}. Note that Eq.~\eqref{LTEZS3} consists of an infinite set of equations, and can be derived from Eq.~\eqref{LTEZS4} directly and vice versa.

Then possible collective modes describing oscillations of the system can be found from the poles of the response function $\nu_{l}/U$, which depends on the variable $s$ only. Moreover, one can solve Eq.~\eqref{LTEZS3} by introducing a truncation in $F_{l}^{s}$, namely, keep finite number of $F_{l}^{s}$ and neglect others. For instance, the truncation up to $l=1$ leads to a two channel model, while the truncation up to $l=2$ gives rise to a three channel model.

\begin{table}[tb]
\caption{A list of first few $\Omega _{ll^{\prime }}(s)$ defined in Eq.~\eqref{eq:omega}. Note that for the physical situation, we always choose $s$ with positive imaginary part (can be infinitesimal). So that for a real value of $s$,
$\Omega _{00}(s)=1-\frac{s}{2}\ln \left\vert \frac{s+1}{s-1}\right\vert +\frac{i\pi s}{2}\theta (1-|s|).$}
\label{tab:omega}
\renewcommand\arraystretch{2}
\begin{tabular}{l}
\hline\hline
$\Omega _{00}=1-\frac{s}{2}\ln \frac{s+1}{s-1}$, \\
$\Omega _{10}=s\Omega _{00}$ \\
$\Omega _{11}=\frac{1}{3}+s^{2}\Omega _{00}$ \\
$\Omega _{20}=\frac{1}{2}+P_{2}(s)\Omega _{00}$ \\
$\Omega _{21}=s\left[ \frac{1}{2}+P_{2}(s)\Omega _{00}\right]$ \\
$\Omega _{22}=\frac{3}{4}s^{2}-\frac{1}{20}+P_{2}(s)^{2}\Omega _{00}=\frac{1}{5}+P_{2}(s)\Omega _{20}$\\
\hline
\end{tabular}
\end{table}

\subsubsection{Two incompressibility conditions}
Notice that the $l=0$ mode, $\nu_0$, represents the coherence motion of the Fermi surface compression and expansion. Therefore, the quantum liquid will become incompressible, i.e., acquire a charge excitation energy gap, if and only if $$\nu_0(s)={}0$$ under some conditions.

In order to find possible incompressibility conditions, let's examine the first equation of type II, say, the one with $l=0$ in Eq.~\eqref{LTEZS4}. Note that second term in the left hand side vanishes automatically, and we have
\begin{equation}
\nu _{0}s-\left( 1+\frac{F_{1}^{s}}{3}\right)\frac{\nu _{1}}{3}=0. \label{eq1:nu}
\end{equation}
This leads to an incompressibility condition:
\begin{equation}\label{eq:incomp1}
1+\frac{F_{1}^{s}}{3}=0,
\end{equation}
which is exactly the QSL condition in Eq.~\eqref{eq:F1s} for $d=3$.

On the other hand, the second equation of type II, say, the one with $l=1$ in Eq.~\eqref{LTEZS4} reads
\begin{equation}\label{eq2:nu}
\nu _{1}s-\left(1+F_{0}^{s}\right)\nu_{0}-\left( 1+\frac{F_{2}^{s}}{5}\right) \frac{2}{5}\nu_{2}=U.
\end{equation}
Therefore, another incompressibility condition is given by
\begin{equation}\label{eq:incomp2}
F_{0}^{s}\rightarrow +\infty. 
\end{equation}
This is nothing but the the strong repulsion limit in which a metal-to-conventional-insulator Mott transition occurs. 


\subsubsection{Three channel model for $m=0$ mode} 

To solve the longitudinal ($m=0$) zero sound mode, we consider a three channel model by keeping $F_{0}^{s}$, $F_{1}^{s}$, and $F_{2}^{s}$, and setting $F_{l}^{s}=0$ for $l\geq 3$. For convenience, we introduce a new set of variables $$\tilde{\nu}_{l}\equiv \frac{\nu_{l}}{2l+1},$$ and simplify Eq.~\eqref{LTEZS3} as follows,
\begin{equation}
\tilde{\nu}_{l}+F_{0}^{s}\Omega _{l0}(s)\tilde{\nu}_{0}+F_{1}^{s}\Omega_{l1}(s)\tilde{\nu}_{1}+F_{2}^{s}\Omega_{l2}(s)\tilde{\nu}_{2}=-\Omega _{l0}(s)U.  \label{ZS21}
\end{equation}
The above equation consists of a series of algebraic equations that can be classified into two categories: (i) $l\geq 3$ and (ii) $l=0,1,2$. The equations in the second category form a close set that is composed of $\tilde{\nu}_{0}$, $\tilde{\nu}_{1}$ and $\tilde{\nu}_{2}$:
\begin{equation}
\left( 
\begin{array}{ccc}
1+F_{0}^{s}\Omega_{00} & F_{1}^{s}\Omega_{01} & F_{2}^{s}\Omega_{02} \\ 
F_{0}^{s}\Omega_{10} & 1+F_{1}^{s}\Omega_{11} & F_{2}^{s}\Omega_{12} \\ 
F_{0}^{s}\Omega_{20} & F_{1}^{s}\Omega_{21} & 1 + F_{2}^{s}\Omega_{22}
\end{array}
\right) \left(
\begin{array}{c}
\tilde{\nu}_{0} \\ 
\tilde{\nu}_{1} \\ 
\tilde{\nu}_{2}
\end{array}
\right)=-U \left(
\begin{array}{c}
\Omega_{00} \\
\Omega_{10} \\ 
\Omega_{20}
\end{array}
\right).
\end{equation}

Then the mode frequency is determined by the following secular equation,
\begin{equation}\label{eq:det}
g_{3}(s)= \det\left( 
\begin{array}{ccc}
1+F_{0}^{s}\Omega_{00} & F_{1}^{s}\Omega_{01} & F_{2}^{s}\Omega_{02} \\ 
F_{0}^{s}\Omega_{10} & 1+F_{1}^{s}\Omega_{11} & F_{2}^{s}\Omega_{12} \\ 
F_{0}^{s}\Omega_{20} & F_{1}^{s}\Omega_{21} & 1 + F_{2}^{s}\Omega_{22}
\end{array}
\right) =0.
\end{equation}
Here the function $g_{3}(s)$ can be written explicitly as follows,
\begin{equation}\label{eq:g3s}
\begin{split}
g_{3}(s) = &\left( 1+\frac{F_{1}^{s}}{3}\right)\left(1+F_{2}^{s}\Omega_{22}\right)\\
&+\left[F_{0}^{s}\left( 1+\frac{F_{1}^{s}}{3}\right)+F_{1}^{s}s^{2}\right]\left[\Omega_{00}+F_{2}^{s}\left(\Omega_{00}\Omega_{22}-\Omega_{02}^2\right)\right].  
\end{split}
\end{equation}
To find a real solution $s$ to the secular equation, we notice that $g_{3}(s\to\infty)=1>0$ and
\begin{equation*}
\begin{split}
g_{3}(s\to{1})=&\left\{\left( 1+\frac{F_{1}^{s}}{3}\right)\left[\frac{3+F_{0}^{s}}{2}+\frac{F_2^{s}}{20}\left(1-3F_{0}^{s}\right)\right]\right.\\
&\left.+\frac{9}{20}\left(F_{2}^{s}-\frac{10}{3}\right)\right\}  \ln\frac{s-1}{2} +O(1).
\end{split}
\end{equation*}
Therefore, the inequality
\begin{equation*}
\left( 1+\frac{F_{1}^{s}}{3}\right)\left[\frac{3+F_{0}^{s}}{2}+\frac{F_2^{s}}{20}\left(1-3F_{0}^{s}\right)\right]+\frac{9}{20}\left(F_{2}^{s}-\frac{10}{3}\right)>0
\end{equation*}
or
\begin{equation}\label{ineq:three-m0}
\left[9+\left(1+\frac{F_{1}^{s}}{3}\right)\left(1-3F_{0}^{s}\right)\right]\frac{F_{2}^{s}}{10}+\left(1+\frac{F_{1}^{s}}{3}\right)\left(3+F_{0}^{s}\right)-3>0    
\end{equation}
gives rise to a sufficient condition for a real solution $s>1$ to the secular equation $g_{3}(s)=0$.
Note that in the limit $F_{2}^{s}\to{}0$ and $1+F_{1}^{s}/3\to{}0$, the inequality in Eq.~\eqref{ineq:three-m0} cannot be satisfied. So that a two channel model does not host a weakly damped $m=0$ zero sound mode under the incompressibility condition $1+F_{1}^{s}/3=0$. 

In general, the secular equation $g_{3}(s)=0$ must be found numerically. Using the limiting forms of $\Omega_{00}$,
\begin{equation}
\Omega_{00}\left(s\right)=\begin{cases}
1+\frac{1}{2}\ln\frac{s-1}{2}, & s\to{}1+0^{+},\\
-\frac{1}{3s^{2}}-\frac{1}{5s^{4}}-\frac{1}{7s^{6}}-\cdots, & s\rightarrow\infty,
\end{cases}
\end{equation}
one can solve the secular equation Eq.~\eqref{eq:det} approximately in two limits: $s\to{}1+0^{+}$ and $s\to\infty$, resulting in the following analytical form: when $0<s-1\ll{}1$, we can obtain
\begin{subequations}\label{eq:sol-3-m0}
\begin{equation}
s\simeq
1+2e^{-\frac{11}{3}-
\frac{50-\left(30+\frac{50F_{0}^{s}+8F_{2}^{s}}{3}\right)\left(1+\frac{F_{1}^{s}}{3}\right)} {F_{2}^{s}\left[9+\left(1-3F_{0}^{s}\right)\left(1+\frac{F_{1}^{s}}{3}\right)\right]+10\left[\left(3+F_{0}^s\right)\left(1+\frac{F_{1}^{s}}{3}\right)-3\right]}},
\end{equation}
and for $s\gg{}1$, we obtain
\begin{equation}
s\simeq\sqrt{\frac{F_{2}^{s}}{25}\left[\frac{9}{7}+\frac{4}{3}\left(1+\frac{F_{1}^{s}}{3}\right)\right]
+\frac{F_{0}^{s}}{3}\left(1+\frac{F_{1}^{s}}{3}\right)+\frac{F_{1}^{s}}{5}}.
\end{equation}
\end{subequations}

Taking into account of the sufficient condition given in Eq.~\eqref{ineq:three-m0}, we find that the solution will take simpler forms in the two incompressible conditions (i) $1+\frac{F_{1}^{s}}{3}=0$ and (ii) $F_{0}^{s}\to+\infty$: 

(i) In the incompressibility condition $1+\frac{F_1^s}{3}=0$, we have
\begin{equation}
s\simeq\begin{cases}
1+2e^{-\frac{11}{3}-
\frac{50}{9}\frac{1} {F_{2}^{s}-10/3}}, & F_{2}^{s}>\frac{10}{3},\\
1+2e^{-\frac{11}{3}}\,\,\mbox{or}\,\,\frac{3}{5}\sqrt{\frac{F_{2}^{s}}{7}-\frac{5}{3}}, & F_{2}^{s}\to+\infty.
\end{cases}
\end{equation}
When $F_{2}^{s}>10/3$, there exists at least one weakly damped zero sound mode with the sound speed $c_0=sv_{F}=v_{F}\left(1+2e^{-\frac{11}{3}-\frac{50}{9}\frac{1}{F_{2}^{s}-10/3}}\right)$; while when $F_{2}^{s}\to{}+\infty$, an additional weakly damped zero sound mode occurs with the sound speed $c_{0}=\frac{3v_{F}}{5}\sqrt{\frac{F_{2}^{s}}{7}-\frac{5}{3}}$.

(ii) In the other incompressibility condition $F_{0}^{s}\to+\infty$, we have two approximate solutions to the secular equation from Eq.~\eqref{eq:sol-3-m0}:
\begin{equation}
s\simeq\begin{cases}
1+2e^{-\frac{11}{3}-
\frac{50}{9}\frac{1} {F_{2}^{s}-10/3}}, & s\to{}1+0^{+},\\
\sqrt{\frac{F_{0}^{s}}{3}\left(1+\frac{F_{1}^{s}}{3}\right)}, & s\to+\infty.
\end{cases}
\end{equation}
Here the constraint $F_{2}^{s}>10/3$ is still imposed by Eq.~\eqref{ineq:three-m0}.

\subsection{$m\neq{}0$ modes in the collisionless regime}

\begin{figure}[tb]
	\centering
	\includegraphics[width=\linewidth]{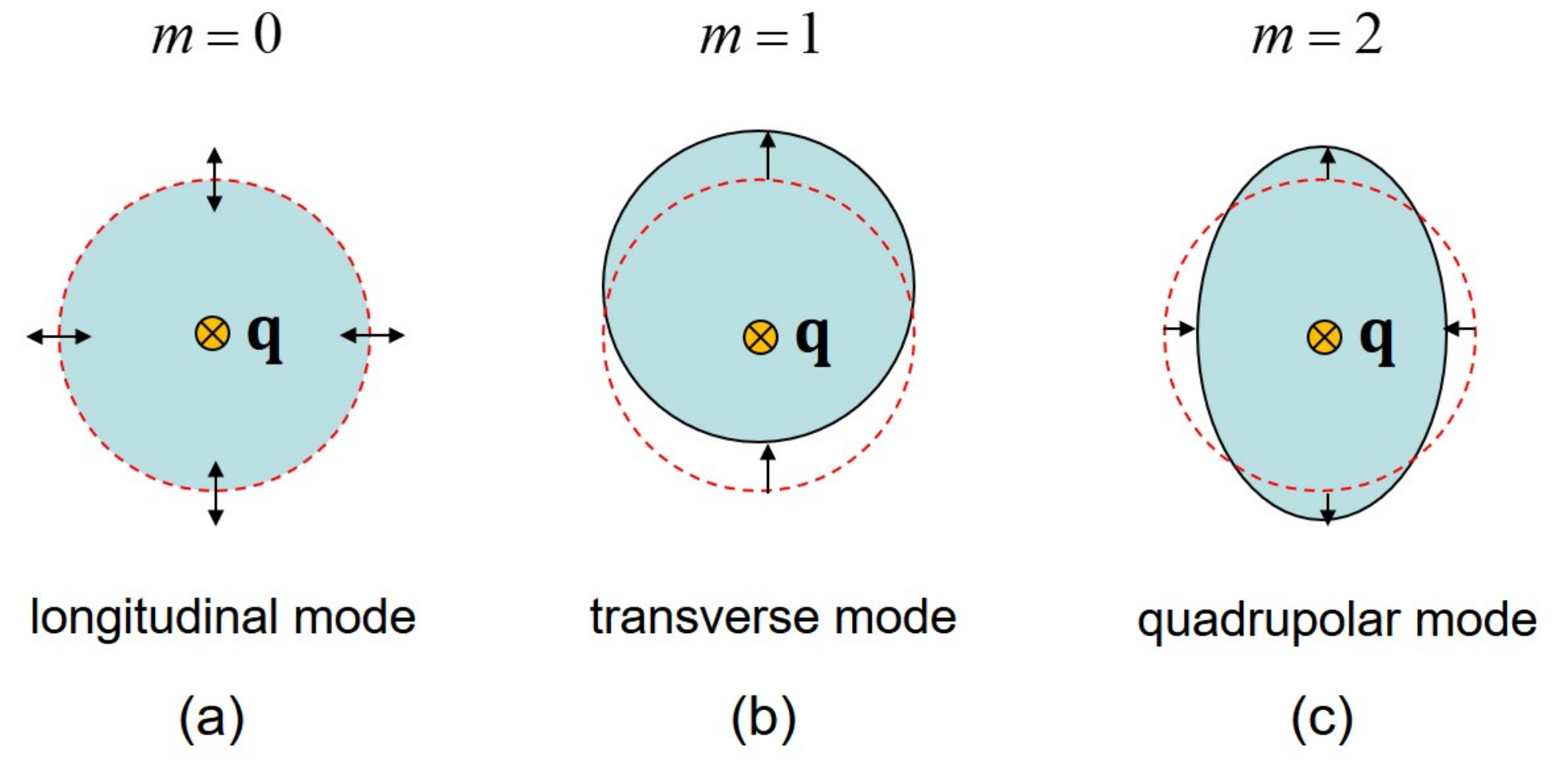}
	\caption{\label{fig1} Illustration of collective modes in 3D. The propagation wave vector $\mathrm{q}$ is normal to the plane.}
\end{figure}

To go beyond the ``longitudinal" $\left( m=0\right) $ mode and study $m\neq{}0$ modes including ``transverse'' $\left(m=1\right) $, ``quadrupolar" $\left(m=2\right) $, and higher angular momentum modes, we consider the more generic expansion for $\nu_{\mathrm{p}}$ given in Eq.~\eqref{eq:nu}. These collective modes are illustrated in Fig.~\ref{fig1}. As mentioned, because of the $O(2)$ rotational symmetry, the collective modes associated with different values of $m$ are decoupled. 
Since an $m\neq{}0$ mode cannot be excited by a $\phi_{\mathrm{p}}$-independent external field, we need to consider a new external field $U_{\mathrm{p}}\left(\mathrm{r},t\right) $ rather than the original one $U\left( \mathrm{r},t\right) $, where
\begin{equation*}
U_{\mathrm{p}}\left( \mathrm{r},t\right) =Ue^{im\phi _{\mathrm{p}}}e^{i(\mathrm{q}\cdot \mathrm{r}-\omega t)}.
\end{equation*}
Similar to the case of $m=0$, the Landau's kinetic equation gives rise to two sets of algebraic equations for arbitrary $m$, i.e., the type I equations as
\begin{subequations}\label{eq:LTEm}
\begin{equation}
\frac{\nu_{l}^{m}}{2l+1}+\sum_{l'=m}^{\infty}F_{l'}^{s}\Omega_{ll'}^{m}\left(s\right)\frac{\nu_{l'}^{m}}{2l'+1}=-\Theta_{l}^{m}\left(s\right)U,  \label{LTEm2}
\end{equation}
and type II equations as
\begin{equation}
\begin{split}
\nu _{l}^{m}s & -\nu _{l-1}^{m}\left( 1+\frac{F_{l-1}^{s}}{2l-1}\right) \frac{\sqrt{l^{2}-m^{2}}}{2l-1}\\
&-\nu _{l+1}^{m}\left( 1+\frac{F_{l+1}^{s}}{2l+3}\right) 
\frac{\sqrt{(l+1)^{2}-m^{2}}}{2l+3}=\alpha _{l}^{m}U.
\end{split}\label{LTEm4}
\end{equation}
\end{subequations}
Here $\Omega_{ll^{\prime }}^{m}(s)$, $\Theta_{l}^{m}$, and $\alpha _{l}^{m}$ are defined as
\begin{subequations}
\begin{equation}
\begin{split}
& \Omega _{ll^{\prime}}^{m}(s) = \Omega _{l^{\prime}l}^{m}(s) \\
=& \frac{1}{2}\sqrt{\frac{(l-m)!(l^{\prime}-m)!}{(l+m)!(l^{\prime}+m)!}}\int_{-1}^{1}d\mu P_{l}^{m}(\mu) \frac{\mu}{\mu-s}P_{l^{\prime}}^{m}(\mu),
\end{split}
\end{equation}
\begin{equation}
\begin{split}
\Theta_{l}^{m}\left(s\right) &=\left(-1\right)^{m}\Theta_{l}^{-m}\left(s\right)\\
&=\frac{1}{2}\sqrt{\frac{\left(l-m\right)!}{\left(l+m\right)!}}\int_{-1}^{1}d\mu P_{l}^{m}\left(\mu\right)\frac{\mu}{\mu-s},
\end{split}
\end{equation}
and 
\begin{equation}
\alpha _{l}^{m}=\left(-1\right)^{m}\alpha_{l}^{-m}=\frac{2l+1}{2}\sqrt{\frac{(l-m)!}{(l+m)!}}\int_{-1}^{1}d\mu P_{l}^{m}(\mu )\mu, \label{eq:alpha}
\end{equation}
\end{subequations}
respectively, where $P_{l}^{m}$ are associated Legendre polynomials. To derive these algebraic equations, we have used some relations for $\Omega_{ll^{\prime}}^{m}$, $\Theta_{l}^{m}$ and $\alpha_{l}^{m}$: (i) $\Omega_{ll'}^{m}\left(s\right)=0$ for $m>l,l'$; (ii) $\Theta_{l}^{m}\left(s\right)=0$ for $m>l$; and (iii) $\alpha _{l}^{m}=0$ when $l-m$ is even and $\alpha _{l}^{0}=0$ for $l>1$. 
The first few nonzero $\Omega_{ll^{\prime}}^{m}$, $\Theta_{l}^{m}$, and $\alpha_{l}^{m}$ can be found in Table \ref{tab:Omega_Theta_alpha}.

Notice that with the help of the identity for associated Legendre polynomials $P_{l}^{-m}(x)=(-1)^{m}\frac{(l-m)!}{(l+m)!}P_{l}^{m}(x)$, one is easy to verify:
(i) $\Omega_{ll^{\prime}}^{-m}=\Omega_{ll^{\prime}}^{m}$, (ii) $\Theta_{l}^{-m}=\left(-1\right)^{m}\Theta_{l}^{m}$, and (iii) $\alpha_{l}^{-m}=\left(-1\right)^{m}\alpha_{l}^{m}$. This means that $m\leftrightarrow-m$ symmetry is respected by both type I and II algebraic equations [Eqs.~\eqref{LTEm2} and \eqref{LTEm4}]. Namely, if $\nu_{l}^{m}$ is a solution, then $\tilde{\nu}_{l}^{m}=\left(-1\right)^{m}\nu_{l}^{-m}$ gives rise to another solution to both type I and II algebraic equations. So that we will focus on $m>0$ cases in this section.

\begin{table}[tb]
\caption{A list of first few nonzero $\Omega_{ll'}^{m}$, $\Theta_{l}^{m}$, and $\alpha_{l}^{m}$. 
Note that $\Omega_{ll'}^{m}=\Omega_{l'l}^{m}=\Omega_{ll'}^{-m}$, $\Theta_{l}^{m}=\left(-1\right)^{m}\Theta_{l}^{-m}$, and $\alpha_{l}^{m}=\left(-1\right)^{m}\alpha_{l}^{-m}$. Besides, we also have (i) $\Omega_{ll'}^{m}=0$ for $m>l,l'$, (ii) $\Theta_{l}^{m}=\alpha_{l}^{m}=0$ for $m>l$, (iii) $\alpha_{l}^{m}=0$ for $l+m=\mathrm{even}$, and (iv) $\alpha_{l}^{0}=0$ for $l>1$.}
\label{tab:Omega_Theta_alpha}
\renewcommand\arraystretch{2}
\setlength{\tabcolsep}{4mm}{
\begin{tabular}{l}
\hline\hline
$\Omega_{00}^{0}=1-\frac{s}{2}\ln\frac{s+1}{s-1}=\Omega_{00}$  \\
$\Omega_{11}^{1}=\frac{1}{4}\left[\frac{4}{3}-2s^{2}+s\left(s^{2}-1\right)\ln\frac{s+1}{s-1}\right]$  \\
$\Omega_{22}^{2}=\frac{3}{16}\left[\frac{2}{15}\left(8-25s^{2}+15s^{4}\right)-s\left(s^{2}-1\right)^{2}\ln\frac{s+1}{s-1}\right]$  \\
$\Theta_{0}^{0}=1-\frac{s}{2}\ln\frac{s+1}{s-1}=\Omega_{00}^{0}$  \\
$\Theta_{1}^{1}=\frac{\pi}{2}\left[1-2s^{2}+2s\left(s-1\right)\sqrt{\frac{s+1}{s-1}}\right]$  \\
$\Theta_{2}^{2}=\sqrt{\frac{3}{2}}\frac{1}{4}\left[\frac{4}{3}-2s^{2}+s\left(s^{2}-1\right)\ln\frac{s+1}{s-1}\right]=\sqrt{\frac{3}{2}}\Omega_{11}^{1}$  \\
$\alpha _{1}^{0} =1$  \\
$\alpha _{2}^{1} =-\frac{5}{16}\sqrt{\frac{3}{2}}\pi$  \\
$\alpha _{3}^{2} =\frac{7}{\sqrt{30}}$  \\
$\alpha _{4}^{1} =-\frac{9\sqrt{5}}{128}\pi$ \\
$\alpha _{4}^{3}=-\frac{9\sqrt{35}}{128}\pi$ \\
\hline
\end{tabular}
}
\end{table}

\subsubsection{A connate two-channel model ($m\geq{}1$)}
To study generic $m\geq{}1$ modes, we keep the Landau parameters up to $F_{m+1}^{s}$, i.e., keep $m+2$ finite $F_{l}^{s}(l=0,\cdots,m+1)$ and set $F_{l}^{s}=0$ for $l>m+2$. An interesting observation on Eq.~\eqref{eq:LTEm} is the following: The Landau parameter $F_{l}^{s}$ does not affect on a mode with $m>l$. For instance, $F_{0}^{s}$ does not affect on the transverse $m=1$ mode. Therefore, for a fixed $m>0$, there are only two relevant Landau parameters, $F_{m}^{s}$ and $F_{m+1}^{s}$, which are active in Eq.~\eqref{eq:LTEm}. These Landau parameters will result in a \emph{connate} two-channel model only consisting of two variables, $\nu_{m}^{m}$ and $\nu_{m+1}^{m}$.

Such a connate two channel-model can be derived from Eq.~\eqref{eq:LTEm}. First, the $l=m$ component of the type I equation, Eq.~\eqref{LTEm2}, reduces to
\begin{subequations}
\begin{equation}
\frac{\nu_{m}^{m}}{2m+1}+F_{m}^{s}\Omega_{mm}^{m}\frac{\nu_{m}^{m}}{2m+1}+F_{m+1}^{s}\Omega_{mm+1}^{m}\frac{\nu_{m+1}^{m}}{2m+3}=-\Theta_{m}^{m}U.
\end{equation}
Second, the $l=m$ component of type II equation, Eq.~\eqref{LTEm4}, is given by
\begin{equation}
\nu_{m}^{m}s-\sqrt{2m+1}\left(1+\frac{F_{m+1}^{s}}{2m+3}\right)\frac{\nu_{m+1}^{m}}{2m+3}=0.
\end{equation}
\end{subequations}
These two equations form a close set and the solution to the response function $\nu_{l}^{m}/U$ is found to be:
\begin{subequations}
\begin{align}
\frac{\nu_{m}^{m}}{U} & =-\frac{\left(1+\frac{F_{m+1}^{s}}{2m+3}\right)\Theta_{m}^{m}}{\frac{\left(1+F_{m}^{s}\Omega_{mm}^{m}\right)\left(1+\frac{F_{m+1}^{s}}{2m+3}\right)}{2m+1}+\frac{F_{m+1}^{s}\Omega_{mm+1}^{m}s}{\sqrt{2m+1}}},\\
\frac{\nu_{m+1}^{m}}{U} & =-\frac{(2m+3)s\Theta_{m}^{m}}{\frac{\left(1+F_{m}^{s}\Omega_{mm}^{m}\right)\left(1+\frac{F_{m+1}^{s}}{2m+3}\right)}{\sqrt{2m+1}}+F_{m+1}^{s}\Omega_{mm+1}^{m}s}.
\end{align}
\end{subequations}
With the help of the identity $\Omega_{mm+1}^{m}\left(s\right)=\sqrt{2m+1}s\Omega_{mm}^{m}\left(s\right)$, the mode frequency can be determined via the pole of the response functions $\nu_{l}^{m}/U$ (see Appendix \ref{app:two-channel-m}),
\begin{equation}
\left(1+F_{m}^{s}\Omega_{mm}^{m}\right)\left(1+\frac{F_{m+1}^{s}}{2m+3}\right)+\left(2m+1\right)F_{m+1}^{s}s^{2}\Omega_{mm}^{m}=0.\label{eq:mode_fre-m}
\end{equation}

\begin{figure}[tb]
	\centering
	\includegraphics[width=0.8\linewidth]{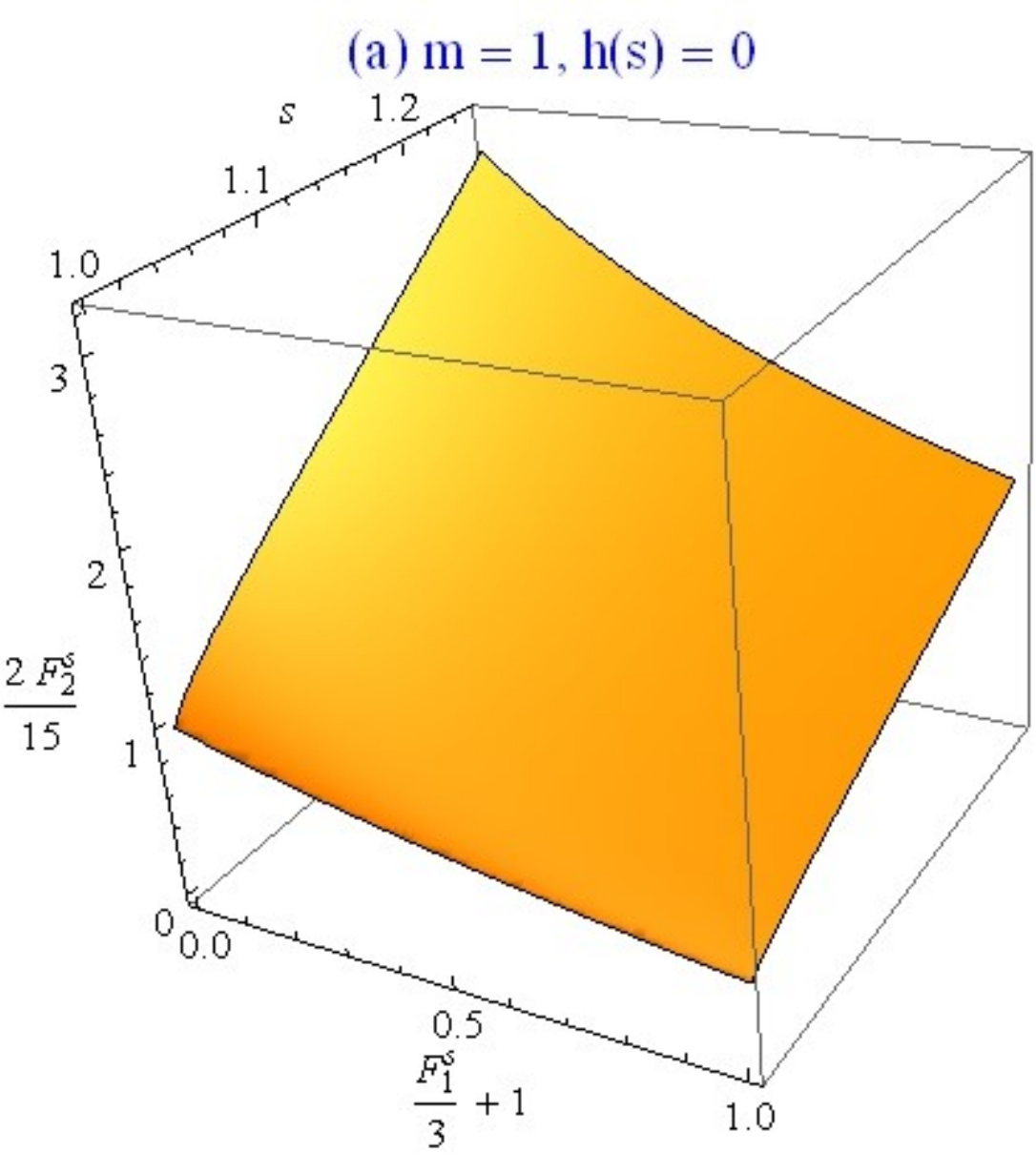}
	\includegraphics[width=0.8\linewidth]{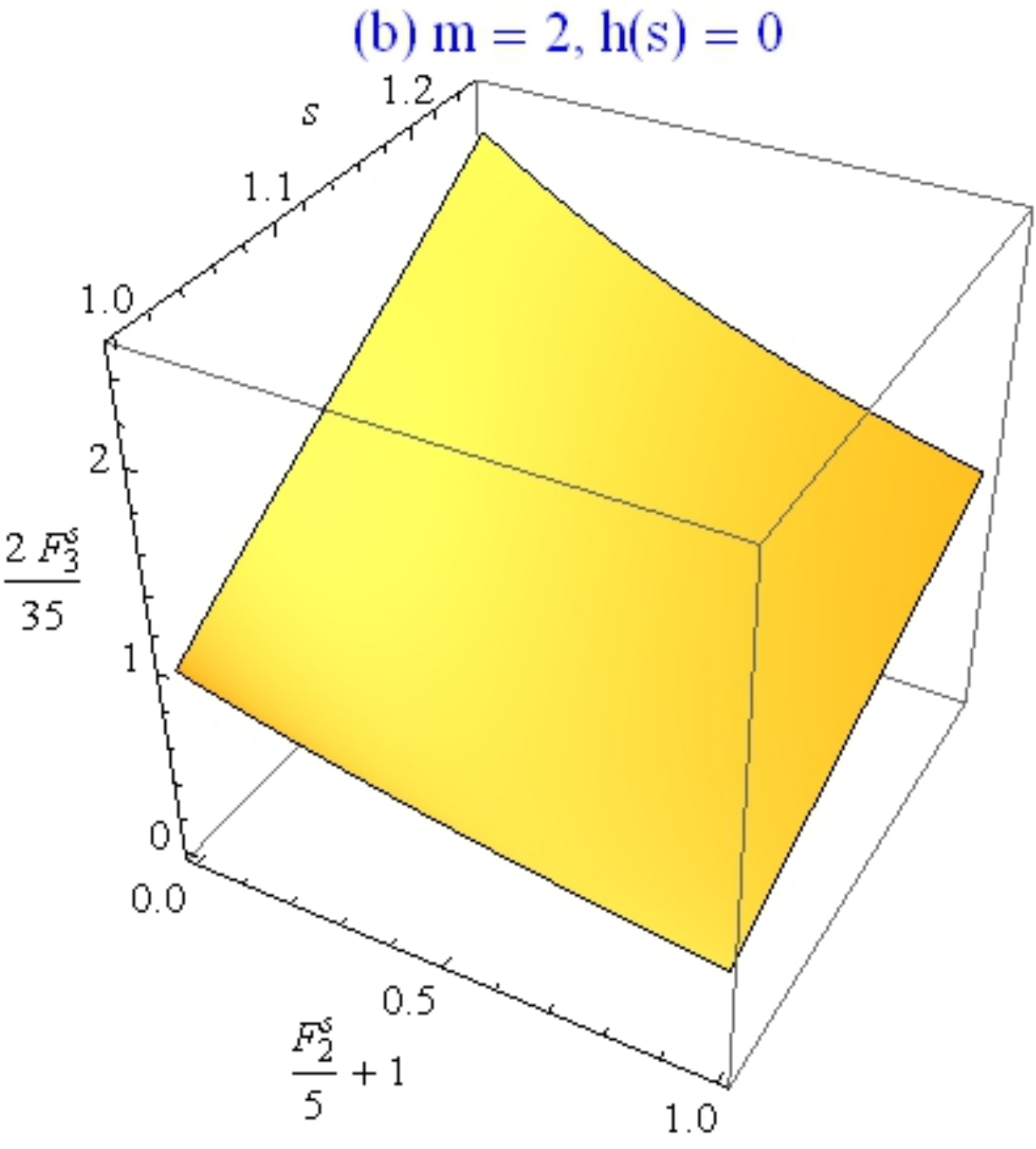}
	\caption{\label{fig:h0} The numerical solution for the frequency equation $h(s)=0$ is plotted in the 3D parameter space $\left(s,1+\frac{F_{m}^{s}}{2m+1},\frac{2F_{m+1}^{s}}{(2m+1)(2m+3)}\right)$, where $h(s)$ is defined in Eq.~\eqref{eq:hs}. (a) $m=1$ and (b) $m=2$ numeric solutions.}
\end{figure}

To solve Eq.~\eqref{eq:mode_fre-m}, we define a function
\begin{equation}
h(s)=\left[1+F_{m}^{s}\Omega_{mm}^{m}(s)\right]\left(1+\frac{F_{m+1}^{s}}{2m+3}\right)+(2m+1)F_{m+1}^{s}s^2\Omega_{mm}^{m}(s). \label{eq:hs}   
\end{equation}
Then the mode frequency can be found as a root of $h(s)$. For $m\geq{}1$, by using Eq.~\eqref{eq:omegammm}, we have
\begin{equation}
h(s\to{}\infty) = 1.    \label{eq:h-infty}
\end{equation}
Meanwhile, the recursive relation Eq.~\eqref{eq:omega-recursive} gives rise to
\begin{equation*}
\Omega_{mm}^{m}(s\to{}1) = -\frac{1}{2m(2m+1)}  
\end{equation*}
and 
\begin{equation}
\begin{split}
h(s\to{}1) = & \left[1-\frac{F_{m}^{s}}{2m(2m+1)}\right]\left(1+\frac{F_{m+1}^{s}}{2m+3}\right)-\frac{F_{m+1}^{s}}{2m}\\
= & -\frac{1}{2m(2m+3)}\left[2+\left(1+\frac{F_{m}^{s}}{2m+1}\right)\right]F_{m+1}^{s}\\
&+\frac{1}{2m}\left[2m+1-\left(1+\frac{F_{m}^{s}}{2m+1}\right)\right].
\end{split}\label{eq:h-1}
\end{equation}
Since $1+\frac{F_{m}^{s}}{2m+1}\geq{}0$ is required by thermodynamic stability of the system\cite{FLBook1,FLBook2}, there must exist a real solution $s>1$ to the frequency equation $h(s)=0$, when
\begin{equation}
F_{m+1}^{s}>(2m+3)\frac{2m+1-\left(1+\frac{F_{m}^{s}}{2m+1}\right)}{2+\left(1+\frac{F_{m}^{s}}{2m+1}\right)}. \label{eq:exist-m} 
\end{equation}
In the limit of $1+\frac{F_{m}^{s}}{2m+1}=0$, the above inequality becomes
\begin{equation}\label{ineq:exist-m-gap}
F_{m+1}^{s}>\frac{(2m+1)(2m+3)}{2},   
\end{equation}
and when
\begin{equation}
F_{m+1}^{s} = \frac{(2m+1)(2m+3)}{2},   
\end{equation}
there is an exact solution, $s=1$.

In particular, we are interested in $m=1$ (transverse) and $m=2$ (quadrupolar) modes. For these modes, we found numerically that Eq.~\eqref{eq:exist-m} is also a necessary condition for a real solution $s>1$ to $h(s)=0$. And these numerical solutions to $h(s)=0$ are shown in Fig.~\ref{fig:h0} as a 3D contour plot in the parameter space $	\left(s,1+\frac{F_{m}^{s}}{2m+1},\frac{2F_{m+1}^{s}}{(2m+1)(2m+3)}\right)$. Note that these solutions are all around $s\to{}1+0^{+}$.

Indeed, another solution for arbitrary $m\geq{}1$ can be found analytically in the large $s$ limit as follows,
\begin{equation}
s=\sqrt{\frac{1}{2m+3}\left[\frac{F_{m}^{s}}{2m+1}\left(1+\frac{F_{m+1}^{s}}{2m+3}\right)+\frac{3F_{m+1}^{s}}{2m+5}\right]}.
\end{equation}
as long as the condition 
\begin{equation}
\frac{F_{m}^{s}}{2m+1}\left(1+\frac{F_{m+1}^{s}}{2m+3}\right)+\frac{3F_{m+1}^{s}}{2m+5}\gg{}2m+3   
\end{equation}
is satisfied (see Appendix \ref{app:two-channel-m}).

\subsubsection{Suppression of $|m|=l\geq{}1$ modes}
First, we examine the $m=l=1$ case. For this mode, $\nu _{1}^{1}$ corresponds to the transverse current, and Eq.~\eqref{LTEm4} leads to
\begin{equation*}
\nu _{1}^{1}s-\nu _{2}^{1}\left( 1+\frac{F_{2}^{s}}{5}\right) \frac{\sqrt{3}}{5}=0.
\end{equation*}
It is interesting to see when $1+\frac{F_{2}^{s}}{5}=0$, the transverse current mode will be completely suppressed, i.e.,
\begin{equation*}
\nu _{1}^{1}=0.
\end{equation*}
In general, when $m=\pm{}l$, Eq.~\eqref{LTEm4} gives rise to
\begin{equation}
\nu _{l}^{\pm{}l}s-\nu _{l+1}^{\pm{}l}\left( 1+\frac{F_{l+1}^{s}}{2l+3}\right) \frac{\sqrt{2l+1}}{2l+3}=0.
\end{equation}
Therefore, when
\begin{equation}
1+\frac{F_{l+1}^{s}}{2l+3}=0,
\end{equation}
the $m=\pm{}l$ modes will be completely suppressed in low energies, i.e., $\nu _{l}^{m=\pm{}l}=0$, thereby acquire an excitation energy gap.

\emph{Nematic QSL state. ---}  It is remarkable that the suppression of transverse current ($l=1$ and $m=\pm{}1$) modes will give rise to a ``nematic" QSL state that cannot be described by the usual $U(1)$ gauge theory, in which the transverse current mode is gapless. However, higher angular momentum modes, i.e., $|m|\geq{}2$ mode can still be gapless, even though $m=\pm{}1$ modes are gapped.

\emph{Hierarchy of QSL states. ---} The possibility of the suppression of $|m|=l$ modes and the inequality in Eq.~\eqref{ineq:exist-m-gap} suggest that there is a hierarchy structure for gapless QSL states:

\noindent
\fbox{\parbox{0.96\linewidth}{
When
\begin{equation}\label{eq:hierarchy}
1+\frac{F_{l}^{s}}{2l+1}\begin{cases}
\, >0, & l=0\,\,\mbox{or}\,\,l\geq{}n+2,\\
\, =0, & l=1,\cdots,n,\\
\, >\frac{2l+1}{2}, & l=n+1,
\end{cases}
\end{equation}
the Landau-type effective theory given by Eq.~\eqref{eq:F} describes a gapless QSL state in 3D, on which $|m|=0,\cdots,n$ zero sound modes
are gapped, while $|m|=n+1$ zero sound modes are gapless. Here $n$ is a positive integer, and $m$ is the ``magnetic" quantum number along the collective mode propagating direction $\hat{\mathrm{q}}$. 
}}

\section{First sound in the hydrodynamic regime: $\omega\tau\ll{}1$}\label{sec:FS3D}

First sound is the ordinary sound in a liquid, i.e., density fluctuations in a liquid that is composed of longitudinal modes in the hydrodynamic regime $\omega\tau \ll 1$. In this regime, the collision integral $I[n_{\mathrm{p}^{\prime}}]\sim \delta n /\tau$ becomes significant, and the system is essentially in local thermodynamic equilibrium.
In such a strong collision regime, following Abrikosov and Khalatnikov~\cite{Abrikosov59}, we assume that the collision integral takes the form:
\begin{subequations}\label{eq:collision}
\begin{equation}
I[n_{\mathrm{p}}]  =  - \frac{1}{\tau_\mathrm{p}}\left[\delta{}n_{\mathrm{p}}-\langle\delta{}n_{\mathrm{p}}\rangle-3\langle\delta{}n_{\mathrm{p}}\cos\theta_{\mathrm{p}}\rangle\cos\theta_{\mathrm{p}}\right],
\end{equation}
and
\begin{equation}
\frac{1}{\tau_\mathrm{p}}  =  \frac{1}{\tau_{i}} + \frac{(\varepsilon_{\mathrm{p}}-\varepsilon_{F})^{2}}{\varepsilon_{a}}+\frac{(k_{\beta}T)^{2}}{\varepsilon_{b}},
\end{equation}
\end{subequations}
where the brackets $\langle\cdots\rangle$ denote the average over the solid angle, $\tau_{\mathrm{p}}$ is the relaxation time and its inverse gives rise to the scattering rate. Here $1/\tau_{i}$ is the impurity scattering rate, and the rest parts correspond to the relaxations due to quasiparticle interactions and thermal fluctuations. And $\varepsilon_{a}$ and $\varepsilon_{b}$ are two characteristic energy scales for quasiparticle interactions and thermal fluctuations respectively.

Notice that the choice of collision integral in Eq.~\eqref{eq:collision} automatically satisfies the conservation laws of energy, momentum and the total number of particles. With the help of the expansion given in Eq.~\eqref{eq:nu0}, the expectation values in the collision integral $I[n_{\mathrm{p}}]$ in Eq.~\eqref{eq:collision} can be computed as follows,
\begin{equation}
\langle{}\delta{}n_{\mathrm{p}}\rangle=-\frac{\partial n_{\mathrm{p}}^{0}}{\partial \varepsilon _{\mathrm{p}}}\nu_{0},\quad{}3\langle{}\delta{}n_{\mathrm{p}}\cos{\theta_{\mathrm{p}}}\rangle=-\frac{\partial n_{\mathrm{p}}^{0}}{\partial \varepsilon _{\mathrm{p}}}\nu_{1}.
\end{equation}
Then the Landau's kinetic equation reads
\begin{equation}
\frac{\partial \delta n_{\mathrm{p}}}{\partial t}+\vec{v}_{\mathrm{p}}\cdot
\nabla _{\mathrm{r}}\left( \delta n_{\mathrm{p}}-\frac{\partial n_{\mathrm{p}
	}^{0}}{\partial \varepsilon _{\mathrm{p}}}\delta \varepsilon _{\mathrm{p}
}\right) =I[n_{\mathrm{p}}].  \label{LTEFS}
\end{equation}
In the presence of a scalar external field $U$ as in the collisionless regime, the Landau's kinetic equation can be rewritten in frequency-momentum space as
\begin{equation}\label{eq:LTEFS1}
(\omega -\mathrm{q}\cdot \vec{v}_{\mathrm{p}})\delta n_{\mathrm{p}}
+\frac{\partial n_{\mathrm{p}}^{0}}{\partial \varepsilon _{\mathrm{p}}}(\mathrm{q}\cdot \vec{v}_{\mathrm{p}})\left( U+\sum_{\mathrm{p}^{\prime }}f_{\mathrm{pp}^{\prime }}^{s}\delta n_{\mathrm{p}^{\prime }}\right)
=iI[n_{\mathrm{p}}].
\end{equation}
Note that in the hydrodynamic limit, $I[n_{\mathrm{p}}]\propto 1/\tau_{\mathrm{p}}$, and $\hbar/\tau_{\mathrm{p}}$ is the largest energy scale in this limit and does not depend on the azimuthal angle $\phi_{\mathrm{p}}$ and corresponding quantum number $m$. Therefore, $m\neq{}0$ modes will be strongly damped, and we shall study longitudinal ($m=0$) mode only.

Putting the expansion in Eq.~\eqref{eq:nu0} into Eq.~\eqref{eq:LTEFS1}, we obtain reduced Landau's kinetic equation in the $U\to{}0$ limit:
\begin{equation}
\frac{\nu _{l}}{2l+1}+\sum_{l^{\prime }}F_{l^{\prime }}^{s}\Omega_{ll^{\prime }}(\tilde{s})\frac{\nu_{l^{\prime }}}{2l^{\prime }+1}
=-i\kappa\left[\Omega_{l0}(\tilde{s}){\nu}_{1}+\Omega_{l}(\tilde{s}){\nu}_{0}\right],  \label{LTEFS3}
\end{equation}
where
\begin{equation}
\kappa=\frac{1}{\tau{}q{}v_{F}}=\frac{s}{\omega\tau},~~\tilde{s}= \kappa(\omega\tau + i)=s\left(1+\frac{i}{\omega\tau}\right), \label{eq:kappa}
\end{equation}
and
\begin{equation}\label{eq:omegal}
\Omega _{l}(\tilde{s})=\frac{1}{2}\int_{-1}^{1}d\mu P_{l}(\mu )\frac{1}{\mu-\tilde{s}}=\frac{1}{\tilde{s}}\left[\Omega _{l0}(\tilde{s})-\delta_{l,0}\right],
\end{equation}
and $\Omega _{ll^{\prime }}$ is defined in Eq.~\eqref{eq:omega}. Here the assumption $\tau_{\mathrm{p}}=\tau$ has been used. In comparison with Eq.~\eqref{LTEZS3} in the collisionless regime, there appear additional $\kappa$-terms in the right hand side in Eq.~\eqref{LTEFS3} that is induced by collisions.

\subsection{Two channel model}
To study first sound mode, we consider a two channel model by keeping $F_{0}^{s}$ and $F_{1}^{s}$ only and setting $F_{l}^{s}=0$ for $l\geq 2$. Then Landau's kinetic equation can be written as
\begin{equation}\label{eq:SCFS}
\left[
\begin{array}{cc}
1+F^{s}_{0}\Omega_{00}+i\kappa\Omega_{0} & \frac{1}{3}F^{s}_{1}\Omega_{10}+i\kappa\Omega_{00} \\
F^{s}_{0}\Omega_{10}+i\kappa\Omega_{1} & \frac{1}{3}(1+F^{s}_{1}\Omega_{11}) + i\kappa\Omega_{10}
\end{array}
\right]\left(
\begin{array}{c} \nu_0 \\ \nu_1 \end{array}
\right) =0.
\end{equation}
The condition for a nontrivial solution allows us to determine the mode frequency and leads to
\begin{equation}
\begin{split}
[1+F^{s}_{0}\Omega_{00}+i\kappa \Omega_{0}]\left[ \frac{1}{3}(1+F^{s}_{1}\Omega_{11}) +i\kappa \Omega_{10}\right]&\\
-\left(\frac{1}{3}F^{s}_{1}\Omega_{10} + i\kappa \Omega_{00}\right)( F^{s}_{0}\Omega_{10}+i\kappa \Omega_{1})&=0.
\end{split}
\end{equation}
In the hydrodynamic limit, $\omega\tau{}\ll1$, we have
\begin{equation}
\kappa^{-1}\rightarrow{}0,~
\tilde{s}\rightarrow{}i\infty+0^{+},~\text{and}~ 
\tilde{s}\kappa^{-1}\rightarrow{}i.
\end{equation}
Thus
\begin{equation*}
\begin{split}
\Omega_{0}(\tilde{s})&=-\frac{1}{2}\ln\left(\frac{\tilde{s}+1}{\tilde{s}-1}\right)\approx-\tilde{s}^{-1}-\frac{1}{3}\tilde{s}^{-3}-\frac{1}{5}\tilde{s}^{-5},\\
\Omega_{00}(\tilde{s})&=1-\frac{1}{2}\tilde{s}\ln\left(\frac{\tilde{s}+1}{\tilde{s}-1}\right)\approx-\frac{1}{3}\tilde{s}^{-2}-\frac{1}{5}\tilde{s}^{-4}.
\end{split}
\end{equation*}
In accordance with Eq.~\eqref{eq:omegal} and Table \ref{tab:omega}, $\Omega_{1}$, $\Omega_{10}$ and $\Omega_{11}$ can be expressed in terms of $\Omega_{00}$.
Thus, from Eq.~\eqref{eq:SCFS}, we can obtain the dispersion relation between $\omega$ and $\mathrm{q}$ for first sound as follows,
\begin{equation}
\left(\frac{\omega}{qv_{F}}\right)^{2}=
\frac{1}{3}(1+F_{0}^{s})\left(1+\frac{1}{3}F_{1}^{s}\right)-\frac{4}{15}i\omega\tau\left(1+\frac{1}{3}F_{1}^{s}\right),
\end{equation}
which is exactly the same as Eq.~(10.9) in Ref.~[\onlinecite{Abrikosov59}]. In the limit of $\omega\tau\ll{}1$, we obtain the first sound speed,
\begin{equation}
c_{1}=\frac{\omega}{q}=v_{F}\sqrt{\frac{1}{3}\left(1+F_{0}^{s}\right)\left(1+\frac{1}{3}F_{1}^{s}\right)}.    
\end{equation}
We would like to emphasize that the first sound speed also exhibits quite different behaviors in the two incompressibility condition, namely, (i) $c_{1}\to{}0$ when $1+\frac{F_{1}^{s}}{3}\to{}0$, while (ii) $c_{1}\to{}\infty$ when $F_{0}^{s}\to{}+\infty$.

\section{Sound modes in 2D}\label{sec:2D}

\begin{figure}[tb]
	\centering
	\includegraphics[width=\linewidth]{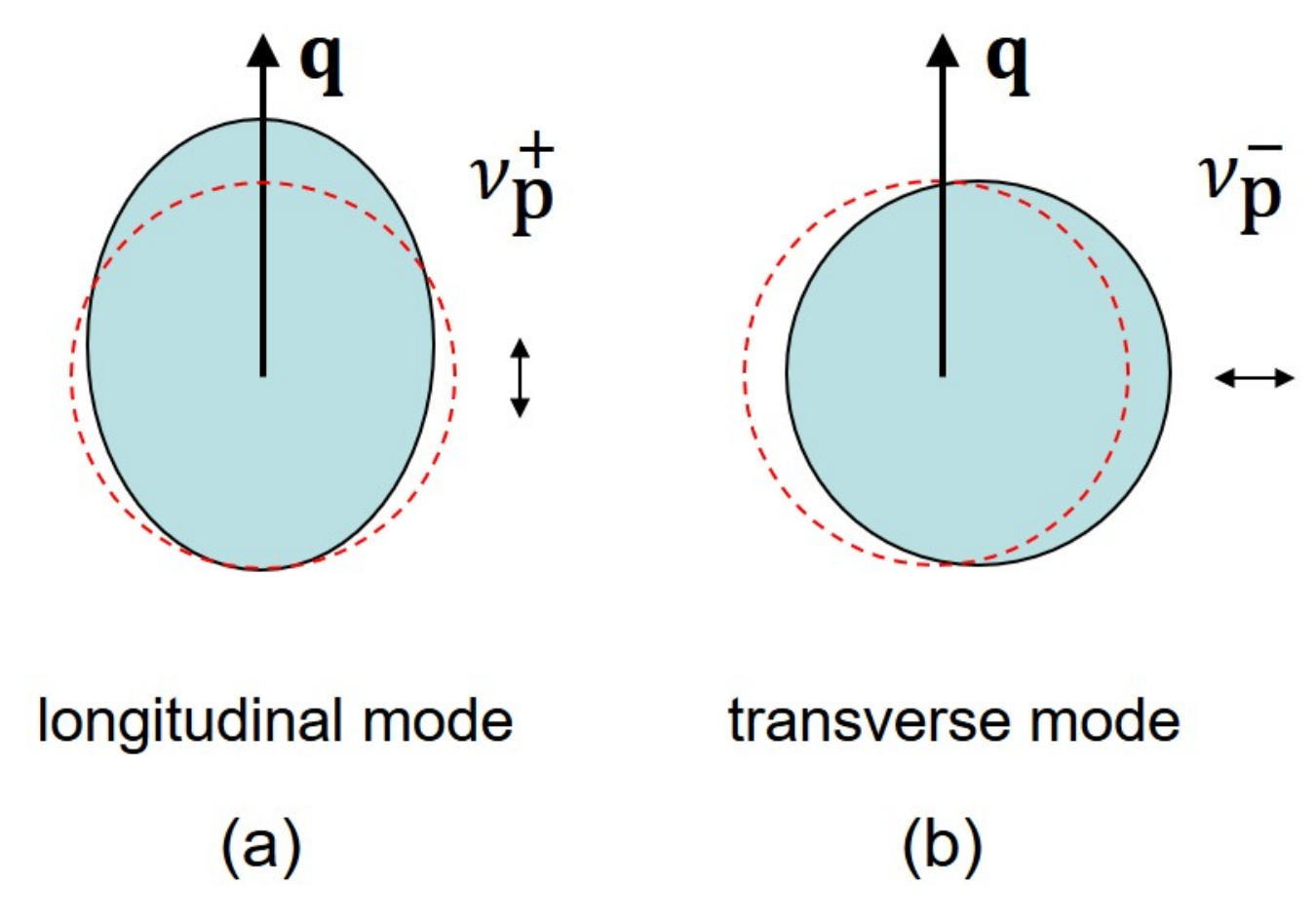}
	\caption{\label{fig:mode2D} Illustration of collective modes in 2D. The density wave is propagating along the wave vector $\mathrm{q}$. Under the reflection about $\mathrm{q}$, i.e., $\theta\to{}-\theta$, (a) the longitudinal mode $\nu_{\mathrm{p}}^{+}$ is symmetric, and (b) the transverse mode $\nu_{\mathrm{p}}^{-}$ is anti-symmetric.}
\end{figure}

In parallel with 3D, we study collective modes in 2D in this section. A significant difference between 2D and 3D is that the $O(3)$ symmetry for a spherical Fermi surface in 3D is reduced to the $O(2)$ symmetry for a circular Fermi surface in 2D. In the presence of a specified propagation wave vector $\mathrm{q}$ for a collective mode, the symmetry of the Landau's kinetic equation will be reduced from $O(2)$ (for 3D) to $Z_{2}$ (for 2D), i.e., the reflection symmetry $\theta\to{}-\theta$.

This symmetry reduction can be viewed from the expansion of the density fluctuation fluctuations $\delta{}n_{\mathrm{p}}$ or $\nu _{\mathrm{p}}$ in 2D:
\begin{equation}\label{eq:nu2D1}
\nu_{\mathrm{p}}=\sum_{l=-\infty }^{\infty}\nu_{l}e^{i{}l\theta_{\mathrm{p}}}=\nu_{0}+\sum_{l=1}^{\infty}u_{l}\cos{}(l\theta_{\mathrm{p}})+\sum_{l=1}^{\infty}v_{l}\sin{}(l\theta_{\mathrm{p}}),
\end{equation}
where $u_{0}=\nu_0$, $u_{l}$ and $v_{l}$ are real numbers, $u_{-l}=u_{l}$ and $v_{-l}=-v_{l}$ are even- and odd-parity components respectively. To ensure a real function $\nu_{\mathrm{p}}$, we have $\nu_{l}=(u_{l}-i{}v_{l})/2$ and $\nu_{-l}=\nu_{l}^{\ast}$. With respect to the reflection symmetry, $\theta\to{}-\theta$, $\nu_{\mathrm{p}}$ can be rewritten as two parts, i.e., the symmetric part $\nu_{\mathrm{p}}^{+}$ and the anti-symmetric part $\nu_{\mathrm{p}}^{-}$, as follows,
\begin{subequations}\label{eq:nu2D2}
\begin{eqnarray}
\nu_{\mathrm{p}} & = & \nu_{\mathrm{p}}^{+} + \nu_{\mathrm{p}}^{-}, \\
\nu_{\mathrm{p}}^{+} & = & \nu_{0}+\sum_{l=1}^{\infty}u_{l}\cos{}(l\theta_{\mathrm{p}}), \\
\nu_{\mathrm{p}}^{-} & = & \sum_{l=1}^{\infty}v_{l}\sin{}(l\theta_{\mathrm{p}}).
\end{eqnarray}
\end{subequations}
As illustrated in Fig.~\ref{fig:mode2D}, the density fluctuations $\nu_{\mathrm{p}}^{+}$ and $\nu_{\mathrm{p}}^{-}$ describe the coherent motion (and/or deformation) of the Fermi surface along and perpendicular to the propagating wave vector $\mathrm{q}$, respectively. So that $\nu_{\mathrm{p}}^{+}$ represents the longitudinal modes and $\nu_{\mathrm{p}}^{-}$ represents the transverse modes.

Note that the longitudinal mode $\nu_{\mathrm{p}}^{+}$ can be excited by an external scalar field in the form of
\begin{subequations}\label{eq:U2D}
\begin{equation}
U_{+}(\mathrm{r},t)=U{}e^{i(\mathrm{q}\cdot \mathrm{r}-\omega{}t)},    
\end{equation}
while the transverse mode $\nu_{\mathrm{p}}^{-}$ can be excited by an scalar field 
\begin{equation}
U_{-}(\mathrm{r},t)=\left[(\hat{\mathrm{n}}\times\hat{\mathrm{q}})\cdot \hat{v}_\mathrm{p}\right]{}U{}e^{i(\mathrm{q}\cdot \mathrm{r}-\omega{}t)}.
\end{equation}
\end{subequations}
Here $\hat{\mathrm{n}}$ is the unit vector normal to the 2D plane and $(\hat{\mathrm{n}}\times\hat{\mathrm{q}})\cdot \hat{v}_\mathrm{p}=\sin\theta_{\mathrm{p}}$.
Below we shall study longitudinal and transverse modes in both collisionless ($\omega\tau\gg{}1$) and hydrodynamic ($\omega\tau\ll{}1$) regimes.

\subsection{Zero sound: $\omega\tau\gg{}1$}

In the collisionless regime, type I and type II algebraic equations can be derived from Landau's kinetic equation, which are similar to those in 3D. 
Indeed, substituting Eqs.~\eqref{eq:U2D} into Eq.~\eqref{LTEZS}, for the longitudinal mode $\nu_{\mathrm{p}}^{+}$, we can obtain 
\begin{subequations}\label{eq:LTEZS2D0}
\begin{equation}
\begin{split}
\nu_{0}&+\sum_{l=1}^{\infty } u_{l}\cos(l\theta _{\mathrm{p}})-\frac{\cos \theta _{\mathrm{p}}}{s-\cos \theta _{\mathrm{p}}}
\left[F_{0}^{s}\nu_{0}+\frac{1}{2}\sum_{l=1}^{\infty } F_{l}^{s}u_{l}
\cos(l\theta_{\mathrm{p}})
\right] \\
&=\frac{\cos \theta _{\mathrm{p}}}{s-\cos \theta _{\mathrm{p}}}U,
\end{split}\label{eq:nu+}
\end{equation}
and for the transverse mode $\nu_{\mathrm{p}}^{-}$,
\begin{equation}
\begin{split}
&\sum_{l=1}^{\infty } v_{l}\sin(l\theta_{\mathrm{p}}) -\frac{1}{2}\frac{\cos\theta_{\mathrm{p}}}{s-\cos \theta_{\mathrm{p}}} \sum_{l=1}^{\infty} F_{l}^{s}v_{l} \sin(l\theta_{\mathrm{p}})\\
=&\frac{\cos \theta _{\mathrm{p}}}{s-\cos \theta_{\mathrm{p}}}U\sin\theta_{\mathrm{p}}. \label{eq:nu-}
\end{split}
\end{equation}
\end{subequations}
After some long but straightforward algebra (see Appendix \ref{app:aEq2D}), we have obtained type I and II algebraic equations for both longitudinal and transverse modes.

First, the type I algebraic equations for longitudinal and transverse modes are obtained as below, respectively,
\begin{equation}\label{eq:aEq2DI}
\begin{split}
\frac{u_{l}+\delta_{l0}u_{0}}{2}+\sum_{l'=0}^{\infty}F_{l'}^{s}\Pi_{ll'}\left(s\right)\frac{u_{l'}+\delta_{l'0}u_{0}}{2}&=-\Pi_{l0}\left(s\right)U, \\
\frac{v_{l}}{2}+\sum_{l'=0}^{\infty}F_{l'}^{s}\Xi_{ll'}\left(s\right)\frac{v_{l'}}{2}&=-\Xi_{l1}\left(s\right)U.
\end{split}
\end{equation}
where $\Pi_{ll'}\left(s\right)=\Pi_{l'l}\left(s\right)$ and $\Xi_{ll'}\left(s\right)=\Xi_{l'l}\left(s\right)$ are defined as follows,
\begin{equation}
\begin{split}
\Pi_{ll'}\left(s\right)&=-\int\frac{d\theta_{\mathrm{p}}}{2\pi}\cos\left(l\theta_{\mathrm{p}}\right)\frac{\cos\theta_{\mathrm{p}}}{s-\cos\theta_{\mathrm{p}}}\cos\left(l'\theta_{\mathrm{p}}\right), \\
\Xi_{ll'}\left(s\right)&=-\int\frac{d\theta_{\mathrm{p}}}{2\pi}\sin\left(l\theta_{\mathrm{p}}\right)\frac{\cos\theta_{\mathrm{p}}}{s-\cos\theta_{\mathrm{p}}}\sin\left(l'\theta_{\mathrm{p}}\right).
\end{split}
\end{equation}
First few nonzero $\Pi_{ll'}$ and $\Xi_{ll'}$ can be found in Table~\ref{tab:Pi_Xi}.

\begin{table}[tb]
\caption{A list of first few nonzero $\Pi_{ll'}$ and $\Xi_{ll'}$. 
Note that $\Pi_{ll'}=\Pi_{l'l}$, $\Xi_{ll'}=\Xi_{l'l}$, and $\Xi_{l0}=0$. Besides, we also have 
$\Pi_{l1}-s\Pi_{l0}=\frac{1}{2}\delta_{l1}$ and $\Xi_{l2}-2s\Xi_{l1}=\frac{1}{2}\delta_{l2}$.
}
\label{tab:Pi_Xi}
\renewcommand\arraystretch{2}
\setlength{\tabcolsep}{4mm}{
\begin{tabular}{l}
\hline\hline
$\Pi_{00}=1-\frac{s}{\sqrt{s^{2}-1}}$  \\
$\Pi_{10}=s\left(1-\frac{s}{\sqrt{s^{2}-1}}\right)=s\Pi_{00}$  \\
$\Pi_{11}=s^{2}\left(1-\frac{s}{\sqrt{s^{2}-1}}\right)+\frac{1}{2}=s\Pi_{10}+\frac{1}{2}$  \\
$\Pi_{20}=1+\left(2s^{2}-1\right)\left(1-\frac{s}{\sqrt{s^{2}-1}}\right)=1+\left(2s^{2}-1\right)\Pi_{00}$  \\
$\Pi_{21}=s\left[1+\left(2s^{2}-1\right)\left(1-\frac{s}{\sqrt{s^{2}-1}}\right)\right]=s\Pi_{20}$  \\
$\Pi_{22}=2s^{2}-\frac{1}{2}+\left(2s^{2}-1\right)^{2}\left(1-\frac{s}{\sqrt{s^{2}-1}}\right)=2s^{2}-\frac{1}{2}+\left(2s^{2}-1\right)^{2}\Pi_{00}$  \\
$\Xi_{11}=\frac{1-2s^{2}}{2}\left(1-\frac{s}{\sqrt{s^{2}-1}}\right)=\frac{1-2s^{2}}{2}\Pi_{00}$  \\
$\Xi_{21}=s\left(1-2s^{2}\right)\left(1-\frac{s}{\sqrt{s^{2}-1}}\right)=2s\Xi_{11}$  \\
$\Xi_{22}=2s^{2}\left(1-2s^{2}\right)\left(1-\frac{s}{\sqrt{s^{2}-1}}\right)+\frac{1}{2}=2s\Xi_{21}+\frac{1}{2}$ \\
\hline
\end{tabular}
}
\end{table}

Second, type II algebraic equations for the longitudinal mode have been found as follows,
\begin{subequations}\label{eq:aEq2DII}
\begin{equation}
u_{l}s-\left(1+\frac{F_{l-1}^{s}+\delta_{l1}F_{0}^{s}}{2}\right)\frac{u_{l-1}+\delta_{l1}u_{0}}{2}-\left(1+\frac{F_{l+1}^{s}}{2}\right)\frac{u_{l+1}}{2}=\delta_{l1}U, 
\end{equation}
where $l=0,1,2,\cdots$ and $u_{-1}=0$ has been set for convention; while type II algebraic equations for the transverse mode read
\begin{equation}
v_{l}s-\left(1+\frac{F_{l-1}^{s}}{2}\right)\frac{v_{l-1}}{2}-\left(1+\frac{F_{l+1}^{s}}{2}\right)\frac{v_{l+1}}{2}=\delta_{l2}\frac{U}{2}
\end{equation}
\end{subequations}
where $l=1,2,3,\cdots$ and $v_0:=0$ for convention. Below we shall study zero sound of longitudinal and transverse modes in 2D with the help of these algebraic equations.

\subsubsection{Incompressibility conditions}
To study the incompressibility, we consider the longitudinal mode and write down $l=0$ and $l=1$ components in the type II algebraic equation Eq.~(\ref{eq:aEq2DII}a) as follows,
\begin{subequations}
\begin{eqnarray}
u_{0}s-\left(1+\frac{F_{1}^{s}}{2}\right)\frac{u_{1}}{2}&=&0, \\
u_{1}s-\left(1+F_{0}^{s}\right)u_{0}-\left(1+\frac{F_{2}^{s}}{2}\right)\frac{u_{2}}{2}&=&U.
\end{eqnarray}
\end{subequations}
These equations give rise two incompressibility conditions similar as those in 3D: (i) $1+\frac{F_{1}^{s}}{d}=0$ and (ii) $F_{0}^{s}\to{}\infty$. Here $d=2$ is the dimensionality.

\subsubsection{Longitudinal mode $\nu_{\mathrm{p}}^{+}$}

We study the longitudinal mode by using a three-channel model, namely, keeping $F_{0}^{s}$, $F_{1}^{s}$ and $F_{2}^{s}$ and setting $F_{l}^{s}=0$ for $l\geq 3$. From type I algebraic equation Eq.~\eqref{eq:aEq2DI}, we obtain the close set of components $u_{0}$, $u_{1}$, and $u_{2}$ as follows,
\begin{equation}
\left(\begin{array}{ccc}
1+F_{0}^{s}\Pi_{00} & F_{1}^{s}\Pi_{01} & F_{2}^{s}\Pi_{02}\\
F_{0}^{s}\Pi_{10} & 1+F_{1}^{s}\Pi_{11} & F_{2}^{s}\Pi_{12}\\
F_{0}^{s}\Pi_{20} & F_{1}^{s}\Pi_{21} & 1+F_{2}^{s}\Pi_{22}
\end{array}\right)\left(\begin{array}{c}
u_{0}\\
\frac{u_{1}}{2}\\
\frac{u_{2}}{2}
\end{array}\right)=-U\left(\begin{array}{c}
\Pi_{00}\\
\Pi_{10}\\
\Pi_{20}
\end{array}\right).
\end{equation}
Then the mode frequency is determined by the following secular equation,
\begin{equation}
\det\left(\begin{array}{ccc}
1+F_{0}^{s}\Pi_{00} & F_{1}^{s}\Pi_{01} & F_{2}^{s}\Pi_{02}\\
F_{0}^{s}\Pi_{10} & 1+F_{1}^{s}\Pi_{11} & F_{2}^{s}\Pi_{12}\\
F_{0}^{s}\Pi_{20} & F_{1}^{s}\Pi_{21} & 1+F_{2}^{s}\Pi_{22}
\end{array}\right)=0.
\end{equation}
It turns out that the secular equation always has at least one real solution $s>1$ as long as the following inequality,
\begin{equation}\label{ineq:three-2D}
\left[1-\frac{F_{0}^{s}}{2}\left(1+\frac{F_{1}^{s}}{2}\right)\right]F_{2}^{s}+\left[F_{0}^{s}\left(1+\frac{F_{1}^{s}}{2}\right)+F_{1}^{s}\right]>0,
\end{equation}
is satisfied (see Appendix \ref{app:ZSL2D}). 

In particular, we are interested in the two incompressible limits: (i) $1+\frac{F_{1}^{s}}{2}=0$ and (ii) $F_{0}^{s}\to+\infty$: 

(i) In the incompressibility condition $1+\frac{F_1^s}{2}=0$, we have
\begin{equation}
s=\begin{cases}
1+\frac{1}{2\left[1+\frac{2F_{2}^{s}}{F_{2}^{s}-2}\right]^{2}}, & F_{2}^{s}>2;\\
\sqrt{\frac{F_{2}^{s}}{8}-\frac{3}{4}}, & F_{2}^{s}\rightarrow\infty.
\end{cases}
\end{equation}
When $F_{2}^{s}>2$, there exists at least one weakly damped zero sound mode with the sound speed 
\begin{equation*}
c_{0}=sv_{F}=v_{F}\left[1+\frac{1}{2\left(1+\frac{2F_{2}^{s}}{F_{2}^{s}-2}\right)^{2}}\right].
\end{equation*}
Moreover, in the limit of $F_{2}^{s}\to{}\infty$, an additional weakly damped zero sound mode occurs with the sound speed $c_{0}=v_{F}\sqrt{\frac{F_{2}^{s}}{8}-\frac{3}{4}}$.

(ii) In the other incompressibility condition $F_{0}^{s}\to+\infty$, we have two approximate solutions to the secular equation
\begin{equation}
s=\begin{cases}
1+\frac{1}{2\left[1+\frac{2F_{2}^{s}}{F_{2}^{s}-2}\right]^{2}} & s\rightarrow1+0^{+}\\
\sqrt{\frac{F_{0}^{s}}{2}\left(1+\frac{F_{1}^{s}}{2}\right)} & s\rightarrow\infty
\end{cases}
\end{equation}
Here the constraint $F_{2}^{s}>2$ is still imposed by Eq.~\eqref{ineq:three-2D}.

\subsubsection{Transverse mode $\nu_{\mathrm{p}}^{-}$}

Similar to the $m>0$ modes in 3D, the transverse zero sound mode can be depicted by a two channel model, on which we keep only $F_{1}^{s}$ and $F_{2}^{s}$ while let $F_{l}^{s}=0$ for $l\geq 3$.
The components of algebraic equation I can be classified into two groups: (i) $l\geq 3$ and (ii) $l=1$ and $l=2$. The components in the second group form a close set that is composed of $v_{1}$ and $v_{2}$:
\begin{subequations}
\begin{align}
\frac{v_{1}}{2}+F_{1}^{s}\Xi_{11}\frac{v_{1}}{2}+F_{2}^{s}\Xi_{12}\frac{v_{2}}{2} & =-\Xi_{11}U,\\
\frac{v_{2}}{2}+F_{1}^{s}\Xi_{21}\frac{v_{1}}{2}+F_{2}^{s}\Xi_{22}\frac{v_{2}}{2} & =-\Xi_{21}U.
\end{align}
\end{subequations}
The solution from algebraic equation I leads to the response functions as follows,
\begin{subequations}
\begin{align}
\frac{v_{1}}{2U} & =-\frac{\left(1+\frac{F_{2}^{s}}{2}\right)\Xi_{11}}{\left(1+F_{1}^{s}\Xi_{11}\right)\left(1+\frac{F_{2}^{s}}{2}\right)+F_{2}^{s}4s^{2}\Xi_{11}},\\
\frac{v_{2}}{2U} & =-\frac{2s\Xi_{11}}{\left(1+F_{1}^{s}\Xi_{11}\right)\left(1+\frac{F_{2}^{s}}{2}\right)+F_{2}^{s}4s^{2}\Xi_{11}}.
\end{align}
\end{subequations}
The zero sound mode frequency can be determined via the pole of the response functions, i.e.,
\begin{equation}\label{eq:mode_fre_2D_two_trans}
\bar{h}_{2}(s)\equiv\left(1+F_{1}^{s}\Xi_{11}(s)\right)\left(1+\frac{F_{2}^{s}}{2}\right)+4s^{2}F_{2}^{s}\Xi_{11}(s)=0.
\end{equation}
To have a real solution $s$ to $\bar{h}_{2}(s)=0$, one requires
\begin{equation}\label{eq:ineq-tmode-2D}
F_{2}^{s}\left[F_{1}^{s}\left(1+\frac{F_{2}^{s}}{2}\right)+4F_{2}^{s}\right]<0.
\end{equation}
In the limit of $0<s-1\ll1$, an approximate solution is found to be
\begin{equation}
s=1+\frac{1}{2\left[1-\frac{2+F_{2}^{s}}{F_{1}^{s}\left(1+\frac{F_{2}^{s}}{2}\right)+4F_{2}^{s}}\right]^{2}}.
\end{equation}
Under the QSL condition $1+\frac{F_{1}^{s}}{2}=0$, Eq.~\eqref{eq:ineq-tmode-2D} gives rise to $0<F_{2}^{s}<\frac{2}{3}$, and the approximate solution is simplified as follows,
\begin{equation}
s=1+\frac{1}{8}\left(\frac{3F_{2}^{s}-2}{F_{2}^{s}-2}\right)^{2}.
\end{equation}

\begin{figure}[tb]
	\centering
	\includegraphics[width=0.8\linewidth]{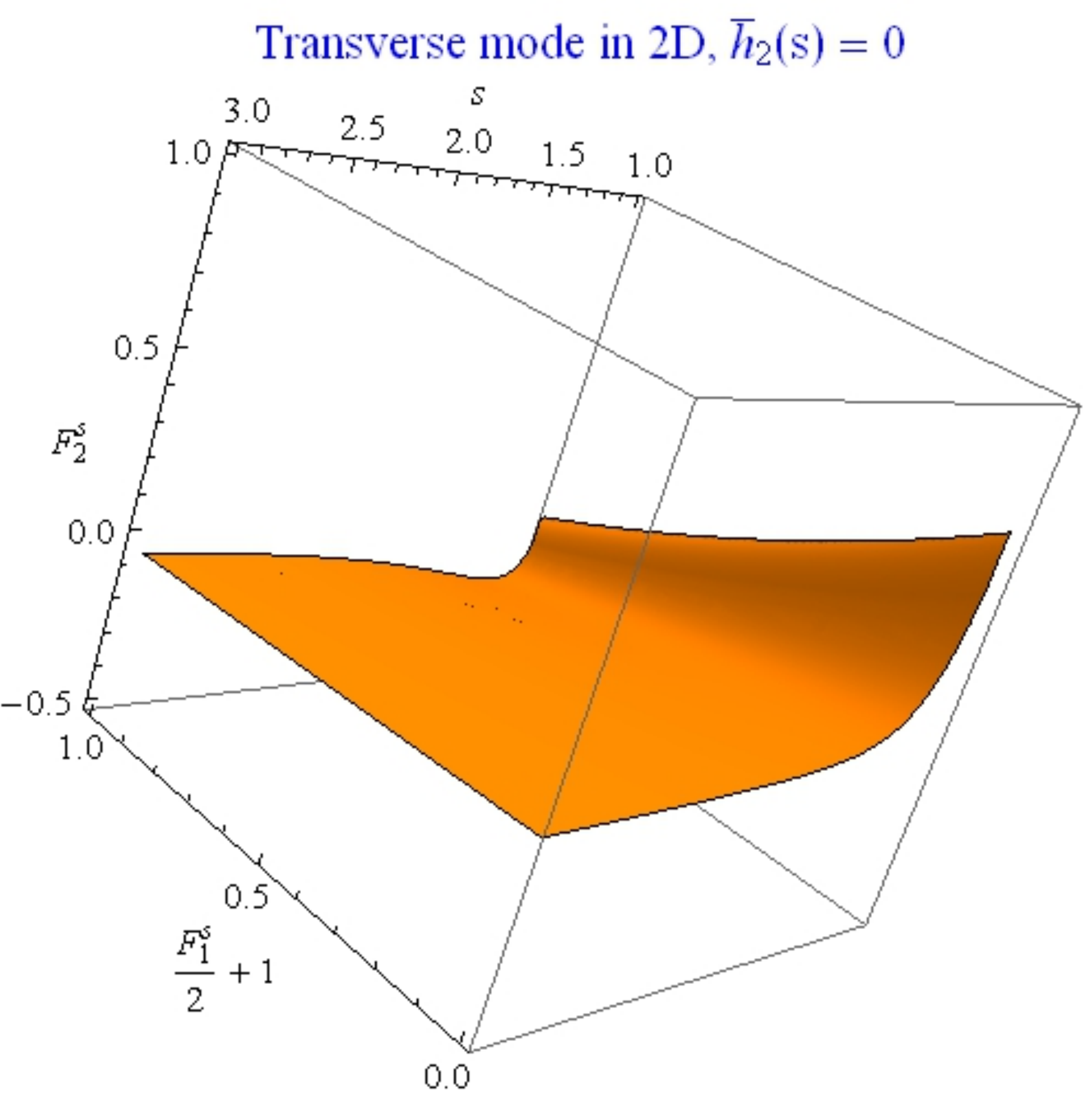}
	\caption{\label{fig:h0t2D} Transverse mode in 2D. The numerical solution for the frequency equation $\bar{h}_{2}(s)=0$ is plotted in the 3D parameter space $\left(s,1+\frac{F_{1}^{s}}{2},F_{2}^{s}\right)$, where $\bar{h}_{2}(s)$ is defined in Eq.~\eqref{eq:mode_fre_2D_two_trans}}
\end{figure}

The numerical solution to Eq.~\eqref{eq:mode_fre_2D_two_trans} can be found in Fig.~\ref{fig:h0t2D}, where the zeros of the function $\bar{h}_{2}(s)$ are plotted as a 3D contour in the parameter space $\left(s,1+\frac{F_{1}^{s}}{2},F_{2}^{s}\right)$. In order to have a real solution $s$ to $\bar{h}_{2}(s)=0$, the Landau parameter $F_{2}^{s}$ should be restricted to a finite region given in Eq.~\eqref{eq:ineq-tmode-2D}. This is quite different from the situation in 3D, in which a positive and sufficiently large $F_{2}^{s}$ always gives rise to a weakly damped transverse mode.

\subsection{First sound: $\omega\tau\ll{}1$}

As mentioned in Section \ref{sec:FS3D}, the transverse sound mode will become strongly damped in the hydrodynamic regime. So that we will discuss the longitudinal mode only. Similar the 3D case, the simplest model for studying first sound mode is the two-channel model with finite $F_{0}^{s}$ and $F_{1}^{s}$ and vanishing $F_{l}^{s}=0$ for $l\geq 2$. Then Landau kinetic equation becomes
\begin{equation}
\left(\begin{array}{cc}
1+F_{0}^{s}\Pi_{00}+i\kappa\Pi_{0} & \frac{1}{2}F_{1}^{s}\Pi_{01}+i\kappa\Pi_{00}\\
F_{0}^{s}\Pi_{10}+i\kappa\Pi_{1} & \frac{1}{2}\left(1+F_{1}^{s}\Pi_{11}\right)+i\kappa\Pi_{10}
\end{array}\right)\left(\begin{array}{c}
\nu_{0}\\
\nu_{1}
\end{array}\right)=0.
\end{equation}
The first sound frequency is determined by the secular equation,
\begin{equation}
\det\left(\begin{array}{cc}
1+F_{0}^{s}\Pi_{00}+i\kappa\Pi_{0} & \frac{1}{2}F_{1}^{s}\Pi_{01}+i\kappa\Pi_{00}\\
F_{0}^{s}\Pi_{10}+i\kappa\Pi_{1} & \frac{1}{2}\left(1+F_{1}^{s}\Pi_{11}\right)+i\kappa\Pi_{10}
\end{array}\right)=0.
\end{equation}
After some simple algebra, we obtain the dispersion relation for first sound in 2D (see Appendix \ref{app:FS2D}), which is given by
\begin{equation}
\left(\frac{\omega}{qv_{F}}\right)^{2}=\frac{1}{2}\left(1+F_{0}^{s}\right)\left(1+\frac{F_{1}^{s}}{2}\right)-\frac{1}{4}i\omega\tau\left(1+\frac{F_{1}^{s}}{2}\right).
\end{equation}
In the limit of $\omega\tau\ll{}1$, the first sound speed reads
\begin{equation}
c_{1}=\frac{\omega}{q}=v_{F}\sqrt{\frac{1}{2}\left(1+F_{0}^{s}\right)\left(1+\frac{1}{2}F_{1}^{s}\right)}.    
\end{equation}

\section{Summary and discussions}\label{sec:sum}

In summary, by using a Landau-type effective theory, we have studied collective modes, i.e., zero and first sound modes for an electronic liquid with spatial dimensionality $d=2,3$. 

\begin{itemize}

\item{}All these collective modes can be classified according to symmetries.
\begin{itemize}
\item{}In 3D, due to the $O(2)$ rotational symmetry along the propagating direction $\hat{\mathrm{q}}$, collective modes can be decoupled in accordance with magnetic quantum number $m$. Namely, $m=0$ labels the longitudinal mode, $m=\pm{}1$ label transverse modes, and $|m|\geq{}2$ denote other higher angular momentum modes.
\item{}In 2D, the $O(2)$ symmetry is reduced to the reflection symmetry $Z_{2}:\theta\to{}-\theta$, such that there are only two types of collective modes, i.e., longitudinal and transverse modes.
\end{itemize}

\item{}By investigating a particular component of longitudinal zero sound mode, namely, compression and expansion mode $\nu_{0}$, we have found two incompressibility conditions as follows:
\begin{itemize}
\item{}(i) $1+\frac{F_{1}^{s}}{d}=0$, which gives rise to a gapless QSL state that is a charge insulator and a thermal conductor.
\item{}(ii) $F_{0}^{s}\to{}+\infty$, which leads to a conventional insulator that is both charge and thermal insulating.
\item{}Other zero and first sound modes have been studied under these two incompressibility conditions respectively.
\end{itemize}

\item{}For zero sounds, we have derived two types (I and II) of algebraic equations from Landau's kinetic equation. These algebraic equations can be solved via Landau parameter truncation, resulting in:
\begin{itemize}
 \item{}For longitudinal modes, there exists at least one longitudinal mode in both incompressibility conditions (i) and (ii), as long as the Landau parameter $F_{2}^{s}$ is large enough, i.e., $F_{2}^s>2$ in 2D or $F_{2}^s>10/3$ in 3D. 
 
 \item{}For transverse modes, the Landau parameter $F_{0}^{s}$ does not have an effect on transverse modes in either 2D or 3D. The existence of a transverse mode is determined by $F_{1}^{s}$ and $F_{2}^{s}$. However, the condition will be quite different in 2D and 3D.
 
 \begin{itemize}
 \item{}For a gapless QSL state in 2D, there exists a weakly damped transverse mode only when $0<F_{2}^{s}<2/3$. 
 \item{}For a gapless QSL state in 3D, a weakly damped transverse mode exists as long as $F_{2}^{s}$ is sufficiently large, i.e., $F_{2}^{s}>15/2$. 
 \end{itemize}

 \item{}In 3D, there allows higher angular momentum modes with $m\geq{}2$. When the Landau parameter $F_{m+1}^{s}$ is positive and sufficiently large, there will exist a weakly damped zero sound mode for a given $m\geq{}2$.
\end{itemize}

\item{}For first sounds, we find that the only weakly damped mode is the longitudinal $m=0$ mode and other first sound modes will be strongly damped. The sound velocity of the longitudinal mode is found to be $c_{1}=v_{F}\sqrt{\frac{1}{d}\left(1+F_{0}^{s}\right)\left(1+\frac{F_{1}^{s}}{d}\right)}$. Thus, when the electronic liquid becomes incompressible in two conditions, the first sound velocity will be quite different in these two conditions:
\begin{itemize}
\item{}For (i) $1+\frac{F_{1}^{s}}{d}=0$, $c_{1}\to{}0$.
\item{}While for (ii) $F_{0}^{s}\to{}+\infty$, $c_{1}\to{}+\infty$.
\end{itemize}

\end{itemize}

All the above results have been summarized in Table~\ref{tab:sound}.

In 3D, we have also found a hierarchy structure for gapless QSL states depicted by the Landau-type effective theory. It remains a open question whether there is a similar hierarchy structure in 2D, and we would like to leave this issue for future studies.

\begin{table*}[htbp]
\caption{A summary for sound modes in $d=2,3$. L: longitudinal mode, T: transverse mode.}
\label{tab:sound}
\renewcommand\arraystretch{2.5}
\newcommand{\rb}[1]{\raisebox{3.2ex}{#1}}
\setlength{\tabcolsep}{1.0ex}
\begin{tabular}{|c|c|l|l|}
\hline\hline
 & Mode & $d=2$ & $d=3$ \\
\hline
Stability & All & $1+F_{0}^{s}\geq{}0$, and $1+\frac{F_{l}^{s}}{2}\geq{}0$ for $l=1,2,3,\cdots$ & $1+\frac{F_{l}^{s}}{2l+1}\geq{}0$ \\
\hline
 &  & \multicolumn{2}{l|}{ $\Leftarrow$ \hspace{0.5ex} (1) \hspace{1ex}  $1+\frac{F_{1}^{s}}{d}=0$ \hspace{2ex} (QSL)}  \\
\cline{3-4}
\rb{Incompressibility} & \rb{$\nu_{0}=0$} & \multicolumn{2}{l|}{$\Leftarrow$ \hspace{0.5ex} (2) \hspace{1ex} $F_{0}^{s}\to{}+\infty$ \hspace{2ex}(conventional insulator)}  \\
\hline\hline
 & & \multicolumn{2}{c|}{Type I and II Algebraic Equations}  \\
\cline{3-4}
 &  & $\frac{u_{l}+\delta_{l0}u_{0}}{2}+\sum_{l'=0}^{\infty}F_{l'}^{s}\Pi_{ll'}\left(s\right)\frac{u_{l'}+\delta_{l'0}u_{0}}{2}=-\Pi_{l0}\left(s\right)U$ & $\frac{\nu _{l}}{2l+1}+\sum_{l'=0}F_{l^{\prime }}^{s}\Omega_{ll^{\prime }}(s)\frac{\nu _{l^{\prime }}}{2l^{\prime }+1}=-\Omega_{l0}(s)U$ \\
\cline{3-4}
 & & $u_{l}s-\left(1+\frac{F_{l-1}^{s}+\delta_{l1}F_{0}^{s}}{2}\right)\frac{u_{l-1}+\delta_{l1}u_{0}}{2}-\left(1+\frac{F_{l+1}^{s}}{2}\right)\frac{u_{l+1}}{2}=\delta_{l1}U$ & $\nu _{l}s-\left( 1+\frac{F_{l-1}^{s}}{2l-1}\right) \frac{l}{2l-1}\nu_{l-1}-\left( 1+\frac{F_{l+1}^{s}}{2l+3}\right) \frac{l+1}{2l+3}\nu_{l+1}=\delta _{l1}U$ \\
\cline{3-4}
 & \rb{L} & \multicolumn{2}{c|}{Solutions in Two Incompressibility Conditions (i) and (ii)}  \\
 \cline{3-4}
 &  & $(1)\Rightarrow{}s\simeq\begin{cases}
1+\frac{1}{2}\left(\frac{F_{2}^{s}-2}{3F_{2}^{s}-2}\right)^2, & F_{2}^{s}>2\\
\sqrt{\frac{F_{2}^{s}}{8}-\frac{3}{4}}, & F_{2}^{s}\rightarrow\infty
\end{cases}$ & $(1)\Rightarrow{}s\simeq\begin{cases}
1+2e^{-\frac{11}{3}-
\frac{50}{9}\frac{1} {F_{2}^{s}-10/3}} & F_{2}^{s}>\frac{10}{3},\\
\frac{3}{5}\sqrt{\frac{F_{2}^{s}}{7}-\frac{5}{3}}, & F_{2}^{s}\to+\infty.
\end{cases}$ \\
\cline{3-4}
 &  & $(2)\Rightarrow{}s\simeq\begin{cases}
1+\frac{1}{2}\left(\frac{F_{2}^{s}-2}{3F_{2}^{s}-2}\right)^2, & F_{2}^{s}\to{}2+0^{+}\\
\sqrt{\frac{F_{0}^{s}}{2}\left(1+\frac{F_{1}^{s}}{2}\right)}, & F_{2}^{s}>2
\end{cases}$ & $(2)\Rightarrow{}s\simeq\begin{cases}
1+2e^{-\frac{11}{3}-
\frac{50}{9}\frac{1} {F_{2}^{s}-10/3}}, & F_{2}^{s}\to{}\frac{10}{3}+0^{+}\\
\sqrt{\frac{F_{0}^{s}}{3}\left(1+\frac{F_{1}^{s}}{3}\right)}, & F_{2}^{s}>\frac{10}{3}
\end{cases}$ \\
\cline{2-4}
\rb{Zero Sound} & & \multicolumn{2}{c|}{Type I and II Algebraic Equations}  \\
\cline{3-4}
 &  &  $\frac{v_{l}}{2}+\sum_{l'=1}^{\infty}F_{l'}^{s}\Xi_{ll'}\left(s\right)\frac{v_{l'}}{2}=-\Xi_{l1}\left(s\right)U$ & $\frac{\nu_{l}^{m}}{2l+1}+\sum_{l'=m}^{\infty}F_{l'}^{s}\Omega_{ll'}^{m}\left(s\right)\frac{\nu_{l'}^{m}}{2l'+1}=-\Theta_{l}^{m}\left(s\right)U$ \\
\cline{3-4}
 & T & $v_{l}s-\left(1+\frac{F_{l-1}^{s}}{2}\right)\frac{v_{l-1}}{2}-\left(1+\frac{F_{l+1}^{s}}{2}\right)\frac{v_{l+1}}{2}=\delta_{l2}\frac{U}{2}$ &  $\begin{array}{ll}
\nu _{l}^{m}s & -\nu _{l-1}^{m}\left( 1+\frac{F_{l-1}^{s}}{2l-1}\right) \frac{\sqrt{l^{2}-m^{2}}}{2l-1}\\
&-\nu _{l+1}^{m}\left( 1+\frac{F_{l+1}^{s}}{2l+3}\right) 
\frac{\sqrt{(l+1)^{2}-m^{2}}}{2l+3}=\alpha _{l}^{m}U\end{array}$ \\
\cline{3-4}
 & & \multicolumn{2}{c|}{Solutions in Incompressibility Condition (i)}  \\
 \cline{3-4}
 & & $(1)\Rightarrow{}s\simeq1+\frac{1}{8}\left(\frac{3F_{2}^{s}-2}{F_{2}^{s}-2}\right)^{2}, \,\mbox{if}\,\, 0<F_{2}^{s}<\frac{2}{3}$ &  $(1)\Rightarrow{}s\to{}1+0^{+},\,\,\mbox{when}\,\, F_{2}^{s}\to\frac{15}{2}+0^{+}$\\
\cline{2-4}
 & $m\geq{}2$ & none & $s\to{}1+0^{+}$, when $1+\frac{F_{m}^{s}}{2m+1}\to{}0^{+}$ \& $\frac{2F_{2m+1}^{s}}{(2m+1)(2m+3)}\to{}1+0^{+}$ \\
\hline\hline
 & & \multicolumn{2}{c|}{Dispersion Relation}  \\
 \cline{3-4}
 First Sound & L & $\left(\frac{\omega}{qv_{F}}\right)^{2}=\frac{1}{2}\left(1+F_{0}^{s}\right)\left(1+\frac{F_{1}^{s}}{2}\right)-\frac{1}{4}i\omega\tau\left(1+\frac{F_{1}^{s}}{2}\right)$ & $\left(\frac{\omega}{qv_{F}}\right)^{2}=\frac{1}{3}(1+F_{0}^{s})\left(1+\frac{F_{1}^{s}}{3}\right)-\frac{4}{15}i\omega\tau\left(1+\frac{F_{1}^{s}}{3}\right)$ \\
\cline{3-4}
 &  & \multicolumn{2}{c|}{$c_{1}=v_{F}\sqrt{\frac{1}{d}\left(1+F_{0}^{s}\right)\left(1+\frac{F_{1}^{s}}{d}\right)}$}\\
\hline\hline
\end{tabular}
\end{table*}

\section*{Acknowledgment}

YZ would like to thank Tai-Kai Ng for the collaboration on relevant topics. This work is supported in part by National Natural Science Foundation of China (No.12034004), 
the K. C. Wong Education Foundation (Grant No. GJTD-2020-01) 
and the Strategic Priority Research Program of Chinese Academy of Sciences (No. XDB28000000). 
JJM is supported by General Research Fund Grant No. 14302021 from Research Grants Council and Direct Grant No. 4053416 from the Chinese University of Hong Kong.
HKJ is supported by the European Research Council (ERC) under the European Unions Horizon 2020 research and innovation program (grant agreement No. 771537).

\clearpage
\appendix

\begin{widetext}
\section{Zero sound: longitudinal ($m=0$) mode in 3D}\label{app:m0}
In this appendix, we shall derive the algebraic equations for the longitudinal model in the collisionless regime.
With the help of the addition theorem,
\begin{equation*}
P_{l}\left(\cos\theta_{\mathrm{\mathrm{pp'}}}\right)=\sum_{m=-l}^{l}Y_{l}^{m}\left(\theta_{\mathrm{p}},\phi_{\mathrm{p}}\right)Y_{l}^{m*}\left(\theta_{\mathrm{p'}},\phi_{\mathrm{p'}}\right),
\end{equation*}
we have
\begin{align*}
\sum_{\mathrm{p'}}f_{\mathrm{pp'}}^{s}\frac{\partial n_{\mathrm{p'}}^{0}}{\partial\varepsilon_{\mathrm{p'}}}\nu_{\mathrm{p'}} & =N\left(0\right)\int d\varepsilon\int\frac{d\mathrm{\hat{p}'}}{4\pi}f_{\mathrm{pp'}}^{s}\frac{\partial n_{\mathrm{\varepsilon}}^{0}}{\partial\varepsilon}\nu_{\mathrm{p'}}\nonumber \\
 & =-N\left(0\right)\int\frac{d\mathrm{\hat{p}'}}{4\pi}\sum_{l=0}^{\infty}P_{l}\left(\cos\theta_{\mathrm{\mathrm{pp'}}}\right)f_{l}^{s}\sum_{l'=0}^{\infty}P_{l'}\left(\cos\theta_{\mathrm{p'}}\right)\nu_{l'}\nonumber \\
 & =-\int\frac{d\mathrm{\hat{p}'}}{4\pi}\sum_{l=0}^{\infty}\sum_{m=-l}^{l}Y_{l}^{m}\left(\theta_{\mathrm{p}},\phi_{\mathrm{p}}\right)Y_{l}^{m*}\left(\theta_{\mathrm{p'}},\phi_{\mathrm{p'}}\right)F_{l}^{s}\sum_{l'=0}^{\infty}P_{l'}\left(\cos\theta_{\mathrm{p'}}\right)\nu_{l'}\nonumber \\
 & =-\sum_{l=0}^{\infty}\sum_{m=-l}^{l}Y_{l}^{m}\left(\theta_{\mathrm{p}},\phi_{\mathrm{p}}\right)F_{l}^{s}\sum_{l'=0}^{\infty}\int\frac{d\mathrm{\hat{p}'}}{4\pi}Y_{l}^{m*}\left(\theta_{\mathrm{p'}},\phi_{\mathrm{p'}}\right)Y_{l'}^{0}\left(\theta_{\mathrm{p'}},\phi_{\mathrm{p'}}\right)\nu_{l'}\nonumber \\
 & =-\sum_{l=0}^{\infty}\sum_{m=-l}^{l}Y_{l}^{m}\left(\theta_{\mathrm{p}},\phi_{\mathrm{p}}\right)F_{l}^{s}\sum_{l'=0}^{\infty}\frac{1}{2l+1}\delta_{ll'}\delta_{m0}\nu_{l'}\nonumber \\
 & =-\sum_{l=0}^{\infty}Y_{l}^{0}\left(\theta_{\mathrm{p}},\phi_{\mathrm{p}}\right)F_{l}^{s}\frac{\nu_{l}}{2l+1}\nonumber \\
 & =-\sum_{l=0}^{\infty}F_{l}^{s}P_{l}\left(\cos\theta_{\mathrm{p}}\right)\frac{\nu_{l}}{2l+1}.
\end{align*}
Putting the above into Eq.~(\ref{LTEZS1}) leads to
\begin{equation}
\sum_{l}\nu _{l}P_{l}(\mu )-\frac{\mu }{s-\mu }\sum_{l}F_{l}^{s}\frac{1}{2l+1}\nu _{l}P_{l}(\mu )=\frac{\mu }{s-\mu }U, \label{LTEZS2}
\end{equation}
where $s\equiv \frac{\omega }{qv_{F}}$ and $\mu =\cos \theta _{\mathrm{p}}$.
Using the orthogonal relations,
\begin{equation*}
\int_{-1}^{1}d\mu P_{l}(\mu )P_{l^{\prime }}(\mu )=\frac{2}{2l+1}\delta_{ll^{\prime }},
\end{equation*}
we find that Eq.~(\ref{LTEZS2}) becomes Eq.~\eqref{LTEZS3}.

On the other hand, by using Bonnet's recursion formula
\begin{equation*}
\mu P_{l}(\mu )=\frac{1}{2l+1}\left[ (l+1)P_{l+1}(\mu )+lP_{l-1}(\mu )\right],
\end{equation*}
we also have
\begin{eqnarray*}
\mu U &=&\sum_{l}\nu _{l}\left[ s-\left( 1+\frac{F_{l}^{s}}{2l+1}\right) \mu \right] P_{l}(\mu ) \\
&=&\sum_{l}\nu _{l}sP_{l}(\mu )-\sum_{l}\nu _{l}\left( 1+\frac{F_{l}^{s}}{2l+1}\right) \frac{1}{2l+1}\left[ (l+1)P_{l+1}(\mu )+lP_{l-1}(\mu )\right] \\
&=&\sum_{l}P_{l}(\mu )\left[ \nu _{l}s-\left( 1+\frac{F_{l-1}^{s}}{2l-1}\right) \frac{l}{2l-1}-\left( 1+\frac{F_{l+1}^{s}}{2l+3}\right) \frac{l+1}{2l+3}\right] .
\end{eqnarray*}
This leads another set of algebraic equations, i.e., Eq.~\eqref{LTEZS4}, as
\begin{equation}
\nu _{l}s-\left( 1+\frac{F_{l-1}^{s}}{2l-1}\right) \frac{l}{2l-1}\nu_{l-1}-\left( 1+\frac{F_{l+1}^{s}}{2l+3}\right) \frac{l+1}{2l+3}\nu_{l+1}=\delta _{l1}U.
\end{equation}

For later use, we write down some useful relations for $\Omega _{ll^{\prime}}(s)$ at first, i.e. $\Omega _{ll^{\prime }}(s)$ can be expressed directly in terms of the first and second kind of Legendre functions, $P_{l}$ and $Q_{l}$, as
\begin{equation}
\Omega _{l^{\prime }l}=\Omega _{ll^{\prime }}=\frac{\delta _{ll^{\prime }}}{2l+1}-sP_{l^{\prime }}(s)Q_{l}(s),\; l^{\prime }\leq l. \label{eq:rec}
\end{equation}

\subsection{Algebraic equations of type I and type II}
In the main text, we obtain the type I algebraic equation [i.e., Eq.~\eqref{LTEZS3}],
\begin{equation}
\frac{\nu_{l}}{2l+1}+\sum_{l'=0}^{\infty}F_{l'}^{s}\Omega_{ll'}\frac{\nu_{l'}}{2l'+1}=-\Omega_{l0}U,
\end{equation}
and type II algebraic equation [i.e., Eq.~\eqref{LTEZS4}],
\begin{equation}
\nu_{l}s-l\left(1+\frac{F_{l-1}^{s}}{2l-1}\right)\frac{\nu_{l-1}}{2l-1}-\left(l+1\right)\left(1+\frac{F_{l+1}^{s}}{2l+3}\right)\frac{\nu_{l+1}}{2l+3}=\delta_{l1}U,
\end{equation}
which are both derived from the Landau kinetic equation in the collisionless regime and thus are equivalent. It's an easy exercise that one can derive algebraic equation type I from algebraic equation type II by multiplying the Legendre polynomials, summing over all components, and using Bonnet's recursion formula for Legendre polynomials. 

The algebraic equations type I and type II are two representations of Landau kinetic equation in the collisionless regime. The former is in the infinitely coupled form and can be solved by truncation in Landau parameters $F_{l}^{s}$, while the later is in the hierarchical from. In order to manifest the equivalence, multiplying the $l=0$ and $l=1$ components of algebraic equation I by $\Omega_{10}$ and $\Omega_{00}$, respectively, and meanwhile using the relations $\Omega_{l1}=s\Omega_{l0}+\frac{1}{3}\delta_{l1}$, we can eliminate the $U$ terms and obtain
\begin{align}
0= & \left(\nu_{0}+\sum_{l'=0}^{\infty}F_{l'}^{s}\Omega_{0l'}\frac{\nu_{l'}}{2l'+1}\right)\Omega_{10}-\left(\frac{\nu_{1}}{3}+\sum_{l'=0}^{\infty}F_{l'}^{s}\Omega_{1l'}\frac{\nu_{l'}}{2l'+1}\right)\Omega_{00}\nonumber \\
= & \left[\left(1+F_{0}^{s}\Omega_{00}\right)\Omega_{10}-F_{0}^{s}\Omega_{10}\Omega_{00}\right]\nu_{0}+\left[F_{1}^{s}\Omega_{01}\Omega_{10}-\left(1+F_{1}^{s}\Omega_{11}\right)\Omega_{00}\right]\frac{\nu_{1}}{3}+\sum_{l'=2}^{\infty}F_{l'}^{s}\left(\Omega_{0l'}\Omega_{10}-\Omega_{1l'}\Omega_{00}\right)\frac{\nu_{l'}}{2l'+1}\nonumber \\
= & \Omega_{10}\nu_{0}-\Omega_{00}\left(1+\frac{F_{1}^{s}}{3}\right)\frac{\nu_{1}}{3}\nonumber \\
= & \Omega_{00}\left[\nu_{0}s-\left(1+\frac{F_{1}^{s}}{3}\right)\frac{\nu_{1}}{3}\right],
\end{align}
which is nothing but the $l=0$ component of type II algebraic equation,
\begin{equation}
\nu_{0}s-\left(1+\frac{F_{1}^{s}}{3}\right)\frac{\nu_{1}}{3}=0.
\end{equation}

\subsection{Two channel model} 
We consider a two channel model by keeping only $F_{0}^{s}$ and $F_{1}^{s}$ and setting $F_{l}^{s}=0$ for $l\geq 2$. Introducing a new variable $\tilde{\nu}_{l}\equiv \frac{\nu_{l}}{2l+1}$ for convenience, one can simplify Eq.~\eqref{LTEZS3} as follows,
\begin{equation}
\tilde{\nu}_{l}+F_{0}^{s}\Omega _{l0}(s)\tilde{\nu}_{0}+F_{1}^{s}\Omega_{l1}(s)\tilde{\nu}_{1}=-\Omega _{l0}(s)U.  \label{ZS21}
\end{equation}	
The above consists of a series of algebraic equations that can be classified into two categories: (i) $l\geq 2$ and (ii) $l=0$ and $l=1$. The equations in the second category form a close set which is composed of $\tilde{\nu}_{0}$ and $\tilde{\nu}_{1}$:
\begin{equation}
\left( 
\begin{array}{cc}
1+F_{0}^{s}\Omega _{00} & F_{1}^{s}\Omega _{01} \\ 
F_{0}^{s}\Omega _{10} & 1+F_{1}^{s}\Omega _{11} \\ 
\end{array}
\right) \left(
\begin{array}{c}
\tilde{\nu}_{0} \\ 
\tilde{\nu}_{1} \\ 
\end{array}
\right)=-U \left(
\begin{array}{c}
\Omega _{00} \\
\Omega _{10}
\end{array}
\right).
\end{equation}
The solution to above equations leads to the response functions as follows,
\begin{subequations}\label{eq:resp1}
\begin{align}
\frac{\tilde{\nu}_{0}}{U} & =-\frac{\left(1+F_{1}^{s}\Omega_{11}\right)\Omega_{00}-F_{1}^{s}\Omega_{01}\Omega_{10}}{\left(1+F_{0}^{s}\Omega_{00}\right)\left(1+F_{1}^{s}\Omega_{11}\right)-F_{0}^{s}\Omega_{10}F_{1}^{s}\Omega_{01}}=-\frac{\left(1+\frac{F_{1}^{s}}{3}\right)\Omega_{00}}{\left(1+F_{0}^{s}\Omega_{00}\right)\left(1+\frac{F_{1}^{s}}{3}\right)+F_{1}^{s}s^{2}\Omega_{00}},\\
\frac{\tilde{\nu}_{1}}{U} & =-\frac{-F_{0}^{s}\Omega_{10}\Omega_{00}+\left(1+F_{0}^{s}\Omega_{00}\right)\Omega_{10}}{\left(1+F_{0}^{s}\Omega_{00}\right)\left(1+F_{1}^{s}\Omega_{11}\right)-F_{0}^{s}\Omega_{10}F_{1}^{s}\Omega_{01}}=-\frac{\Omega_{10}}{\left(1+F_{0}^{s}\Omega_{00}\right)\left(1+\frac{F_{1}^{s}}{3}\right)+F_{1}^{s}s^{2}\Omega_{00}}.
\end{align}
\end{subequations}

From the above, we find that a complete suppression of $\tilde{\nu}_{0}$ component requires that
\begin{equation}
\left(1+\frac{F_{1}^{s}}{3}\right)\Omega_{00}=0.
\end{equation}
As the solution to $\Omega_{00}\left(s\right)=0$ is $s=\infty$, $1+\frac{1}{3}F_{1}^{s}=0$ is nothing but the QSL condition given in Eq.~\eqref{eq:F1s}.

We proceed to consider the possibility of the complete suppression of $\tilde{\nu}_{1}$ component, which leads to
\begin{equation}
\Omega_{10}=s\Omega_{00}=0.
\end{equation}
The solutions are $s=0$ and $s=\infty$. This means that it is impossible to suppress the $l=1$ component entirely, and such a mode is generally allowed.

The mode frequency can be determined via the pole of $\tilde{\nu}_{l}$ from Eqs.~\eqref{eq:resp1}, i.e.,
\begin{equation}
1+\frac{F_{1}^{s}}{3}+\Omega _{00}
\left[F_{0}^{s}\left( 1+\frac{F_{1}^{s}}{3}\right)+F_{1}^{s}s^{2}\right] = 0.\label{eq:two-m0}
\end{equation}
To find a real solution $s$ to the above equation, we define a function
\begin{equation*}
g_{2}(s)=1+\frac{F_{1}^{s}}{3}+\Omega _{00}
\left[F_{0}^{s}\left( 1+\frac{F_{1}^{s}}{3}\right)+F_{1}^{s}s^{2}\right],
\end{equation*}
and have $g_{2}(s\to{}\infty)=1>0$. 
Meanwhile, when $s\to{}1$, the leading term in $g_{2}(s)$ reads
\begin{equation*}
g_{2}\left(s\to{}1\right)=\frac{1}{2}\left[ F_{0}^{s}\left(1+\frac{F_{1}^{s}}{3}\right)+F_{1}^{s}\right]\ln\frac{s-1}{2} + O(1).   
\end{equation*}
Therefore, to have a real solution for $g_{2}(s)=0$, one requires that
\begin{equation}\label{ineq:two-m0}
F_{0}^{s}\left(1+\frac{F_{1}^{s}}{3}\right)+F_{1}^{s} > 0.
\end{equation}

Eq.~\eqref{eq:two-m0} must generally be found numerically. Using asymptotic expressions for $\Omega_{00}(s)$ as
\begin{equation*}
\Omega_{00}\left(s\right)=\begin{cases}
1+\frac{1}{2}\ln\frac{s-1}{2}, & s\to{}1+0^{+},\\
-\frac{1}{3s^{2}}-\frac{1}{5s^{4}}-\frac{1}{7s^{6}}-\cdots, & s\rightarrow\infty,
\end{cases}
\end{equation*}
one can find that Eq.~\eqref{eq:two-m0} can be approximately  solved in two limits
\begin{equation}
s=\begin{cases}
1+2e^{-2\left[1+\frac{1+\frac{F_{1}^{s}}{3}}{F_{0}^{s}\left(1+\frac{F_{1}^{s}}{3}\right)+F_{1}^{s}}\right]}, & 0<s-1\ll1,\\
\sqrt{\frac{F_{0}^{s}}{3}\left(1+\frac{F_{1}^{s}}{3}\right)+\frac{F_{1}^{s}}{5}}, & s\rightarrow\infty.
\end{cases}
\end{equation}
Note that in the weak coupling limit $F_{1}^{s}\ll F_{0}^{s}\ll1$, the zero sound velocity is $c_{0}=sv_{F}\approx v_{F}$, and meanwhile the first sound velocity is $c_{1}=v_{F}\sqrt{\frac{1}{3}\left(1+F_{0}^{s}\right)\left(1+\frac{F_{1}^{s}}{3}\right)}\approx v_{F}/\sqrt{3}\approx c_{0}/\sqrt{3}$.

Under the condition of QSL in 2D, Eq.~\eqref{eq:two-m0} reduces to
\begin{equation}
s^{2}\Pi_{00}\left(s\right)=0.
\end{equation}
There is no nontrivial real solution, indicating the absence of weakly damped zero sound mode in the two channel QSL model. 
Also, when $1+\frac{F_{1}^{s}}{3}\to{}0^{+}$, the above inequality Eq.~\eqref{ineq:two-m0} can not be satisfied unless $F_0^{s}\to{}+\infty$.
In order to obtain a weakly damped longitudinal zero sound mode in a finite $F_{0}^{s}$ and under the QSL condition $1+\frac{F_{1}^{s}}{3}=0^{+}$,
one need to involve more Landau parameters $F_{l}^{s}$ with $l\geq{}2$.

\subsection{Three channel model}
Here we provide some details for solving the secular equation $g_{3}(s)=0$, i.e., Eq.~\eqref{eq:det} in the main text. 
To do this, we expand $g_3(s)$ around $s=1$ and $s=\infty$. With the help of the following relations:
\begin{eqnarray*}
\Omega_{02}(s) & = & \frac{1}{2} +P_{2}(s)\Omega_{00}(s), \\
\Omega_{22}(s) & = &  \frac{1}{5} + P_{2}(s) \Omega_{02}, \\
P_{2}(s) & = & \frac{1}{2}(3s^2-1),
\end{eqnarray*}
we can rewrite Eq.~\eqref{eq:g3s} as 
\begin{equation}
\begin{split}
g_{3}(s) & = \left( 1+\frac{F_{1}^{s}}{3}\right)\left(1+F_{2}^{s}\Omega_{22}\right) +\left[F_{0}^{s}\left( 1+\frac{F_{1}^{s}}{3}\right)+F_{1}^{s}s^{2}\right]\left[\Omega_{00}+F_{2}^{s}\left(\Omega_{00}\Omega_{22}-\Omega_{02}^2\right)\right]\\
& = A(s) + B(s)\Omega_{00}(s)\\
& = \tilde{A}(s)+\tilde{B}(s)\ln\frac{s-1}{s+1}.  
\end{split}
\end{equation}
Here
\begin{eqnarray}
A(s) & = & A_{2}s^2 + A_{0}\nonumber\\
& = &\frac{3}{4}F_{2}^{s}s^2+\left(1+\frac{F_{1}^s}{3}\right)\left[1-\frac{F_{2}^{s}}{20}\left(1+5F_{0}^{s}\right)\right], \\
B(s) & = & B_{4}s^4+B_{2}s^2+B_{0}\nonumber\\
& = & \frac{9}{4}F_{2}^{s}s^4
-3s^2\left\{\frac{F_{2}^{s}}{20}\left[9+\left(1+5F_{0}^{s}\right)\left(1+\frac{F_{1}^s}{3}\right)\right]-\frac{F_{1}^s}{3}\right\}
+\left[F_{2}^{s}\left(\frac{1}{4}+\frac{9F_{0}^{s}}{20}\right)+F_{0}^{s}\right]\left(1+\frac{F_{1}^s}{3}\right),
\end{eqnarray}
and
\begin{eqnarray}
\tilde{A}(s) & = & A(s) + B(s), \\
\tilde{B}(s) & = & \frac{s}{2}B(s),
\end{eqnarray}
where $A_{0},A_{2},B_{0},B_{2}$ and $B_{4}$ are coefficients that does not depend on $s$.
Thus the secular equation becomes
\begin{equation}
A(s)+B(s)\Omega_{00}(s) =\tilde{A}(s)+\tilde{B}(s)\ln\frac{s-1}{s+1}=0.    
\end{equation}

First, we consider the solution $s\to{}1+0^{+}$. When
\begin{equation}
\frac{\tilde{A}(s\to{}1+0^{+})}{\tilde{B}(s\to{}1+0^{+})} =\frac{11}{3}+\frac{50-\left(30+\frac{50F_{0}^{s}+8F_{2}^{s}}{3}\right)\left(1+\frac{F_{1}^{s}}{3}\right)} {F_{2}^{s}\left[9+\left(1-3F_{0}^{s}\right)\left(1+\frac{F_{1}^{s}}{3}\right)\right]+10\left[\left(3+F_{0}^s\right)\left(1+\frac{F_{1}^{s}}{3}\right)-3\right]} \gg{} 1,
\end{equation}
the secular equation has an approximate solution
\begin{equation}
s\simeq 1 + 2\exp\left[-\frac{\tilde{A}(s=1)}{\tilde{B}(s=1)}\right]=1+2\exp\left\{-\frac{11}{3}-
\frac{50-\left(30+\frac{50F_{0}^{s}+8F_{2}^{s}}{3}\right)\left(1+\frac{F_{1}^{s}}{3}\right)} {F_{2}^{s}\left[9+\left(1-3F_{0}^{s}\right)\left(1+\frac{F_{1}^{s}}{3}\right)\right]+10\left[\left(3+F_{0}^s\right)\left(1+\frac{F_{1}^{s}}{3}\right)-3\right]}
\right\}. 
\end{equation}
Second, we look for the solution $s\to{}+\infty$, which is available when \begin{equation*}
A_{2}=\frac{B_{4}}{3},
\end{equation*}
and
\begin{equation*}
\frac{\frac{B_4}{7}+\frac{B_2}{5}+\frac{B_0}{3}}{A_{0}-\frac{B_{2}}{3}-\frac{B_{4}}{5}} = \frac{B_4}{7}+\frac{B_2}{5}+\frac{B_0}{3} \gg{}1.    
\end{equation*}
In this situation, we have another solution
\begin{equation}
s \simeq \sqrt{\frac{B_4}{7}+\frac{B_2}{5}+\frac{B_0}{3}}=\sqrt{\frac{F_{2}^{s}}{25}\left[\frac{9}{7}+\frac{4}{3}\left(1+\frac{F_{1}^{s}}{3}\right)\right]
+\frac{F_{0}^{s}}{3}\left(1+\frac{F_{1}^{s}}{3}\right)+\frac{F_{1}^{s}}{5}}. 
\end{equation}

\section{Zero sound: generic $m$ modes in 3D}\label{app:mneq0}
In consistence with the $m=0$ case, we choose the following definition for spherical harmonics,
\begin{eqnarray*}
Y_{l}^{m}(\theta ,\phi ) &=&\sqrt{\frac{(l-m)!}{(l+m)!}}P_{l}^{m}(\cos\theta )e^{im\phi }, \\
\int \frac{d\mathrm{\hat{n}}}{4\pi }Y_{l}^{m}(\theta ,\phi )Y_{l^{\prime}}^{m^{\prime }\ast }(\theta ,\phi ) &=&\frac{\delta _{ll^{\prime }}\delta_{mm^{\prime }}}{2l+1},
\end{eqnarray*}
where $P_{l}^{m}$ are associated Legendre functions satisfying the relation 
\begin{equation*}
P_{l}^{-m}(x)=(-1)^{m}\frac{(l-m)!}{(l+m)!}P_{l}^{m}(x).    
\end{equation*}

Similarly, we have
\begin{equation*}
\sum_{\mathrm{p}^{\prime }}f_{\mathrm{pp}^{\prime }}^{s}\frac{\partial n_{\mathrm{p}^{\prime }}^{0}}{\partial \varepsilon _{\mathrm{p}^{\prime }}}\nu_{\mathrm{p}^{\prime }}=-\sum_{l}F_{l}^{s}\frac{1}{2l+1}\sum_{m=-l}^{l}Y_{l}^{m}(\theta _{\mathrm{p}},\phi _{\mathrm{p}})\nu
_{l}^{m},
\end{equation*}
and
\begin{equation}
\sum_{l}\nu _{l}^{m}Y_{l}^{m}(\theta _{\mathrm{p}},\phi _{\mathrm{p}})-\frac{\cos \theta _{\mathrm{p}}}{s-\cos \theta _{\mathrm{p}}}\sum_{l}F_{l}^{s}\frac{1}{2l+1}\nu _{l}^{m}Y_{l}^{m}(\theta _{\mathrm{p}},\phi _{\mathrm{p}})=\frac{\cos \theta _{\mathrm{p}}}{s-\cos \theta _{\mathrm{p}}}Ue^{im\phi _{\mathrm{p}}}.  \label{LTEm1}
\end{equation}
Then, by defining
\begin{align}
\Omega_{ll'}^{m}\left(s\right)=\Omega_{l'l}^{m}\left(s\right) & =-\int\frac{d\mathrm{\hat{p}}}{4\pi}Y_{l}^{m*}\left(\theta_{\mathrm{p}},\phi_{\mathrm{p}}\right)\frac{\cos\theta_{\mathrm{p}}}{s-\cos\theta_{\mathrm{p}}}Y_{l'}^{m}\left(\theta_{\mathrm{p}},\phi_{\mathrm{p}}\right)\nonumber\\
 & =\frac{1}{2}\sqrt{\frac{\left(l-m\right)!}{\left(l+m\right)!}\frac{\left(l'-m\right)!}{\left(l'+m\right)!}}\int_{-1}^{1}d\mu P_{l}^{m}\left(\mu\right)\frac{\mu}{\mu-s}P_{l'}^{m}\left(\mu\right),
\end{align}
as well as
\begin{align}
\Theta_{l}^{m}\left(s\right)=\left(-1\right)^{m}\Theta_{l}^{-m}\left(s\right)= & -\int\frac{d\mathrm{\hat{p}}}{4\pi}Y_{l}^{m*}\left(\theta_{\mathrm{p}},\phi_{\mathrm{p}}\right)\frac{\cos\theta_{\mathrm{p}}}{s-\cos\theta_{\mathrm{p}}}e^{im\phi_{\mathrm{p}}}\nonumber \\
= & -\int\frac{d\mathrm{\hat{p}}}{4\pi}\sqrt{\frac{\left(l-m\right)!}{\left(l+m\right)!}}P_{l}^{m}\left(\cos\theta_{\mathrm{p}}\right)e^{-im\phi_{\mathrm{p}}}\frac{\cos\theta_{\mathrm{p}}}{s-\cos\theta_{\mathrm{p}}}e^{im\phi_{\mathrm{p}}}\nonumber \\
= & \frac{1}{2}\sqrt{\frac{\left(l-m\right)!}{\left(l+m\right)!}}\int_{-1}^{1}d\mu P_{l}^{m}\left(\mu\right)\frac{\mu}{\mu-s},
\end{align}
we obtain type I algebraic equations similar to Eq.~(\ref{LTEZS3}) as
\begin{equation}
\frac{\nu_{l}^{m}}{2l+1}+\sum_{l'=m}^{\infty}F_{l'}^{s}\Omega_{ll'}^{m}\left(s\right)\frac{\nu_{l'}^{m}}{2l'+1}=-\Theta_{l}^{m}\left(s\right)U.
\end{equation}
Here we have used the facts that $\Omega_{ll'}^{m}\left(s\right)=0$ for $m>l,l'$ and $\Theta_{l}^{m}\left(s\right)=0$ for $m>l$.

We can further write Eq.~(\ref{LTEm1}) in terms of associated Legendre functions as
\begin{equation}
\sum_{l}\nu _{l}^{m}\sqrt{\frac{(l-m)!}{(l+m)!}}P_{l}^{m}(\mu )-\frac{\mu }{s-\mu }\sum_{l}F_{l}^{s}\frac{1}{2l+1}\nu _{l}^{m}\sqrt{\frac{(l-m)!}{(l+m)!}}P_{l}^{m}(\mu )=\frac{\mu }{s-\mu }U.  \label{LTEm3}
\end{equation}
Using the following formula,
\begin{eqnarray*}
(2l+1)\mu P_{l}^{m}(\mu ) &=&(l-m+1)P_{l+1}^{m}(\mu )+(l+m)P_{l-1}^{m}(\mu ), \\
\int_{-1}^{1}d\mu P_{l}^{m}(\mu )P_{l^{\prime }}^{m}(\mu ) &=&\frac{2}{2l+1}\frac{(l+m)!}{(l-m)!}\delta _{ll^{\prime }},
\end{eqnarray*}
we can obtain
\begin{eqnarray*}
\mu U &=&\sum_{l}\nu _{l}^{m}\left[ s-\left( 1+\frac{F_{l}^{s}}{2l+1}\right)\mu \right] \sqrt{\frac{(l-m)!}{(l+m)!}}P_{l}^{m}(\mu ) \\
&=&\sum_{l}\nu _{l}^{m}s\sqrt{\frac{(l-m)!}{(l+m)!}}P_{l}^{m}(\mu)-\sum_{l}\nu _{l}^{m}\left( 1+\frac{F_{l}^{s}}{2l+1}\right) \sqrt{\frac{(l-m)!}{(l+m)!}}\frac{1}{2l+1}\left[ (l-m+1)P_{l+1}^{m}(\mu)+(l+m)P_{l-1}^{m}(\mu )\right] \\
&=&\sum_{l}\nu _{l}^{m}s\sqrt{\frac{(l-m)!}{(l+m)!}}P_{l}^{m}(\mu)-\sum_{l}\nu _{l}^{m}\left( 1+\frac{F_{l}^{s}}{2l+1}\right) \frac{\sqrt{(l+1)^{2}-m^{2}}}{2l+1}\sqrt{\frac{(l-m+1)!}{(l+m+1)!}}P_{l+1}^{m}(\mu ) \\
&&-\sum_{l}\nu _{l}^{m}\left( 1+\frac{F_{l}^{s}}{2l+1}\right) \frac{\sqrt{l^{2}-m^{2}}}{2l+1}\sqrt{\frac{(l-m-1)!}{(l+m-1)!}}P_{l-1}^{m}(\mu ) \\
&=&\sum_{l}\sqrt{\frac{(l-m)!}{(l+m)!}}P_{l}^{m}(\mu )\left[ \nu_{l}^{m}s-\nu _{l-1}^{m}\left( 1+\frac{F_{l-1}^{s}}{2l-1}\right) \frac{\sqrt{l^{2}-m^{2}}}{2l-1}-\nu _{l+1}^{m}\left( 1+\frac{F_{l+1}^{s}}{2l+3}\right)\frac{\sqrt{(l+1)^{2}-m^{2}}}{2l+3}\right] .
\end{eqnarray*}
This leads to the type II algebraic equations which are similar to Eq.~(\ref{LTEZS4})
\begin{equation}
\nu _{l}^{m}s-\nu _{l-1}^{m}\left( 1+\frac{F_{l-1}^{s}}{2l-1}\right) \frac{\sqrt{l^{2}-m^{2}}}{2l-1}-\nu _{l+1}^{m}\left( 1+\frac{F_{l+1}^{s}}{2l+3}\right) \frac{\sqrt{(l+1)^{2}-m^{2}}}{2l+3}=\alpha _{l}^{m}U, 
\end{equation}
where
\begin{equation*}
\alpha _{l}^{m}=\left(-1\right)^{m}\alpha_{l}^{-m}=\frac{2l+1}{2}\sqrt{\frac{(l-m)!}{(l+m)!}}\int_{-1}^{1}d\mu P_{l}^{m}(\mu )\mu.
\end{equation*}
Note that $P_{l}^{m}(\mu )=(-1)^{m}(1-\mu ^{2})^{m/2}\frac{d^{m}}{d\mu^{m}} P_{l}(\mu )$ are even (odd) functions of $\mu $ when $l-m$ are even (odd). So that $\alpha _{l}^{m}=0$ when $m>l$ or $l-m$ is even. Moreover, we also have $\alpha _{l}^{0}=0$ for $l>1$, and  $\alpha _{l}^{-m}=\left(-1\right)^{m}\alpha _{l}^{m}$.

\subsection{Some useful formula}
In the collisionless regime, we obtain the type I and type II algebraic equations from Landau kinetic for $m>0$, in which we introduce following three functions
\begin{align}
\Omega_{ll'}^{m}\left(s\right) & =\frac{1}{2}\sqrt{\frac{\left(l-m\right)!}{\left(l+m\right)!}\frac{\left(l'-m\right)!}{\left(l'+m\right)!}}\int_{-1}^{1}d\mu P_{l}^{m}\left(\mu\right)\frac{\mu}{\mu-s}P_{l'}^{m}\left(\mu\right),\\
\Theta_{l}^{m}\left(s\right) & =\frac{1}{2}\sqrt{\frac{\left(l-m\right)!}{\left(l+m\right)!}}\int_{-1}^{1}d\mu P_{l}^{m}\left(\mu\right)\frac{\mu}{\mu-s},\\
\alpha_{l}^{m} & =\frac{2l+1}{2}\sqrt{\frac{\left(l-m\right)!}{\left(l+m\right)!}}\int_{-1}^{1}d\mu P_{l}^{m}\left(\mu\right)\mu,
\end{align}
where the associated Legendre functions are given by
\begin{equation}
P_{l}^{m}\left(\mu\right)=\frac{\left(-1\right)^{m}}{2^{l}l!}\left(1-\mu^{2}\right)^{m/2}\frac{d^{l+m}}{d\mu^{l+m}}\left(\mu^{2}-1\right)^{l}.
\end{equation}
Since $P_{l}^{m}\left(\mu\right)=0$ for $m>l$, we have $\Omega_{ll'}^{m}\left(s\right)=0$ for $m>l,l'$, $\Theta_{l}^{m}\left(s\right)=\alpha_{l}^{m}=0$ for $m>l$.
From the relations between $P_{l}^{-m}\left(\mu\right)$ and $P_{l}^{m}\left(\mu\right)$,
\begin{equation}
P_{l}^{-m}\left(\mu\right)=\left(-1\right)^{m}\frac{\left(l-m\right)!}{\left(l+m\right)!}P_{l}^{m}\left(\mu\right),
\end{equation}
we can obtain
\begin{align}
\Omega_{ll'}^{m}\left(s\right) & =\Omega_{ll'}^{-m}\left(s\right),\\
\Theta_{l}^{m}\left(s\right) & =\left(-1\right)^{m}\Theta_{l}^{-m}\left(s\right),\\
\alpha_{l}^{m} & =\left(-1\right)^{m}\alpha_{l}^{-m}.
\end{align}

By definition, $\Omega_{ll'}^{m}$ is symmetric in terms of two subscripts $l$ and $l'$ as
\begin{equation}
\Omega_{ll'}^{m}\left(s\right)=\Omega_{l'l}^{m}\left(s\right).
\end{equation}
Using the identities of associated Legendre functions as
\begin{align}
P_{m}^{m}\left(\mu\right)=P_{m}^{m}\left(-\mu\right) & =\left(-1\right)^{m}\left(2m-1\right)!!\left(1-\mu^{2}\right)^{m/2},\\
P_{m+1}^{m}\left(\mu\right) & =\mu\left(2m+1\right)P_{m}^{m}\left(\mu\right),\\
P_{m+1}^{m+1}\left(\mu\right) & =-\left(2m+1\right)\sqrt{1-\mu^{2}}P_{m}^{m}\left(\mu\right),
\end{align}
the orthogonality of associated Legendre functions for fixed $m$ as
\begin{align}
\int_{-1}^{1}d\mu P_{k}^{m}\left(\mu\right)P_{l}^{m}\left(\mu\right) & =\frac{2\left(l+m\right)!}{\left(2l+1\right)\left(l-m\right)!}\delta_{kl},
\end{align}
and the integrals of associated Legendre functions and polynomials as
\begin{align}
\frac{1}{2}\frac{1}{\left(2m\right)!}\int_{-1}^{1}d\mu\left[P_{m}^{m}\left(\mu\right)\right]^{2}\mu^{2n} & =\frac{\left(2m-1\right)!!\left(2n-1\right)!!}{\left(2m+2n+1\right)!!},
\end{align}
we can obtain useful relations of $\Omega_{ll'}^{m}$ function
\begin{align}
\Omega_{m+1,l}^{m}\left(s\right) & =s\sqrt{2m+1}\Omega_{ml}^{m}\left(s\right)+\frac{\delta_{m+1,l}}{\left(2m+3\right)},\label{eq:omega-delta}\\
\Omega_{m+1,m+1}^{m+1}\left(s\right) & =\frac{2m+1}{2m+2}\left(1-s^{2}\right)\Omega_{mm}^{m}\left(s\right)-\frac{1}{\left(2m+2\right)\left(2m+3\right)}.\label{eq:omega-recursive}
\end{align}
For later use, the large $s$ limit of $\Omega_{mm}^{m}\left(s\right)$ is given by
\begin{align}
\lim_{s\rightarrow\infty}\Omega_{mm}^{m}\left(s\right)= & \lim_{s\rightarrow\infty}\frac{1}{2}\frac{1}{\left(2m\right)!}\int_{-1}^{1}d\mu\left[P_{m}^{m}\left(\mu\right)\right]^{2}\frac{\mu}{\mu-s}\nonumber \\
= & -\frac{1}{2}\frac{1}{\left(2m\right)!}\int_{-1}^{1}d\mu\left[P_{m}^{m}\left(\mu\right)\right]^{2}\sum_{n=1}^{\infty}\left(\frac{\mu}{s}\right)^{2n}\nonumber \\
= & -\sum_{n=1}^{\infty}\frac{\left(2m-1\right)!!\left(2n-1\right)!!}{\left(2m+2n+1\right)!!}\frac{1}{s^{2n}}.\label{eq:omegammm}
\end{align}

For $\Theta_{l}^{m}\left(s\right)$, we similarly obtain the useful relations as follows
\begin{align}
\Theta_{m+1}^{m}\left(s\right) & =s\sqrt{2m+1}\Theta_{m}^{m}\left(s\right),\\
\Theta_{2m}^{2m}\left(s\right) & =\sqrt{\frac{\left(4m-1\right)!!}{\left(4m\right)!!}}\frac{\left(2m\right)!!}{\left(2m-1\right)!!}\Omega_{mm}^{m}\left(s\right).
\end{align}

For the $\alpha_{l}^{m}$ function, from the parity of associated Legendre functions
\begin{equation}
P_{l}^{m}\left(-\mu\right)=\left(-1\right)^{l+m}P_{l}^{m}\left(\mu\right),
\end{equation}
we have $\alpha_{l}^{m}=0$ for $l+m=\mathrm{even}$. From the relation between Legendre polynomials and associated Legendre polynomials as
\begin{align}
P_{l}^{0}\left(\mu\right) & =P_{l}\left(\mu\right),
\end{align}
 and the orthogonality of Legendre polynomials as
\begin{align}
\int_{-1}^{1}d\mu P_{l}\left(\mu\right)P_{l'}\left(\mu\right) & =\frac{2}{2l+1}\delta_{ll'},
\end{align}
we can obtain that $\alpha_{l}^{0}=0$ for $l>1$.

We can use these relations to prove that the $l=m$ and $l=m+1$ components of type I algebraic equation can lead to the $l=m$ component of type II algebraic equation as follows
\begin{align}
0= & -\Theta_{m}^{m}U\Theta_{m+1}^{m}+\Theta_{m+1}^{m}U\Theta_{m}^{m}\nonumber \\
= & \left(\frac{\nu_{m}^{m}}{2m+1}+\sum_{l'=m}^{\infty}F_{l'}^{s}\Omega_{ml'}^{m}\frac{\nu_{l'}^{m}}{2l'+1}\right)\Theta_{m+1}^{m}-\left(\frac{\nu_{m+1}^{m}}{2m+3}+\sum_{l'=m}^{\infty}F_{l'}^{s}\Omega_{m+1,l'}^{m}\frac{\nu_{l'}^{m}}{2l'+1}\right)\Theta_{m}^{m}\nonumber \\
= & \left[\left(1+F_{m}^{s}\Omega_{mm}^{m}\right)\Theta_{m+1}^{m}-F_{m}^{s}\Omega_{m+1m}^{m}\Theta_{m}^{m}\right]\frac{\nu_{m}^{m}}{2m+1}+\left[F_{m+1}^{s}\Omega_{mm+1}^{m}\Theta_{m+1}^{m}-\left(1+F_{m+1}^{s}\Omega_{m+1m+1}^{m}\right)\Theta_{m}^{m}\right]\frac{\nu_{m+1}^{m}}{2m+3}\nonumber \\
 & +\sum_{l'=m+2}^{\infty}F_{l'}^{s}\left(\Omega_{ml'}^{m}\Theta_{m+1}^{m}-\Omega_{m+1,l'}^{m}\Theta_{m}^{m}\right)\frac{\nu_{l'}^{m}}{2l'+1}\nonumber \\
= & \left[\left(1+F_{m}^{s}\Omega_{mm}^{m}\right)s\sqrt{2m+1}-F_{m}^{s}\Omega_{m+1m}^{m}\right]\Theta_{m}^{m}\frac{\nu_{m}^{m}}{2m+1}+\left[F_{m+1}^{s}\Omega_{mm+1}^{m}s\sqrt{2m+1}-\left(1+F_{m+1}^{s}\Omega_{m+1m+1}^{m}\right)\right]\Theta_{m}^{m}\frac{\nu_{m+1}^{m}}{2m+3}\nonumber \\
= & \left\{ \left[s\sqrt{2m+1}+F_{m}^{s}\Omega_{mm}^{m}\left(s\sqrt{2m+1}-s\sqrt{2m+1}\right)\right]\frac{\nu_{m}^{m}}{2m+1}+\left[-1+F_{m+1}^{s}\left(\Omega_{mm+1}^{m}s\sqrt{2m+1}-\Omega_{m+1m+1}^{m}\right)\right]\frac{\nu_{m+1}^{m}}{2m+3}\right\} \Theta_{m}^{m}\nonumber \\
= & \left\{ \frac{\nu_{m}^{m}s}{\sqrt{2m+1}}+\left[-1-\frac{F_{m+1}^{s}}{2m+3}\right]\frac{\nu_{m+1}^{m}}{2m+3}\right\} \Theta_{m}^{m}\nonumber \\
\Longrightarrow & \nu_{m}^{m}s-\sqrt{2m+1}\left(1+\frac{F_{m+1}^{s}}{2m+3}\right)\frac{\nu_{m+1}^{m}}{2m+3}=0.
\end{align}

\section{Zero sound: two channel model for $m>0$ modes in 3D} \label{app:two-channel-m}
The type I algebraic equations for generic $m$ modes are 
\begin{equation*}
\frac{\nu_{l}^{m}}{2l+1}+\sum_{l'=0}^{\infty}F_{l'}^{s}\Omega_{ll'}^{m}\left(s\right)\frac{\nu_{l'}^{m}}{2l'+1}=-\Theta_{l}^{m}\left(s\right)U,
\end{equation*}
while the type II algebraic equation read
\begin{equation*}
\nu_{l}^{m}s-\sqrt{l^{2}-m^{2}}\left(1+\frac{F_{l-1}^{s}}{2l-1}\right)\frac{\nu_{l-1}^{m}}{2l-1}-\sqrt{\left(l+1\right)^{2}-m^{2}}\left(1+\frac{F_{l+1}^{s}}{2l+3}\right)\frac{\nu_{l+1}^{m}}{2l+3}=\alpha_{l}^{m}U.
\end{equation*}
It turns out that there exists a specific solution for a given good quantum number $m$ and the two channel model.

To find this specific solution, we consider a two channel model by keeping the Landau parameters up to $F_{m+1}^{s}$, i.e., $F_{l}^{s}=0$ for $l>m+2$. Then the $l=m$ component of the type I equation reduces to
\begin{equation}
\frac{\nu_{m}^{m}}{2m+1}+F_{m}^{s}\Omega_{mm}^{m}\frac{\nu_{m}^{m}}{2m+1}+F_{m+1}^{s}\Omega_{mm+1}^{m}\frac{\nu_{m+1}^{m}}{2m+3}=-\Theta_{m}^{m}U.
\end{equation}
The $l=m$ component of type II equation is given by
\begin{equation}
\nu_{m}^{m}s-\sqrt{2m+1}\left(1+\frac{F_{m+1}^{s}}{2m+3}\right)\frac{\nu_{m+1}^{m}}{2m+3}=0.
\end{equation}
Those two equations (similar to the two channel model for $m=0$ modes) form a close set and can be easily solved as follows,
\begin{align}
\frac{\nu_{m}^{m}}{U} & =-\frac{\left(1+\frac{F_{m+1}^{s}}{2m+3}\right)\Theta_{m}^{m}}{\frac{\left(1+F_{m}^{s}\Omega_{mm}^{m}\right)\left(1+\frac{F_{m+1}^{s}}{2m+3}\right)}{2m+1}+\frac{F_{m+1}^{s}\Omega_{mm+1}^{m}s}{\sqrt{2m+1}}},\\
\frac{\nu_{m+1}^{m}}{2m+3} & =\frac{\nu_{m}^{m}s}{\sqrt{2m+1}\left(1+\frac{F_{m+1}^{s}}{2m+3}\right)}=-\frac{s\Theta_{m}^{m}U}{\frac{\left(1+F_{m}^{s}\Omega_{mm}^{m}\right)\left(1+\frac{F_{m+1}^{s}}{2m+3}\right)}{\sqrt{2m+1}}+F_{m+1}^{s}\Omega_{mm+1}^{m}s}.
\end{align}
The mode frequency is determined by the pole of the response function
\begin{equation}
\left(1+F_{m}^{s}\Omega_{mm}^{m}\right)\left(1+\frac{F_{m+1}^{s}}{2m+3}\right)+\sqrt{2m+1}F_{m+1}^{s}\Omega_{mm+1}^{m}s=0.
\end{equation}
Using the relation derived from Eq.~\eqref{eq:omega-delta}
\begin{equation}
\Omega_{mm+1}^{m}\left(s\right)=\sqrt{2m+1}s\Omega_{mm}^{m}\left(s\right),
\end{equation}
the equation determining mode frequency is simplified as
\begin{equation}
\left[1+F_{m}^{s}\Omega_{mm}^{m}(s)\right]\left(1+\frac{F_{m+1}^{s}}{2m+3}\right)+(2m+1)F_{m+1}^{s}s^2\Omega_{mm}^{m}(s)=0.
\end{equation}

\subsection{$m=1$ modes}
For the $m=1$ modes, we obtain
\begin{equation}
\frac{\nu_{1}^{1}}{U}=-\frac{3\left(1+\frac{F_{2}^{s}}{5}\right)\Theta_{1}^{1}}{\left(1+F_{1}^{s}\Omega_{11}^{1}\right)\left(1+\frac{F_{2}^{s}}{5}\right)+3F_{2}^{s}\Omega_{11}^{1}s^{2}}.
\end{equation}
The $\nu_{1}^{1}$ mode is completely suppressed when
\begin{equation}
\left(1+\frac{F_{2}^{s}}{5}\right)\Theta_{1}^{1}\left(s\right)=0.
\end{equation}
As there is no real solution to $\Theta_{1}^{1}\left(s\right)=0$ except at $s\rightarrow\infty$, thus we obtain $1+\frac{F_{2}^{s}}{5}=0$, which is nothing but the Pomeranchuk instability in $l=2$ channel.

The frequency of $m=1$ mode is determined by
\begin{equation}\label{eq:mode_fre_m=1}
\left(1+F_{1}^{s}\Omega_{11}^{1}\right)\left(1+\frac{F_{2}^{s}}{5}\right)+3F_{2}^{s}\Omega_{11}^{1}s^{2}=0.
\end{equation}
The above equation must generally be found numerically. Using the limiting form $\lim_{s\rightarrow\infty}\Omega_{11}^{1}\left(s\right)=-\frac{1}{15s^{2}}-\frac{1}{35s^{4}}$, Eq.~\eqref{eq:mode_fre_m=1}  can approximately be solved in the large $s$ limit 
\begin{equation}
s=\sqrt{\frac{F_{1}^{s}}{15}\left(1+\frac{F_{2}^{s}}{5}\right)+\frac{3F_{2}^{s}}{35}}.
\end{equation}
Note that the above solution is well defined only when $\frac{F_{1}^{s}}{15}\left(1+\frac{F_{2}^{s}}{5}\right)+\frac{3F_{2}^{s}}{35}\gg{}1$.

From Eq.~\eqref{eq:h-1} and the inequality $1+\frac{F_{1}^{s}}{3}\geq 0$, we have the following condition for a real solution $s>1$ to the frequency equation,
\begin{equation}
F_{2}^{s} >\frac{5}{2+\left(1+\frac{F_{1}^{s}}{3}\right)}\left[3-\left(1+\frac{F_{1}^{s}}{3}\right)\right].    
\end{equation}
In the limit of $1+\frac{F_{1}^{s}}{3}=0$, the above inequality becomes
\begin{equation}
F_{2}^{s} >\frac{15}{2}.
\end{equation}
For a solution $s\to{}1+0^{+}$, it is hard to obtain a simple analytical form, since the convergence of $\lim_{s\to{}1}(s-1)\ln(s-1)=0$ is very slow. The numerical solution to $h(s)=0$ can be found in Fig.~\ref{fig:h0}~(b) as a 3D contour plot in the parameter space $\left(s,1+\frac{F_{1}^{s}}{3},\frac{2F_{2}^{s}}{15}\right)$.

\subsection{$m=2$ modes}
For the $m=2$ modes, we obtain
\begin{equation}
\frac{\nu_{2}^{2}}{U}=-\frac{5\left(1+\frac{F_{3}^{s}}{7}\right)\Theta_{2}^{2}}{\left(1+F_{2}^{s}\Omega_{22}^{2}\right)\left(1+\frac{F_{3}^{s}}{7}\right)+5F_{3}^{s}\Omega_{22}^{2}s^{2}}.
\end{equation}
The $\nu_{2}^{2}$ mode is completely suppressed when
\begin{equation}
\left(1+\frac{F_{3}^{s}}{7}\right)\Theta_{2}^{2}\left(s\right)=0.
\end{equation}
As there is no real solution to $\Theta_{2}^{2}\left(s\right)=0$ except at $s\rightarrow\infty$, thus we obtain $1+\frac{F_{3}^{s}}{7}=0$, which is nothing but the Pomeranchuk instability for $l=3$ channel.

The mode frequency of $\nu_{2}^{2}$ mode is determined by
\begin{equation}
\left(1+F_{2}^{s}\Omega_{22}^{2}\right)\left(1+\frac{F_{3}^{s}}{7}\right)+5F_{3}^{s}\Omega_{22}^{2}s^{2}=0.
\end{equation}
The above equation must generally be found numerically. Using the limiting form $\lim_{s\rightarrow\infty}\Omega_{22}^{2}\left(s\right)=-\frac{1}{35s^{2}}-\frac{1}{105s^{4}}$,
Eq.~\eqref{eq:mode_fre_m=1}  can approximately be solved in the large $s$ limit 
\begin{equation}
s=\sqrt{\frac{F_{2}^{s}}{35}\left(1+\frac{F_{3}^{s}}{7}\right)+\frac{F_{3}^{s}}{21}},
\end{equation}
when the condition
\begin{equation}
\frac{F_{2}^{s}}{35}\left(1+\frac{F_{3}^{s}}{7}\right)+\frac{F_{3}^{s}}{21s^{2}}\gg{}1    
\end{equation}
is satisfied.

From Eq.~\eqref{eq:h-1} and the inequality $1+\frac{F_{2}^{s}}{5}\geq 0$, we obtain the following condition for a real solution $s>1$ to the frequency equation,
\begin{equation}
F_{2}^{s} >\frac{7}{2+\left(1+\frac{F_{2}^{s}}{5}\right)}\left[5-\left(1+\frac{F_{1}^{s}}{5}\right)\right].    
\end{equation}
In the limit of $1+\frac{F_{2}^{s}}{5}=0$, the above inequality becomes
\begin{equation}
F_{3}^{s} >\frac{35}{2}.
\end{equation}
For a solution $s\to{}1+0^{+}$, it is hard to obtain a simple analytical form, since the convergence of $\lim_{s\to{}1}(s-1)\ln(s-1)=0$ is very slow. The numerical solution to $h(s)=0$ can be found in Fig.~\ref{fig:h0}~(b) as a 3D contour plot in the parameter space $	\left(s,1+\frac{F_{2}^{s}}{5},\frac{2F_{3}^{s}}{35}\right)$.

\subsection{Arbitrary $m\geq{}1$ modes}
The mode frequency of $\nu_{m}^{m}$ mode is determined by
\begin{equation}
\left(1+F_{m}^{s}\Omega_{mm}^{m}\right)\left(1+\frac{F_{m+1}^{s}}{2m+3}\right)+\left(2m+1\right)F_{m+1}^{s}\Omega_{mm}^{m}s^{2}=0.
\end{equation}
Using the limiting form $\lim_{s\rightarrow\infty}\Omega_{mm}^{m}\left(s\right)=-\frac{1}{\left(2m+1\right)\left(2m+3\right)s^{2}}-\frac{3}{\left(2m+1\right)\left(2m+3\right)\left(2m+5\right)}\frac{1}{s^{4}}$, the solution to the above equation in the large $s$ limit is
\begin{equation}
s=\sqrt{\frac{F_{m}^{s}}{\left(2m+1\right)\left(2m+3\right)}\left(1+\frac{F_{m+1}^{s}}{2m+3}\right)+\frac{3F_{m+1}^{s}}{\left(2m+3\right)\left(2m+5\right)}}.
\end{equation}
when the condition 
\begin{equation}
\frac{F_{m}^{s}}{\left(2m+1\right)\left(2m+3\right)}\left(1+\frac{F_{m+1}^{s}}{2m+3}\right)+\frac{3F_{m+1}^{s}}{\left(2m+3\right)\left(2m+5\right)}\gg{}1    
\end{equation}
is satisfied.

\section{Zero sound in 2D: Algebraic equations for longitudinal and transverse modes}\label{app:aEq2D}

To derive algebraic equations for collective modes in 2D, we consider the expansion for the quasi-particle interaction $f_{\mathrm{pp}^{\prime }}^{s}$ in Eq.~(\ref{eq:flsa}b). With the expressions for $\nu_{\mathrm{p}}$ in Eq.~\eqref{eq:nu2D1} and \eqref{eq:nu2D2}, we have
\begin{eqnarray*}
\sum_{\mathrm{p}^{\prime }}f_{\mathrm{pp}^{\prime }}^{s}\frac{\partial n_{\mathrm{p}^{\prime }}^{0}}{\partial \varepsilon _{\mathrm{p}^{\prime }}}\nu_{\mathrm{p}^{\prime }} &=&N(0)\int d\varepsilon \frac{\partial n^{0}(\varepsilon )}{\partial \varepsilon }\int \frac{d\theta _{\mathrm{p}^{\prime }}}{2\pi }f_{\mathrm{pp}^{\prime }}^{s}\nu _{\mathrm{p}^{\prime }} \\
&=&-N(0)\int \frac{d\theta _{\mathrm{p}^{\prime }}}{2\pi }\sum_{l=0}^{\infty}f_{l}^{s}\cos (l\theta _{\mathrm{pp}^{\prime }})\sum_{l^{\prime }=-\infty}^{\infty }\nu _{l^{\prime }}e^{il^{\prime }\theta _{\mathrm{p}^{\prime }}} \\
&=&-\frac{1}{2}\int \frac{d\theta _{\mathrm{p}^{\prime }}}{2\pi} \sum_{l=0}^{\infty }F_{l}^{s}\left[ e^{il(\theta _{\mathrm{p}}-\theta _{\mathrm{p}^{\prime }})}+e^{-il(\theta _{\mathrm{p}}-\theta _{\mathrm{p}^{\prime }})}\right] \sum_{l^{\prime }=-\infty }^{\infty }\nu _{l^{\prime}}e^{il^{\prime }\theta _{\mathrm{p}^{\prime }}} \\
&=&-\frac{1}{2}\sum_{l=0}^{\infty }F_{l}^{s}\left( \nu _{l}e^{il\theta _{\mathrm{p}}}+\nu _{-l}e^{-il\theta _{\mathrm{p}}}\right).
\end{eqnarray*}
Putting the above into Eq.~(\ref{LTEZS1}) leads to
\begin{equation}
\sum_{m=-\infty }^{\infty }\nu _{m}e^{im\theta _{\mathrm{p}}}-\frac{1}{2}\frac{\cos \theta _{\mathrm{p}}}{s-\cos \theta _{\mathrm{p}}}\sum_{l=0}^{\infty }F_{l}^{s}\left( \nu _{l}e^{il\theta _{\mathrm{p}}}+\nu_{-l}e^{-il\theta _{\mathrm{p}}}\right) =\frac{\cos \theta _{\mathrm{p}}}{s-\cos \theta _{\mathrm{p}}}U. \label{eq:LTE2D1}
\end{equation}
For the longitudinal mode $\nu_{\mathrm{p}}^{+}$, Eq.~\eqref{eq:LTE2D1} becomes
\begin{equation}\label{eq:Landau_2D_longi}
u_{0}+\sum_{l=1}^{\infty } u_{l}\cos(l\theta _{\mathrm{p}})-\frac{\cos \theta _{\mathrm{p}}}{s-\cos \theta _{\mathrm{p}}}
\left[F_{0}^{s}u_{0}+\frac{1}{2}\sum_{l=1}^{\infty } F_{l}^{s}u_{l}
\cos(l\theta_{\mathrm{p}})
\right]
=\frac{\cos \theta _{\mathrm{p}}}{s-\cos \theta _{\mathrm{p}}}U,
\end{equation}
while for the transverse mode $\nu_{\mathrm{p}}^{-}$, by using Eqs.~\eqref{LTEZS} and \eqref{eq:U2D}, we have
\begin{equation}\label{eq:Landau_2D_trans}
\sum_{l=1}^{\infty } v_{l}\sin(l\theta _{\mathrm{p}})-\frac{1}{2}\frac{\cos \theta _{\mathrm{p}}}{s-\cos \theta _{\mathrm{p}}}
\sum_{l=1}^{\infty } F_{l}^{s}v_{l}
\sin(l\theta_{\mathrm{p}})
=\frac{\cos \theta _{\mathrm{p}}}{s-\cos \theta_{\mathrm{p}}}U\sin\theta_{\mathrm{p}}.
\end{equation}

Using the orthogonal relations for cosine functions with $l,l'\geq0$,
\begin{equation}
\int\frac{d\theta_{\mathrm{p}}}{2\pi}\cos\left(l\theta_{\mathrm{p}}\right)\cos\left(l'\theta_{\mathrm{p}}\right)=\int\frac{d\theta_{\mathrm{p}}}{2\pi}\frac{1}{2}\left[\cos\left(l\theta_{\mathrm{p}}+l'\theta_{\mathrm{p}}\right)+\cos\left(l\theta_{\mathrm{p}}-l'\theta_{\mathrm{p}}\right)\right]=\frac{1}{2}\left(\delta_{ll'}+\delta_{l0}\delta_{l'0}\right),
\end{equation}
the type I algebraic equation for longitudinal mode is derived from Eq.~\eqref{eq:Landau_2D_longi} as follows
\begin{equation}
\frac{u_{l}+\delta_{l0}u_{0}}{2}+\sum_{l'=0}^{\infty}F_{l'}^{s}\Pi_{ll'}\left(s\right)\frac{u_{l'}+\delta_{l'0}u_{0}}{2}=-\Pi_{l0}\left(s\right)U,
\end{equation}
where 
\begin{equation}
\Pi_{ll'}\left(s\right)=\Pi_{l'l}\left(s\right)=-\int\frac{d\theta_{\mathrm{p}}}{2\pi}\cos\left(l\theta_{\mathrm{p}}\right)\frac{\cos\theta_{\mathrm{p}}}{s-\cos\theta_{\mathrm{p}}}\cos\left(l'\theta_{\mathrm{p}}\right).
\end{equation}
Using orthogonal relations for sine functions with $l,l'\geq0$ 
\begin{equation}
\int\frac{d\theta_{\mathrm{p}}}{2\pi}\sin\left(l\theta_{\mathrm{p}}\right)\sin\left(l'\theta_{\mathrm{p}}\right)=\int\frac{d\theta_{\mathrm{p}}}{2\pi}\frac{1}{2}\left[\cos\left(l\theta_{\mathrm{p}}-l'\theta_{\mathrm{p}}\right)-\cos\left(l\theta_{\mathrm{p}}+l'\theta_{\mathrm{p}}\right)\right]=\frac{1}{2}\left(\delta_{ll'}-\delta_{l0}\delta_{l'0}\right),
\end{equation}
the type I algebraic equation  for transverse mode is derived from Eq.~\eqref{eq:Landau_2D_trans} as follows
\begin{equation}
\frac{v_{l}}{2}+\sum_{l'=0}^{\infty}F_{l'}^{s}\Xi_{ll'}\left(s\right)\frac{v_{l'}}{2}=-\Xi_{l1}\left(s\right)U,
\end{equation}
where
\begin{equation}
\Xi_{ll'}\left(s\right)=\Xi_{l'l}\left(s\right)=-\int\frac{d\theta_{\mathrm{p}}}{2\pi}\sin\left(l\theta_{\mathrm{p}}\right)\frac{\cos\theta_{\mathrm{p}}}{s-\cos\theta_{\mathrm{p}}}\sin\left(l'\theta_{\mathrm{p}}\right).
\end{equation}
From Eq.~\eqref{eq:Landau_2D_longi}, we obtain
\begin{align}
\cos\theta_{\mathrm{p}}U & =\left(s-\cos\theta_{\mathrm{p}}\right)\sum_{l=0}^{\infty}\cos\left(l\theta_{\mathrm{p}}\right)u_{l}-\cos\theta_{\mathrm{p}}\sum_{l=0}^{\infty}F_{l}^{s}\cos\left(l\theta_{\mathrm{p}}\right)\frac{u_{l}+\delta_{l0}u_{0}}{2}\nonumber \\
 & =\sum_{l=0}^{\infty}\cos\left(l\theta_{\mathrm{p}}\right)u_{l}s-\cos\theta_{\mathrm{p}}\left(1+F_{0}^{s}\right)u_{0}-\sum_{l=1}^{\infty}\left(1+\frac{F_{l}^{s}}{2}\right)\cos\theta_{\mathrm{p}}\cos\left(l\theta_{\mathrm{p}}\right)u_{l}\nonumber \\
 & =\sum_{l=0}^{\infty}\cos\left(l\theta_{\mathrm{p}}\right)u_{l}s-\cos\theta_{\mathrm{p}}\left(1+F_{0}^{s}\right)u_{0}-\sum_{l=1}^{\infty}\left(1+\frac{F_{l}^{s}}{2}\right)\cos\left(l\theta_{\mathrm{p}}+\theta_{\mathrm{p}}\right)\frac{u_{l}}{2}-\sum_{l=1}^{\infty}\left(1+\frac{F_{l}^{s}}{2}\right)\cos\left(l\theta_{\mathrm{p}}-\theta_{\mathrm{p}}\right)\frac{u_{l}}{2}\nonumber \\
 & =\tilde{u}_{0}s-\left(1+\frac{F_{1}^{s}}{2}\right)\frac{u_{1}}{2}+\sum_{l=1}^{\infty}\cos\left(l\theta_{\mathrm{p}}\right)\left[u_{l}s-\left(1+\frac{F_{l-1}^{s}+\delta_{l1}F_{0}^{s}}{2}\right)\frac{u_{l-1}+\delta_{l1}u_{0}}{2}-\left(1+\frac{F_{l+1}^{s}}{2}\right)\frac{u_{l+1}}{2}\right].
\end{align}
Thus the type II algebraic equation for the longitudinal mode is
\begin{equation}
u_{l}s-\left(1+\frac{F_{l-1}^{s}+\delta_{l1}F_{0}^{s}}{2}\right)\frac{u_{l-1}+\delta_{l1}u_{0}}{2}-\left(1+\frac{F_{l+1}^{s}}{2}\right)\frac{u_{l+1}}{2}=\delta_{l1}U.
\end{equation}
In particular, for $l=0$, we have
\begin{equation}
u_{0}s-\left(1+\frac{F_{1}^{s}}{2}\right)\frac{u_{1}}{2}=0.
\end{equation}
In general dimensions, we conjecture that the $l=0$ component of type II algebraic equation for longitudinal mode is
\begin{equation}
\nu_{0}s-\left(1+\frac{F_{1}^{s}}{d}\right)\frac{\nu_{1}}{d}=0.
\end{equation}
From Eq.~\eqref{eq:Landau_2D_trans}, we obtain
\begin{align}
\cos\theta_{\mathrm{p}}\sin\theta_{\mathrm{p}}U & =\sin2\theta_{\mathrm{p}}\frac{U}{2}\nonumber \\
 & =\left(s-\cos\theta_{\mathrm{p}}\right)\sum_{l=1}^{\infty}v_{l}\sin\left(l\theta_{\mathrm{p}}\right)-\cos\theta_{\mathrm{p}}\left[\sum_{l=1}^{\infty}F_{l}^{s}\frac{v_{l}}{2}\sin\left(l\theta_{\mathrm{p}}\right)\right]\nonumber \\
 & =\sum_{l=1}^{\infty}\sin\left(l\theta_{\mathrm{p}}\right)v_{l}s-\sum_{l=1}^{\infty}\left(1+\frac{F_{l}^{s}}{2}\right)\cos\theta_{\mathrm{p}}\sin\left(l\theta_{\mathrm{p}}\right)v_{l}\nonumber \\
 & =\sum_{l=1}^{\infty}\sin\left(l\theta_{\mathrm{p}}\right)v_{l}s-\sum_{l=1}^{\infty}\left(1+\frac{F_{l}^{s}}{2}\right)\frac{1}{2}\left[\sin\left(\theta_{\mathrm{p}}+l\theta_{\mathrm{p}}\right)-\sin\left(\theta_{\mathrm{p}}-l\theta_{\mathrm{p}}\right)\right]v_{l}\nonumber \\
 & =\sum_{l=1}^{\infty}\sin\left(l\theta_{\mathrm{p}}\right)\left[v_{l}s-\left(1+\frac{F_{l-1}^{s}}{2}\right)\frac{v_{l-1}}{2}-\left(1+\frac{F_{l+1}^{s}}{2}\right)\frac{v_{l+1}}{2}\right].
\end{align}
By noting that $v_{0}=0$,
the type II algebraic equation for transverse mode is
\begin{equation}
v_{l}s-\left(1+\frac{F_{l-1}^{s}}{2}\right)\frac{v_{l-1}}{2}-\left(1+\frac{F_{l+1}^{s}}{2}\right)\frac{v_{l+1}}{2}=\delta_{l2}\frac{U}{2}.
\end{equation}

\section{Zero sound: longitudinal mode in 2D} \label{app:ZSL2D}
The algebraic equations type I and II for longitudinal mode in the collisionless regime in 2D are given by
\begin{equation}
\frac{u_{l}+\delta_{l0}u_{0}}{2}+\sum_{l'=0}^{\infty}F_{l'}^{s}\Pi_{ll'}\left(s\right)\frac{u_{l'}+\delta_{l'0}u_{0}}{2}=-\Pi_{l0}\left(s\right)U,
\end{equation}
and
\begin{equation}
u_{l}s-\left(1+\frac{F_{l-1}^{s}+\delta_{l1}F_{0}^{s}}{2}\right)\frac{u_{l-1}+\delta_{l1}u_{0}}{2}-\left(1+\frac{F_{l+1}^{s}}{2}\right)\frac{u_{l+1}}{2}=\delta_{l1}U,
\end{equation}
respectively. These two algebraic equations are equivalent.
In order to manifest the equivalence, by using the following relations
\begin{align}
\Pi_{l1}-s\Pi_{l0} & =-\int\frac{d\theta_{\mathrm{p}}}{2\pi}\cos\left(l\theta_{\mathrm{p}}\right)\frac{\cos\theta_{\mathrm{p}}}{s-\cos\theta_{\mathrm{p}}}\cos\theta_{\mathrm{p}}+\int\frac{d\theta_{\mathrm{p}}}{2\pi}\cos\left(l\theta_{\mathrm{p}}\right)\frac{s\cos\theta_{\mathrm{p}}}{s-\cos\theta_{\mathrm{p}}}\nonumber \\
 & =\int\frac{d\theta_{\mathrm{p}}}{2\pi}\cos\left(l\theta_{\mathrm{p}}\right)\frac{s-\cos\theta_{\mathrm{p}}}{s-\cos\theta_{\mathrm{p}}}\cos\theta_{\mathrm{p}}\nonumber \\
 & =\int\frac{d\theta_{\mathrm{p}}}{2\pi}\cos\left(l\theta_{\mathrm{p}}\right)\cos\theta_{\mathrm{p}}\nonumber \\
 & =\frac{1}{2}\delta_{l1}.
\end{align}
and then multiplying the $l=0$ and $l=1$ components of algebraic equation type I by $\Pi_{10}$ and $\Pi_{00}$, respectively, we can eliminate the $U$ terms and obtain
\begin{align}
0 & =\left(u_{0}+\sum_{l'=0}^{\infty}F_{l'}^{s}\Pi_{0l'}\left(s\right)\frac{u_{l'}+\delta_{l'0}u_{0}}{2}\right)\Pi_{10}-\left(\frac{u_{1}}{2}+\sum_{l'=0}^{\infty}F_{l'}^{s}\Pi_{1l'}\left(s\right)\frac{u_{l'}+\delta_{l'0}u_{0}}{2}\right)\Pi_{00}\nonumber \\
 & =\left[\left(1+F_{0}^{s}\Pi_{00}\right)\Pi_{10}-F_{0}^{s}\Pi_{10}\Pi_{00}\right]u_{0}+\left[F_{1}^{s}\Pi_{01}\Pi_{10}-\left(1+F_{1}^{s}\Pi_{11}\right)\Pi_{00}\right]\frac{u_{1}}{2}+\sum_{l'=2}^{\infty}F_{l'}^{s}\left(\Pi_{0l'}\Pi_{10}-\Pi_{1l'}\Pi_{00}\right)\frac{u_{l'}}{2}\nonumber \\
 & =\Pi_{10}u_{0}-\Pi_{00}\left(1+\frac{F_{1}^{s}}{2}\right)\frac{u_{1}}{2}\nonumber \\
 & =\Pi_{00}\left[u_{0}s-\left(1+\frac{F_{1}^{s}}{2}\right)\frac{u_{1}}{2}\right],
\end{align}
which is nothing but the $l=0$ component of algebraic equation type II.

\subsection{Two channel model}
Now we consider a two channel model by keeping only $F_{0}^{s}$ and $F_{1}^{s}$ and setting $F_{l}^{s}=0$ for $l\geq 2$. The components of algebraic equation I can be classified into two categories: (i) $l\geq 2$ and (ii) $l=0$ and $l=1$. The components in the second category form a close set that is composed of $u_{0}$ and $u_{1}$:
\begin{align}
u_{0}+F_{0}^{s}\Pi_{00}u_{0}+F_{1}^{s}\Pi_{01}\frac{u_{1}}{2} & =-\Pi_{00}U,\\
\frac{u_{1}}{2}+F_{0}^{s}\Pi_{10}u_{0}+F_{1}^{s}\Pi_{11}\frac{u_{1}}{2} & =-\Pi_{10}U.
\end{align}
The solution from algebraic equation I leads to the response functions as follows,
\begin{align}
\frac{u_{0}}{U} & =-\frac{\left(1+F_{1}^{s}\Pi_{11}\right)\Pi_{00}-F_{1}^{s}\Pi_{01}\Pi_{10}}{\left(1+F_{0}^{s}\Pi_{00}\right)\left(1+F_{1}^{s}\Pi_{11}\right)-F_{0}^{s}\Pi_{10}F_{1}^{s}\Pi_{01}}=-\frac{\left(1+\frac{F_{1}^{s}}{2}\right)\Pi_{00}}{\left(1+F_{0}^{s}\Pi_{00}\right)\left(1+\frac{F_{1}^{s}}{2}\right)+F_{1}^{s}s^{2}\Pi_{00}},\\
\frac{u_{1}}{2U} & =-\frac{-F_{0}^{s}\Pi_{10}\Pi_{00}+\left(1+F_{0}^{s}\Pi_{00}\right)\Pi_{10}}{\left(1+F_{0}^{s}\Pi_{00}\right)\left(1+F_{1}^{s}\Pi_{11}\right)-F_{0}^{s}\Pi_{10}F_{1}^{s}\Pi_{01}}=-\frac{s\Pi_{00}}{\left(1+F_{0}^{s}\Pi_{00}\right)\left(1+\frac{F_{1}^{s}}{2}\right)+F_{1}^{s}s^{2}\Pi_{00}},
\end{align}
where we have used the relations $\Pi_{l1}=s\Pi_{l0}+\frac{1}{2}\delta_{l1}$.

From the above solution, we find that the complete suppression of $u_{0}$ component requires that
\begin{equation}
\left(1+\frac{F_{1}^{s}}{2}\right)\Pi_{00}\left(s\right)=0.
\end{equation}
As there is no real solution to $\Pi_{00}\left(s\right)=0$ except at $s\rightarrow\infty$, thus we obtain $1+\frac{F_{1}^{s}}{2}=0$, which is nothing but the QSL condition in 2D as given in Eq.~\eqref{eq:F1s}.

We proceed to consider the possibility of the complete suppression of $u_{1}$ component, which leads to
\begin{equation}
s\Pi_{00}\left(s\right)=0.
\end{equation}
Again there is no real solution except at $s=0$ or $s\rightarrow\infty$. This means that it is impossible to suppress the $l=1$ component entirely, and such a mode is generally allowed.

The zero sound mode frequency can be determined via the pole of the response functions, i.e.,
\begin{equation}\label{eq:mode_fre_2D_two}
\left(1+F_{0}^{s}\Pi_{00}\right)\left(1+\frac{F_{1}^{s}}{2}\right)+F_{1}^{s}s^{2}\Pi_{00}=0.
\end{equation}
To find a real solution s to the above equation, we define function
\begin{equation}
h_{2}\left(s\right)=\left(1+\frac{F_{1}^{s}}{2}\right)+\Pi_{00}\left[F_{0}^{s}\left(1+\frac{F_{1}^{s}}{2}\right)+F_{1}^{s}s^{2}\right].
\end{equation}
Note that $h_{2}\left(s\rightarrow\infty\right)=1>0$ and meanwhile
\begin{equation}
h_{2}\left(s\rightarrow1\right)=-\frac{F_{0}^{s}\left(1+\frac{F_{1}^{s}}{2}\right)+F_{1}^{s}}{\sqrt{2\left(s-1\right)}}+O\left(1\right).
\end{equation}
To have a real solution for $h_{2}\left(s\right)=0$, one requires
\begin{equation}\label{eq:mode_fre_2D_condi}
F_{0}^{s}\left(1+\frac{F_{1}^{s}}{2}\right)+F_{1}^{s}>0.
\end{equation}

Eq.~\eqref{eq:mode_fre_2D_two} must generally be found numerically. Using the limiting forms of $\Pi_{00}$,
\begin{equation}
\Pi_{00}\left(s\right)=\begin{cases}
1-\frac{1}{\sqrt{2\left(s-1\right)}}, & 0<s-1\ll1,\\
-\frac{1}{2s^{2}}-\frac{3}{8s^{4}}-\frac{5}{16s^{6}}-\cdots, & s\rightarrow\infty,
\end{cases}
\end{equation}
Eq.~\eqref{eq:mode_fre_2D_two} can approximately be solved in two limits
\begin{equation}
s=\begin{cases}
1+\frac{1}{2\left[1+\frac{\left(1+\frac{F_{1}^{s}}{2}\right)}{F_{0}^{s}\left(1+\frac{F_{1}^{s}}{2}\right)+F_{1}^{s}}\right]^{2}}, & 0<s-1\ll1,\\
\sqrt{\frac{F_{0}^{s}}{2}\left(1+\frac{F_{1}^{s}}{2}\right)+\frac{3F_{1}^{s}}{8}}, & s\rightarrow\infty.
\end{cases}
\end{equation}

Under the condition of QSL in 2D, Eq.~\eqref{eq:mode_fre_2D_two} reduces to
\begin{equation}
s^{2}\Pi_{00}\left(s\right)=0.
\end{equation}
There is no nontrivial real solution, indicating the absence of weakly damped zero sound mode in the two channel QSL model. 
Also, when $1+\frac{F_{1}^{s}}{2}\rightarrow0^{+}$, the inequality Eq.~\eqref{eq:mode_fre_2D_condi} can not be satisfied unless $F_{0}^{s}\rightarrow\infty$. In order to obtain a weakly damped longitudinal zero sound mode in a finite $F_{0}^{s}$ and under the QSL condition $1+\frac{F_{1}^{s}}{2}\rightarrow0^{+}$, one need to involve more $F_{l}^{s}$ with $l\geq2$.

\subsection{Three channel model}
Consider a three channel model by keeping $F_{0}^{s}$, $F_{1}^{s}$ and $F_{2}^{s}$ and setting $F_{l}^{s}=0$ for $l\geq 3$, we obtain the close set of components $\nu_{0}$, $\nu_{1}$, and $\nu_{2}$ from algebraic equation I as follows,
\begin{equation}
\left(\begin{array}{ccc}
1+F_{0}^{s}\Pi_{00} & F_{1}^{s}\Pi_{01} & F_{2}^{s}\Pi_{02}\\
F_{0}^{s}\Pi_{10} & 1+F_{1}^{s}\Pi_{11} & F_{2}^{s}\Pi_{12}\\
F_{0}^{s}\Pi_{20} & F_{1}^{s}\Pi_{21} & 1+F_{2}^{s}\Pi_{22}
\end{array}\right)\left(\begin{array}{c}
u_{0}\\
\frac{u_{1}}{2}\\
\frac{u_{2}}{2}
\end{array}\right)=-U\left(\begin{array}{c}
\Pi_{00}\\
\Pi_{10}\\
\Pi_{20}
\end{array}\right).
\end{equation}
Then the mode frequency is determined by the following secular equation,
\begin{equation}
h_{3}\left(s\right)=\det\left(\begin{array}{ccc}
1+F_{0}^{s}\Pi_{00} & F_{1}^{s}\Pi_{01} & F_{2}^{s}\Pi_{02}\\
F_{0}^{s}\Pi_{10} & 1+F_{1}^{s}\Pi_{11} & F_{2}^{s}\Pi_{12}\\
F_{0}^{s}\Pi_{20} & F_{1}^{s}\Pi_{21} & 1+F_{2}^{s}\Pi_{22}
\end{array}\right)=0.
\end{equation}
Here the function $h_{3}(s)$ can be written explicitly as follows,
\begin{equation}
h_{3}\left(s\right)=\left(1+\frac{F_{1}^{s}}{2}\right)\left(1+F_{2}^{s}\Pi_{22}\right)+\left[F_{0}^{s}\left(1+\frac{F_{1}^{s}}{2}\right)+F_{1}^{s}s^{2}\right]\left[\Pi_{00}+F_{2}^{s}\left(\Pi_{00}\Pi_{22}-\Pi_{20}^{2}\right)\right].
\end{equation}
To find a real solution $s$ to the secular equation, we notice that $h_{3}\left(s\rightarrow\infty\right)=1>0$ and
\begin{equation}
h_{3}\left(s\rightarrow1\right)=-\left\{ \left(1+\frac{F_{1}^{s}}{2}\right)\left[2+F_{0}^{s}-\frac{F_{2}^{s}}{2}F_{0}^{s}\right]+2\left(\frac{F_{2}^{s}}{2}-1\right)\right\} \frac{1}{\sqrt{2\left(s-1\right)}}.
\end{equation}
Therefore, the inequality
\begin{equation}\label{ineq:three_condi_2D}
\left(1+\frac{F_{1}^{s}}{2}\right)\left[2+F_{0}^{s}-\frac{F_{2}^{s}}{2}F_{0}^{s}\right]+2\left(\frac{F_{2}^{s}}{2}-1\right)>0,
\end{equation}
or
\begin{equation}
\left[1-\frac{F_{0}^{s}}{2}\left(1+\frac{F_{1}^{s}}{2}\right)\right]F_{2}^{s}+\left[F_{0}^{s}\left(1+\frac{F_{1}^{s}}{2}\right)+F_{1}^{s}\right]>0,
\end{equation}
gives rise to a sufficient condition for a real solution $s>1$ to the secular equation $h_{3}(s)=0$.

In general, the secular equation $h_{3}(s)=0$ must be found numerically.
Nevertheless we can solve the secular equation analytically around $s=1$ and $s=\infty$.
To do this, we expand $h_3(s)$ around $s=1$ and $s=\infty$. With the help of the following relations:
\begin{align}
\Pi_{20} & =1+\left(2s^{2}-1\right)\Pi_{00,}\\
\Pi_{22} & =2s^{2}-\frac{1}{2}+\left(2s^{2}-1\right)^{2}\Pi_{00},
\end{align}
we can rewrite $h_{3}(s)$ as follows,
\begin{align}
h_{3}\left(s\right) & =\left(1+\frac{F_{1}^{s}}{2}\right)\left(1+F_{2}^{s}\Pi_{22}\right)+\left[F_{0}^{s}\left(1+\frac{F_{1}^{s}}{2}\right)+F_{1}^{s}s^{2}\right]\left[\Pi_{00}+F_{2}^{s}\left(\Pi_{00}\Pi_{22}-\Pi_{20}^{2}\right)\right]\nonumber \\
 & =C\left(s\right)+D\left(s\right)\Pi_{00}\left(s\right)\nonumber \\
 & =\tilde{C}\left(s\right)+\tilde{D}\left(s\right)\frac{1}{\sqrt{\left(s+1\right)\left(s-1\right)}}.
\end{align}
Here
\begin{align}
C\left(s\right) & =C_{2}s^{2}+C_{0}\nonumber \\
 & =2F_{2}^{s}s^{2}+\left(1+\frac{F_{1}^{s}}{2}\right)\left[1-\frac{F_{2}^{s}}{2}\left(1+2F_{0}^{s}\right)\right],\\
D\left(s\right) & =D_{4}s^{4}+D_{2}s^{2}+D_{0}\nonumber \\
 & =4F_{2}^{s}s^{4}-2s^{2}\left\{ F_{2}^{s}\left[\left(2+F_{0}^{s}\right)\left(1+\frac{F_{1}^{s}}{2}\right)-\frac{3}{4}F_{1}^{s}\right]-\frac{F_{1}^{s}}{2}\right\} +\left[F_{2}^{s}\left(1+\frac{3}{2}F_{0}^{s}\right)+F_{0}^{s}\right]\left(1+\frac{F_{1}^{s}}{2}\right),
\end{align}
and
\begin{align}
\tilde{C}\left(s\right) & =C\left(s\right)+D\left(s\right),\\
\tilde{D}\left(s\right) & =-sD\left(s\right),
\end{align}
where $C_{0},C_{2},D_{0},D_{2}$ and $D_{4}$ are coefficients that does not depend on $s$.
Thus the secular equation becomes
\begin{equation}
C\left(s\right)+D\left(s\right)\Pi_{00}\left(s\right)=\tilde{C}\left(s\right)+\tilde{D}\left(s\right)\frac{1}{\sqrt{\left(s+1\right)\left(s-1\right)}}=0.
\end{equation}

First, we consider the solution $s\to{}1+0^{+}$. When
\begin{equation}
\lim_{s\rightarrow1+0^{+}}\frac{\tilde{C}\left(s\right)}{\tilde{D}\left(s\right)}=-1-\frac{F_{2}^{s}\left[2-\left(1+\frac{F_{1}^{s}}{2}\right)\frac{1}{2}\left(1+2F_{0}^{s}\right)\right]+\left(1+\frac{F_{1}^{s}}{2}\right)}{F_{2}^{s}\left[1-\frac{F_{0}^{s}}{2}\left(1+\frac{F_{1}^{s}}{2}\right)\right]+F_{1}^{s}+F_{0}^{s}\left(1+\frac{F_{1}^{s}}{2}\right)}\gg1,
\end{equation}
the secular equation has an approximate solution
\begin{equation}
s\simeq1+\frac{1}{2\left[\frac{\tilde{C}\left(s=1\right)}{\tilde{D}\left(s=1\right)}\right]^{2}}=1+\frac{1}{2\left[1+\frac{F_{2}^{s}\left[2-\left(1+\frac{F_{1}^{s}}{2}\right)\frac{1}{2}\left(1+2F_{0}^{s}\right)\right]+\left(1+\frac{F_{1}^{s}}{2}\right)}{F_{2}^{s}\left[1-\frac{F_{0}^{s}}{2}\left(1+\frac{F_{1}^{s}}{2}\right)\right]+F_{1}^{s}+F_{0}^{s}\left(1+\frac{F_{1}^{s}}{2}\right)}\right]^{2}}.
\end{equation}
Second, we look for the solution $s\to{}+\infty$, which is available when
\begin{align}
C_{2} & =\frac{D_{4}}{2}\\
\frac{\frac{D_{0}}{2}+\frac{3D_{2}}{8}+\frac{5D_{4}}{16}}{C_{0}-\frac{D_{2}}{2}-\frac{3D_{4}}{8}} & =\frac{D_{0}}{2}+\frac{3D_{2}}{8}+\frac{5D_{4}}{16}\gg1.
\end{align}
In this situation, we have another solution
\begin{equation}
s=\sqrt{\frac{D_{0}}{2}+\frac{3D_{2}}{8}+\frac{5D_{4}}{16}}=\sqrt{\frac{F_{2}^{s}}{8}\left[1+\left(1+\frac{F_{1}^{s}}{2}\right)\right]+\frac{F_{0}^{s}}{2}\left(1+\frac{F_{1}^{s}}{2}\right)+\frac{3F_{1}^{s}}{8}}.
\end{equation}

Taking into account of the sufficient condition given in Eq.~\eqref{ineq:three_condi_2D}, we find that the solution will take simpler forms in the two incompressible conditions (i) $1+
\frac{F_{1}^{s}}{2}=0$ and (ii) $F_{0}^{s}\to+\infty$: 

(i) In the incompressibility condition $1+\frac{F_1^s}{2}=0$, we have
\begin{equation}
s=\begin{cases}
1+\frac{1}{2\left[1+\frac{2F_{2}^{s}}{F_{2}^{s}-2}\right]^{2}}, & F_{2}^{s}>2,\\
\sqrt{\frac{F_{2}^{s}}{8}-\frac{3}{4}}, & F_{2}^{s}\rightarrow\infty.
\end{cases}
\end{equation}
When $F_{2}^{s}>2$, there exists at least one weakly damped zero sound mode with the sound speed $c_{0}=sv_{F}=v_{F}\left(1+\frac{1}{2\left[1+\frac{2F_{2}^{s}}{F_{2}^{s}-2}\right]^{2}}\right)$
, an additional weakly damped zero sound mode occurs with the sound speed $c_{0}=v_{F}\sqrt{\frac{F_{2}^{s}}{8}-\frac{3}{4}}$.

(ii) In the other incompressibility condition $F_{0}^{s}\to+\infty$, we have two approximate solutions to the secular equation
\begin{equation}
s=\begin{cases}
1+\frac{1}{2\left[1+\frac{2F_{2}^{s}}{F_{2}^{s}-2}\right]^{2}}, & s\rightarrow1+0^{+},\\
\sqrt{\frac{F_{0}^{s}}{2}\left(1+\frac{F_{1}^{s}}{2}\right)}, & s\rightarrow\infty.
\end{cases}
\end{equation}
Here the constraint $F_{2}^{s}>2$ is still imposed by Eq.~\eqref{ineq:three_condi_2D}.

\section{Zero sound: transverse mode in 2D}
The algebraic equations type I and II for transverse mode in the collisionless regime in 2D are given by
\begin{equation}
\frac{v_{l}}{2}+\sum_{l'=0}^{\infty}F_{l'}^{s}\Xi_{ll'}\left(s\right)\frac{v_{l'}}{2}=-\Xi_{l1}\left(s\right)U,
\end{equation}
and
\begin{equation}
v_{l}s-\left(1+\frac{F_{l-1}^{s}}{2}\right)\frac{v_{l-1}}{2}-\left(1+\frac{F_{l+1}^{s}}{2}\right)\frac{v_{l+1}}{2}=\delta_{l2}\frac{U}{2}.
\end{equation}
These two algebraic equations are equivalent.
To manifest the equivalence, using the following relations
\begin{align}
\Xi_{l2}-2s\Xi_{l1} & =-\int\frac{d\theta_{\mathrm{p}}}{2\pi}\sin\left(l\theta_{\mathrm{p}}\right)\frac{\cos\theta_{\mathrm{p}}}{s-\cos\theta_{\mathrm{p}}}\sin\left(2\theta_{\mathrm{p}}\right)+\int\frac{d\theta_{\mathrm{p}}}{2\pi}\sin\left(l\theta_{\mathrm{p}}\right)\frac{2s\cos\theta_{\mathrm{p}}}{s-\cos\theta_{\mathrm{p}}}\sin\left(\theta_{\mathrm{p}}\right)\nonumber \\
 & =\int\frac{d\theta_{\mathrm{p}}}{2\pi}\sin\left(l\theta_{\mathrm{p}}\right)\frac{s-\cos\theta_{\mathrm{p}}}{s-\cos\theta_{\mathrm{p}}}\sin\left(2\theta_{\mathrm{p}}\right)\nonumber \\
 & =\int\frac{d\theta_{\mathrm{p}}}{2\pi}\sin\left(l\theta_{\mathrm{p}}\right)\sin\left(2\theta_{\mathrm{p}}\right)\nonumber \\
 & =\frac{1}{2}\delta_{l2},
\end{align}
and then multiplying the $l=1$ and $l=2$ components of algebraic equation type I by $\Xi_{21}$ and $\Xi_{11}$, respectively,  we can eliminate the $U$ terms and obtain
\begin{align}
0 & =\left(\frac{v_{1}}{2}+\sum_{l'=1}^{\infty}F_{l'}^{s}\Xi_{1l'}\frac{v_{l'}}{2}\right)\Xi_{21}-\left(\frac{v_{2}}{2}+\sum_{l'=1}^{\infty}F_{l'}^{s}\Xi_{2l'}\frac{v_{l'}}{2}\right)\Xi_{11}\nonumber \\
 & =\left[\left(1+F_{1}^{s}\Xi_{11}\right)\Xi_{21}-F_{1}^{s}\Xi_{21}\Xi_{11}\right]\frac{v_{1}}{2}+\left[F_{2}^{s}\Xi_{12}\Xi_{21}-\left(1+F_{2}^{s}\Xi_{22}\right)\Xi_{11}\right]\frac{v_{2}}{2}+\sum_{l'=3}^{\infty}F_{l'}^{s}\left(\Xi_{1l'}\Xi_{21}-\Xi_{2l'}\Xi_{11}\right)\frac{v_{l'}}{2}\nonumber \\
 & =\Xi_{21}\frac{v_{1}}{2}-\Xi_{11}\left(1+\frac{F_{2}^{s}}{2}\right)\frac{v_{2}}{2}\nonumber \\
 & =\Xi_{11}\left[v_{1}s-\left(1+\frac{F_{2}^{s}}{2}\right)\frac{v_{2}}{2}\right],
\end{align}
which is nothing but the $l=1$ component of algebraic equation of type II. Note here we have used that $\Xi_{l0}\left(s\right)=0$.

\subsection{Two channel model}
Now we consider a two channel model by keeping only $F_{1}^{s}$ and $F_{2}^{s}$ and setting $F_{l}^{s}=0$ for $l\geq 3$. The components of algebraic equation I can be classified into two categories: (i) $l\geq 3$ and (ii) $l=1$ and $l=2$. The components in the second category form a close set that is composed of $v_{1}$ and $v_{2}$:
\begin{align}
\frac{v_{1}}{2}+F_{1}^{s}\Xi_{11}\frac{v_{1}}{2}+F_{2}^{s}\Xi_{12}\frac{v_{2}}{2} & =-\Xi_{11}U,\\
\frac{v_{2}}{2}+F_{1}^{s}\Xi_{21}\frac{v_{1}}{2}+F_{2}^{s}\Xi_{22}\frac{v_{2}}{2} & =-\Xi_{21}U.
\end{align}
The solution from algebraic equation I leads to the response functions as follows,
\begin{align}
\frac{v_{1}}{2U} & =-\frac{\left(1+\frac{F_{2}^{s}}{2}\right)\Xi_{11}}{\left(1+F_{1}^{s}\Xi_{11}\right)\left(1+\frac{F_{2}^{s}}{2}\right)+F_{2}^{s}4s^{2}\Xi_{11}},\\
\frac{v_{2}}{2U} & =-\frac{2s\Xi_{11}}{\left(1+F_{1}^{s}\Xi_{11}\right)\left(1+\frac{F_{2}^{s}}{2}\right)+F_{2}^{s}4s^{2}\Xi_{11}}.
\end{align}

The zero sound mode frequency can be determined via the pole of the response functions, i.e.,
\begin{equation}
\left(1+F_{1}^{s}\Xi_{11}\right)\left(1+\frac{F_{2}^{s}}{2}\right)+F_{2}^{s}4s^{2}\Xi_{11}=0.
\end{equation}
To find a real solution s to the above equation, we define a function
\begin{align}
\bar{h}_{2}\left(s\right) & =\left(1+\frac{F_{2}^{s}}{2}\right)+\Xi_{11}\left[F_{1}^{s}\left(1+\frac{F_{2}^{s}}{2}\right)+F_{2}^{s}4s^{2}\right]\nonumber \\
 & =\left(1+\frac{F_{2}^{s}}{2}\right)+\frac{1-2s^{2}}{2}\Pi_{00}\left[F_{1}^{s}\left(1+\frac{F_{2}^{s}}{2}\right)+F_{2}^{s}4s^{2}\right],
\end{align}
we have $\bar{h}_{2}\left(s\rightarrow\infty\right)=2F_{2}^{s}s^{2}$ and meanwhile
\begin{equation}
\bar{h}_{2}\left(s\rightarrow1\right)=\frac{1}{2}\frac{F_{1}^{s}\left(1+\frac{F_{2}^{s}}{2}\right)+4F_{2}^{s}}{\sqrt{2\left(s-1\right)}}+O\left(1\right).
\end{equation}
To have a real solution for $\bar{h}_{2}\left(s\right)=0$, one requires
\begin{equation}
F_{2}^{s}\left[F_{1}^{s}\left(1+\frac{F_{2}^{s}}{2}\right)+4F_{2}^{s}\right]<0.
\end{equation}
Under the condition of QSL in 2D $1+\frac{F_{1}^{s}}{2}=0$, it reduces to
\begin{equation}
F_{2}^{s}\left(3F_{2}^{s}-2\right)<0
\end{equation}
Thus we have transverse mode for $0<F_{2}^{s}<2/3$ under the condition of QSL in 2D.

Eq.~\eqref{eq:mode_fre_2D_two_trans} must generally be found numerically. In the limit $0<s-1\ll1$, we can obtain the approximately analytical solution
\begin{equation}
s=1+\frac{1}{2\left[1-\frac{2\left(1+\frac{F_{2}^{s}}{2}\right)}{F_{1}^{s}\left(1+\frac{F_{2}^{s}}{2}\right)+4F_{2}^{s}}\right]^{2}}.
\end{equation}
Under the condition of QSL in 2D $1+\frac{F_{1}^{s}}{2}=0$, it simplifies to
\begin{equation}
s=1+\frac{1}{8}\left(\frac{3F_{2}^{s}-2}{F_{2}^{s}-2}\right)^{2}.
\end{equation}

\section{First sound: longitudinal mode in 2D}\label{app:FS2D}

In 2D, we assume that the collision integral takes the form of
\begin{equation}
I\left[n_{\mathrm{p}}\right]=-\frac{1}{\tau}\left[\delta n_{\mathrm{p}}-\left\langle \delta n_{\mathrm{p}}\right\rangle -2\left\langle \delta n_{\mathrm{p}}\cos\theta_{\mathrm{p}}\right\rangle \cos\theta_{\mathrm{p}}\right].
\end{equation}
The density fluctuation is given by
\begin{equation}
\delta n_{\mathrm{p}}=-\frac{\partial n_{\mathrm{p}}^{0}}{\partial\varepsilon_{\mathrm{p}}}\nu_{\mathrm{p}},
\end{equation}
where $\nu_{\mathrm{p}}$ can be expanded in 2D as follows
\begin{equation}
\nu_{\mathrm{p}}=\sum_{l=0}^{\infty}\cos\left(l\theta_{\mathrm{p}}\right)\nu_{l}.
\end{equation}
The expectation values in the collision integral $I[n_{\mathrm{p}}]$ can be computed as
\begin{align}
\left\langle \delta n_{\mathrm{p}}\right\rangle  & =-\frac{\partial n_{\mathrm{p}}^{0}}{\partial\varepsilon_{\mathrm{p}}}\nu_{0},\\
2\left\langle \delta n_{\mathrm{p}}\cos\theta_{\mathrm{p}}\right\rangle  & =-\frac{\partial n_{\mathrm{p}}^{0}}{\partial\varepsilon_{\mathrm{p}}}\nu_{1}.
\end{align}
The Landau kinetic equation in the collision regime reads
\begin{equation}
\frac{\partial \delta n_{\mathrm{p}}}{\partial t}+\vec{v}_{\mathrm{p}}\cdot
\nabla _{\mathrm{r}}\left( \delta n_{\mathrm{p}}-\frac{\partial n_{\mathrm{p}
	}^{0}}{\partial \varepsilon _{\mathrm{p}}}\delta \varepsilon _{\mathrm{p}
}\right) =I[n_{\mathrm{p}}].  \label{LTEFS}
\end{equation}
Taking into account of a scalar external field $U\left(q,\omega\right)=Ue^{i\left(q\cdot r-\omega t\right)}$, in the collisionless regime, the quasiparticle excitation energy reads
\begin{equation}
\delta\varepsilon_{\mathrm{p}}=U+\sum_{\mathrm{p}'}f_{\mathrm{p}\mathrm{p}'}^{s}\delta n_{\mathrm{p'}},
\end{equation}
and the Landau kinetic equation can be rewritten in frequency-momentum space as
\begin{equation}
\left(\omega+\frac{i}{\tau}-q\cdot\vec{\nu}_{\mathrm{p}}\right)\nu_{\mathrm{p}}+\left(q\cdot\vec{\nu}_{\mathrm{p}}\right)\left(-U+\sum_{\mathrm{p}'}f_{\mathrm{p}\mathrm{p}'}^{s}\frac{\partial n_{\mathrm{p'}}^{0}}{\partial\varepsilon_{\mathrm{p'}}}\nu_{\mathrm{p}'}\right)=\frac{i}{\tau}\left(\nu_{0}+\nu_{1}\cos\theta_{\mathrm{p}}\right).
\end{equation}
Putting the expansion of $\nu_{\mathrm{p}}$ into the above equation, we obtain reduced Landau kinetic equation in the $U=0$ limit:
\begin{equation}
\sum_{l=0}^{\infty}\cos\left(l\theta_{\mathrm{p}}\right)\nu_{l}-\frac{q\cdot\vec{\nu}_{\mathrm{p}}}{\omega+\frac{i}{\tau}-q\cdot\vec{\nu}_{\mathrm{p}}}\sum_{l=0}^{\infty}F_{l}^{s}\cos\left(l\theta_{\mathrm{p}}\right)\frac{\nu_{l}+\delta_{l0}\nu_{0}}{2}=\frac{1}{\omega+\frac{i}{\tau}-q\cdot\vec{\nu}_{\mathrm{p}}}\frac{i}{\tau}\left(\nu_{0}+\nu_{1}\cos\theta_{\mathrm{p}}\right).
\end{equation}
Using the orthogonal relations for cosine functions,
\begin{equation}
\int\frac{d\theta_{\mathrm{p}}}{2\pi}\cos\left(l\theta_{\mathrm{p}}\right)\cos\left(l'\theta_{\mathrm{p}}\right)=\int\frac{d\theta_{\mathrm{p}}}{2\pi}\frac{1}{2}\left[\cos\left(l\theta_{\mathrm{p}}+l'\theta_{\mathrm{p}}\right)+\cos\left(l\theta_{\mathrm{p}}-l'\theta_{\mathrm{p}}\right)\right]=\frac{1}{2}\left(\delta_{ll'}+\delta_{l0}\delta_{l'0}\right),
\end{equation}
we obtain the algebraic equation I in the collision regime as
\begin{equation}
\frac{\nu_{l}+\delta_{l0}\nu_{0}}{2}+\sum_{l'=0}^{\infty}F_{l'}^{s}\Pi_{ll'}\left(\tilde{s}\right)\frac{\nu_{l'}+\delta_{l'0}\nu_{0}}{2}=-i\kappa\left[\Pi_{l0}\left(\tilde{s}\right)\nu_{1}+\Pi_{l}\left(\tilde{s}\right)\nu_{0}\right],
\end{equation}
where
\begin{align}
s & =\frac{\omega}{qv_{F}},\\
\kappa & =\frac{1}{\tau qv_{F}}=\frac{s}{\omega\tau},\\
\tilde{s} & =\frac{\omega}{qv_{F}}+\frac{i}{\tau qv_{F}}=s+i\kappa,
\end{align}
and
\begin{align}
\Pi_{ll'}\left(\tilde{s}\right) & =\Pi_{l'l}\left(\tilde{s}\right)=-\int\frac{d\theta_{\mathrm{p}}}{2\pi}\cos\left(l\theta_{\mathrm{p}}\right)\frac{\cos\theta_{\mathrm{p}}}{\tilde{s}-\cos\theta_{\mathrm{p}}}\cos\left(l'\theta_{\mathrm{p}}\right),\\
\Pi_{l}\left(\tilde{s}\right) & =\frac{1}{\tilde{s}}\left(\Pi_{l0}\left(\tilde{s}\right)-\delta_{l0}\right).
\end{align}
Indeed, the algebraic equation type I now contains $\kappa$-terms which are induced by collisions.

\subsection{Two channel model}
To study first sound mode, we consider a two channel model by keeping $F_{0}^{s}$ and $F_{1}^{s}$ and setting $F_{l}^{s}=0$ for $l\geq 2$. Then Landau kinetic equation becomes
\begin{equation}
\left(\begin{array}{cc}
1+F_{0}^{s}\Pi_{00}+i\kappa\Pi_{0} & \frac{1}{2}F_{1}^{s}\Pi_{01}+i\kappa\Pi_{00}\\
F_{0}^{s}\Pi_{10}+i\kappa\Pi_{1} & \frac{1}{2}\left(1+F_{1}^{s}\Pi_{11}\right)+i\kappa\Pi_{10}
\end{array}\right)\left(\begin{array}{c}
\nu_{0}\\
\nu_{1}
\end{array}\right)=0.
\end{equation}
The condition for a nontrivial solution allows us to determine the frequency and leads to
\begin{align}
0 & =\det\left(\begin{array}{cc}
1+F_{0}^{s}\Pi_{00}+i\kappa\Pi_{0} & \frac{1}{2}F_{1}^{s}\Pi_{01}+i\kappa\Pi_{00}\\
F_{0}^{s}\Pi_{10}+i\kappa\Pi_{1} & \frac{1}{2}\left(1+F_{1}^{s}\Pi_{11}\right)+i\kappa\Pi_{10}
\end{array}\right)\nonumber \\
 & =\left(1+F_{0}^{s}\Pi_{00}+i\kappa\Pi_{0}\right)\left(\frac{1}{2}\left(1+F_{1}^{s}\Pi_{11}\right)+i\kappa\Pi_{10}\right)-\left(\frac{1}{2}F_{1}^{s}\Pi_{01}+i\kappa\Pi_{00}\right)\left(F_{0}^{s}\Pi_{10}+i\kappa\Pi_{1}\right)\nonumber \\
 & =\frac{1}{2}\left\{ \left(1-\frac{i\kappa}{\tilde{s}}\right)\left(1+\frac{F_{1}^{s}}{2}\right)+\Pi_{00}\left[\left(F_{0}^{s}+\frac{i\kappa}{\tilde{s}}\right)\left(1+\frac{F_{1}^{s}}{2}\right)+2\tilde{s}^{2}\left(1-\frac{i\kappa}{\tilde{s}}\right)\left(\frac{F_{1}^{s}}{2}+\frac{i\kappa}{\tilde{s}}\right)\right]\right\},
\end{align}
where we have used
\begin{align}
\Pi_{10}\left(\tilde{s}\right) & =\tilde{s}\Pi_{00}\left(\tilde{s}\right),\\
\Pi_{11}\left(\tilde{s}\right) & =\tilde{s}^{2}\Pi_{00}\left(\tilde{s}\right)+\frac{1}{2},\\
\Pi_{0}\left(\tilde{s}\right) & =\frac{1}{\tilde{s}}\left(\Pi_{00}\left(\tilde{s}\right)-1\right),\\
\Pi_{1}\left(\tilde{s}\right) & =\frac{1}{\tilde{s}}\Pi_{10}\left(\tilde{s}\right)=\Pi_{00}\left(\tilde{s}\right).
\end{align}
In the hydrodynamic limit of $\omega\tau{}\ll1$, we have $\tilde{s}\rightarrow i\infty$ and
\begin{equation}
\Pi_{00}\left(\tilde{s}\right)=1-\frac{\tilde{s}}{\sqrt{\tilde{s}^{2}-1}}\approx-\frac{1}{2\tilde{s}^{2}}-\frac{3}{8\tilde{s}^{4}}.
\end{equation}
To the first order of $i\omega\tau$, we have
\begin{align}
1-\frac{i\kappa}{\tilde{s}} & =\frac{s}{\tilde{s}},\\
\frac{i\kappa}{\tilde{s}} & =\frac{1}{1-i\omega\tau}\approx1+i\omega\tau.
\end{align}
Thus in the hydrodynamic limit, we obtain
\begin{align}
 & \left(1-\frac{i\kappa}{\tilde{s}}\right)\left(1+\frac{F_{1}^{s}}{2}\right)+\Pi_{00}\left[\left(F_{0}^{s}+\frac{i\kappa}{\tilde{s}}\right)\left(1+\frac{F_{1}^{s}}{2}\right)+2\tilde{s}^{2}\left(1-\frac{i\kappa}{\tilde{s}}\right)\left(\frac{F_{1}^{s}}{2}+\frac{i\kappa}{\tilde{s}}\right)\right]\nonumber \\
\approx & \frac{1}{\tilde{s}^{2}}\left[s^{2}-\frac{1}{2}\left(1+F_{0}^{s}\right)\left(1+\frac{F_{1}^{s}}{2}\right)+\frac{1}{4}i\omega\tau\left(1+\frac{F_{1}^{s}}{2}\right)\right]=0.
\end{align}
The dispersion relation between $\omega$ and $\mathrm{q}$ for first sound in 2D now reads
\begin{equation}
\left(\frac{\omega}{qv_{F}}\right)^{2}=\frac{1}{2}\left(1+F_{0}^{s}\right)\left(1+\frac{F_{1}^{s}}{2}\right)-\frac{1}{4}i\omega\tau\left(1+\frac{F_{1}^{s}}{2}\right).
\end{equation}

\end{widetext}

\bibliography{QSLMode}
\end{document}